\documentclass[paper=a4, fontsize=11pt]{report}
\usepackage{color}
\usepackage{mathtools}
\usepackage{indentfirst}
\usepackage{natbib}
\usepackage{algorithm}
\usepackage{graphicx}
\usepackage{subcaption}
\usepackage{subfloat}
\usepackage{multirow}
\usepackage[T1]{fontenc} % Use 8-bit encoding that has 256 glyphs
\usepackage{fourier} % Use the Adobe Utopia font for the document - comment this line to return to the LaTeX default
\usepackage[english]{babel} % English language/hyphenation
\usepackage{amsmath,amsfonts,amsthm} % Math packages
\usepackage{lipsum} % Used for inserting dummy 'Lorem ipsum' text into the template
\usepackage{sectsty} % Allows customizing section commands
\allsectionsfont{\normalfont\scshape} % Make all sections centered, the default font and small caps
\usepackage{fancyhdr} % Custom headers and footers                                   
\numberwithin{equation}{section} % Number equations within sections (i.e. 1.1, 1.2, 2.1, 2.2 instead of 1, 2, 3, 4)
\numberwithin{figure}{section} % Number figures within sections (i.e. 1.1, 1.2, 2.1, 2.2 instead of 1, 2, 3, 4)
\numberwithin{table}{section} % Number tables within sections (i.e. 1.1, 1.2, 2.1, 2.2 instead of 1, 2, 3, 4)
\newcommand{\bbE}{\ensuremath{\mathbb{E}}}

\newcommand{\norm}[1]{\left\lVert#1\right\rVert}

{%
\begin{list}{#1}{
\vspace{-\topsep}
\vspace{-\partopsep}
\setlength{\itemindent}{0cm}
\setlength{\rightmargin}{0cm}
\setlength{\listparindent}{0cm}
\settowidth{\labelwidth}{#1}
\setlength{\leftmargin}{\labelwidth}
\addtolength{\leftmargin}{\labelsep}
\setlength{\itemsep}{0cm}
}%
}%
{%
\end{list}
\vspace{-\topsep}
\vspace{-\partopsep}
}

{\begin{enumerate}%
}%
{\end{enumerate}}

\newcommand{\KLop}{\operatorname{D_{KL}}}
\newcommand{\KL}[2]{\KLop\left(#1\lVert#2\right)} % Kullback-Leibler divergence
\newcommand{\diagop}{\operatorname{diag}}
\newcommand{\diag}[1]{\ensuremath{\diagop\left(#1\right)}}
\newcommand{\traceop}{\operatorname{tr}}
\newcommand{\trace}[1]{\ensuremath{\traceop\left(#1\right)}}

\theoremstyle{plain}% default

\newtheorem*{prop*}{Proposition}

\theoremstyle{definition}

\newtheorem*{defn*}{Definition}

\newtheorem*{exmp*}{Example}

\newtheorem*{conj*}{Conjecture}

\theoremstyle{remark}

\newtheorem*{rmk*}{Remark}

\DeclareMathOperator*{\argmax}{arg\,max}

\DeclarePairedDelimiter{\floor}{\lfloor}{\rfloor}

\usepackage{xcolor}
\newcommand{\weiran}[1]{\textcolor{blue}{ #1 (Weiran)}}
\newcommand{\karen}[2]{\textcolor{red}{ #2 (Karen)}}
\newcommand{\qingming}[3]{\textcolor{green}{ #3 (Qingming)}}
\renewcommand{\weiran}[1]{}
\renewcommand{\karen}[2]{}
\renewcommand{\qingming}[3]{}

\newcommand{\horrule}[1]{\rule{\linewidth}{#1}} % Create horizontal rule command with 1 argument of height
\title{	
\normalfont \normalsize
\horrule{0.4pt} \\[0.2cm] % Thin top horizontal rule
\huge Speech representation learning:  \\
\large Learning bidirectional encoders with single-view, multi-view, \\
and multi-task methods 
\horrule{2pt} \\[0.05cm] % Thick bottom horizontal rule
}
\author{\small{By} \\ Qingming Tang \\
\\
\\
A thesis submitted \\
in partial fulfillment of the requirements for \\
the degree of \\
\\
Doctor of Philosophy in Computer Science \\
\\
\small{at the}
\\
TOYOTA TECHNOLOGICAL INSTITUTE AT CHICAGO\\
Chicago, Illinois \\
\\
July, 2023 \\
\\
\\
\\
Thesis Committee: \\
Karen Livescu (Thesis Advisor) \\
Kevin Gimpel \\
Sam Wiseman \\
Weiran Wang} % Your name
\date{}
\begin{document}

\maketitle % Print the title

\begin{abstract}
Representation learning refers to the problem of transforming high-dimensional and possibly noisy raw inputs to compact and informative representations of the data. Unlike traditional hand-engineered features, representation learning can automatically capture the underlying structure and patterns hidden in the surface features, which can be useful for improving performance on downstream tasks and improving generalization.

This thesis focuses on representation learning for sequence data over time or space, aiming to improve downstream sequence prediction tasks by using the learned representations. This problem is challenging because the sequences can be very diverse. For example, if the series is a continuous signal like a speech recording, sources of variation include (but are not limited to) linguistic material, different speakers, and distinct recording environments. Supervised learning has been the most dominant approach for training deep neural networks for learning good sequential representations. However, one limiting factor to scale supervised learning is the lack of enough annotated data.

Motivated by this challenge, it is natural to explore representation learning methods that can utilize large amounts of unlabeled and weakly labeled data, as well as an additional data modality. I describe my broad study of representation learning for speech data. Unlike most other works that focus on a single learning setting, this thesis studies multiple settings:  supervised learning with auxiliary losses, unsupervised learning, semi-supervised learning, and multi-view learning. Besides different learning problems, I also explore multiple approaches for representation learning. In this thesis, I focus on speech data, but the methods can also be applied to other domains.

In Chapter 3, I systematically study autoencoders for acoustic representation learning, and I show that speech recognition can benefit from the representations learned by a variational autoencoder (VAE) with a feedforward encoder when the VAE is trained using a much larger unlabeled dataset. To the best of my knowledge, this is the first work to experimentally show that VAEs learn representations that significantly improve speech recognition and outperform autoencoding approaches that do not use variational inference. I also explore in depth what makes VAEs powerful for representation learning. I find that their KL divergence term provides regularization effects which may play a critical role in representation learning.

In Chapter 4, I contribute new techniques for variational representation learning models in a multi-view learning setting. Specifically, I extend variational canonical correlation analysis (VCCA), an unsupervised multi-view representation learning method, with more informative sample-specific priors. My proposed extension of VCCA learns representations that improve speech recognition performance, when we have access to multi-view training data consisting of audio paired with articulatory measurements. I also explore two specific problems in multi-view learning — cross-domain multi-view learning and label embedding. I investigate learning an acoustic encoder using an additional modality (view) when we don’t have access to paired in-domain multi-view data. I explore multiple transfer learning techniques and optimization techniques to exploit the learned acoustic-articulatory mapping in the dataset where acoustic-articulatory pairs are accessible. I show that such acoustic-articulatory mapping information can improve speech recognition, especially with small training datasets. For label embedding, I directly treat the label as a second view and use the structural information hidden in the label to assist the representation learning.

In Chapter 5, I study multitask recurrent representation learning, combining supervised and unsupervised losses, and present several technical improvements to variational sequential models. Unlike other variational sequential models developed prior to my work that focus on generation tasks and typically use unidirectional recurrent layers as their encoders, I propose to use variational bidirectional recurrent layers as an encoder to learn better-performing representations for downstream tasks. I also propose to factor the representation into a discriminative component and a reconstruction-specific component. I also propose to update the priors dynamically, discuss the resulting benefits and limitations in depth, and propose potential extensions. My experimental results show that the proposed multitask representation learning framework works well for speech recognition, and also text tasks like entity recognition and text chunking.

In Chapter 6 I focus on unsupervised sequential representation learning. I compare multiple unsupervised learning approaches, including autoencoding, contrastive learning, and masked reconstruction. I also contribute a new approach, multi-view masked reconstruction. My experimental study shows that making representations invariant to different domains and robust to distortions can improve speech recognition performance and generalization to unseen data. 

Overall, the field of representation learning is developing rapidly. State-of-the-art results on speech related tasks are typically based on Transformers pre-trained with large-scale self-supervised learning, which aims to learn generic representations that can benefit multiple downstream tasks. Since 2020, large-scale pre-training has been the de facto choice to achieve good performance. This delayed thesis does not attempt to summarize and compare with the latest results on speech representation learning; Instead, it presents a unique study on speech representation learning before the Transformer era, that covers multiple learning settings. I believe some of the findings in this thesis can still be useful today, and in Chapter 7 I conclude those findings and discuss potential future work.

\end{abstract}

\chapter*{Acknowledgement}

I would like to thank my thesis advisor Karen Livescu. Words cannot express my gratitude for her endless effort in mentoring me.

I also appreciate the constructive feedback and invaluable encouragement from my thesis advisor and also the thesis committee on helping me finish my thesis. It’s not possible to finish this thesis without your support.

I also appreciate all the collaboration and friendship from TTIC colleagues and my research collaborators, I learned a lot from all of them.

Finally but not the least, I appreciate the strong support from my family. Without their support, it’s not possible for me to complete my Ph.D.

\tableofcontents

%----------------------------------------------------------------------------------------
%	Introduction
%       1. High-level introduction to representation learning and sequence prediction tasks.
%       2. Reconstruction-based representation learning.
%       3. Summary of my contributions.
%----------------------------------------------------------------------------------------
\chapter{Introduction}
\label{cha:introduction}

%----------------------------------------------------------------------------------------
% High-level introduction to representation learning and sequence prediction tasks.
%----------------------------------------------------------------------------------------

Representation learning ~\citep{bengio2013representation, goodfellow2016deep} refers to the problem of transforming high-dimensional and possibly noisy raw inputs to better organized and typically low-dimensional forms (e.g., fixed-length vectorial representations). 
Supervised learning has been the most dominant approach for training deep neural networks for learning good representations.
However, one limiting factor to scale supervised learning is the lack of enough annotated data. \par

Motivated by the challenge, it is natural to explore methods that learn generic information from a large amount of unlabeled data, such that the representation reveals explanatory factors hidden in the original noisy high-dimensional data manifold.
%that is possible to learn generic representations without using labels, or with only limited amount of labels.
Actually, there is a long history of unsupervised representation learning.
For example, 
principal component analysis (PCA) ~\citep{jolliffe2011principal}, 
canonical correlation analysis (CCA)~\citep{hotelling1936relations}, 
and autoencoders ~\citep{hinton2006reducing} are some well-known techniques for obtaining low-dimensional representations from data.
% List some methods Like CPC v1, CMC, MoCo, SimCLR etc for self-supervised learning
More recent works have shown that the generic representations learned from a large amount of data can be a good starting point for different downstream tasks to achieve good performance in multiple application domains.
In natural language processing, BERT ~\citep{devlin2018bert} has achieved great success in enhancing performance on many tasks.
BERT learns contextual representations from large amount of unstructured text, via filling in a masked portion of the input from the available context.
In computer vision, image inpainting ~\citep{Pathak2016ContextEF} is a similar task where a context encoder is trained to fill in the missing part of the image. 
Such pretext tasks, where the labels are directly derived from the data, are widely used in computer vision for self-supervised learning.
Many recent works on self-supervised learning in computer vision either combine clustering with pretext tasks, like ~\citep{Caron2018DeepCF, Li2021PrototypicalCL}, or rely on contrastive learning ~\citep{Becker1992SelforganizingNN, Bromley1993SignatureVU} given data-driven labels, like ~\citep{Chen2020ASF, Chen2020BigSM, Grill2020BootstrapYO}.
\newline
I addition, semi-supervised learning, which only requires a small amount of labels, is another popular trend of research. These sort of methods typically can yield better performance than self-supervised learning for the task at hand.
Pseudo labeling ~\citep{Lee2013PseudoLabelT, Xie2020SelfTrainingWN}, where a trained model is used to infer either soft or hard labels of the large amount of unlabeled data, has been shown to be an efficient way of improving model performance for the target task, though the framework is simple.
Instead of inferring the labels using a trained model, consistency regularization ~\citep{Laine2017TemporalEF, Miyato2019VirtualAT} enforces that an augmented sample should have the same label as the original sample. 
Among many works that employ the idea of consistency regularization, unsupervised data augmentation for consistency training ~\citep{Xie2020UnsupervisedDA} has been shown to be significantly helpful for classification tasks on ImageNet and CIFAR, and sentiment analysis on IMDb. 
One challenge to apply this idea to more tasks relies on how to conduct the consistency training given more complex labels. Recently, this simple yet powerful technique has been extended to automatic speech recognition ~\citep{Chen2020SemisupervisedAB}, natural language processing ~\citep{lowell-etal-2021-unsupervised} and acoustic event classification ~\citep{Zharmagambetov2022ImprovedRL}.

%it is possible to learn from huge amount of data without using label information, such that the representation is robust to reveal explanatory factors hidden in the original noisy high-dimensional data manifold, and can be widely used for different downstream tasks.
% List some methods on semi-supervised representation learning.
% We then say we would focus on both semi-supervised and self-supervised learning, and also more on sequence for this thesis.
%Representation learning can be semi-supervised, 
%where labels of some samples are provided, 
%and we hope the representation helps the specific discriminative task, 
%or it can be unsupervised, 
%where the goal is to discover a generic representation that can be widely applicable to many tasks. 
%For example, 
%principal component analysis (PCA) ~\citep{jolliffe2011principal}, 
%canonical correlation analysis (CCA)~\citep{hotelling1936relations}, 
%and autoencoders ~\citep{hinton2006reducing} 
%are widely used approaches for generic representation learning without labels. 
%Besides learning representations in an unsupervised manner,
%it is also possible to discover good representation with a limited amount of labeled data and a large amount of unlabeled data.
%For example, we can also use the examples of generic representation learning we just mentioned to combine discriminative learning objectives to achieve semi-supervised representation learning. \par

This thesis focuses on representation learning for sequence data over time or space,
aiming to improve downstream sequence prediction tasks by using the learned representations.
This problem is challenging because the sequences can be very diverse.
For example, 
if the series is a continuous signal like a speech recording, 
sources of variation include (but are not limited to) linguistic material, different speakers, and distinct recording environments.
If the sequence is discrete like text, it can comprise a vast vocabulary of input symbols. 
Besides the variations in sequence data, the diverse \textit{context and pattern} further complicate the problem of sequence representation learning.
For example, some sequence data (e.g., human speech) may exhibit a smoother structure than others (e.g., cardiogram). 
Some data (e.g., smoke alarm sound) may have a clear periodic pattern while others (e.g., glass breaking sound) do not. 
Such diverse structural information is hidden in the sequence data, making it challenging to design universally applicable representation learning algorithms. 
In this thesis, we focus on speech data, and also show applications of our work to text data. 
\par

We are interested in developing an intermediate variable-length representation of surface features and capturing the information needed for downstream sequence prediction tasks.
If some amount of supervision from downstream tasks is available, 
we also expect to filter out information unrelated to the downstream tasks from the learned representation. 
In this thesis, we consider four settings: 
1) Representation learning with supervision from downstream tasks,
specifically joint training of a representation model and the relevant tasks in a multitask learning framework,
2) Semi-supervised representation learning, 
where we have extra (typically a large amount of) unlabeled sequences not present in the multitask learning scenario,
3) Unsupervised representation learning, where we do not use labels of downstream tasks when training the feature extractors.
In the third setting, the feature extractor is either fixed or fine-tuned when learning models for the downstream tasks with label information,
and 4) Multi-view representation learning, where in addition to the data we are targeting, we also have access to aligned data from another ``view" (e.g. another measurement modality). 
In addition, we consider both feedforward and recurrent neural models.  One feature that sets this work apart from much other work on representation learning is that we explore these multiple settings, whereas much of the existing work tends to focus on a single setting (say, recurrent unsupervised models).
\par

We are interested in computing a representation sequence $z_{1:T}$ that can be used as input for downstream sequence prediction tasks, 
given an input sequence $x_{1:T}$. 
Due to the benefits of multitask learning or training on a large amount of unlabeled samples,
we expect that making predictions based on the acquired representation $z_{1:T}$ will be 
more accurate than making predictions based directly on the raw input $x_{1:T}$. 

%\begin{figure*}[htbp]
%  \centering
%  \includegraphics[width=0.9\textwidth]{./Figures/Framework.eps}
%  \caption{High-level illustration of the learning structure for our representation learning approaches. Basically, time-dependent latent representations $z_{1:T}$ are extracted for each index of the input sequence (e.g., $z_{t-1:t+1}$ for $x_{t-1:t+1}$). Each latent representation is encouraged to reconstruct a time-dependent target $u_t$. Table ~\ref{tab:reconstruction-target} lists a few possible designs of the time-dependent targets.}
%  \label{fig:framework}
%\end{figure*}

%As illustrated in Figure ~\ref{fig:framework},
Much of our work focuses on reconstruction-based approaches.  
We first learn a compressed (e.g., lower-dimensional) time-dependent latent representation $z_t=f(x_{1:T},t)$,
a function of the input sequence and time step $t$,
and use $z_t$ to reconstruct its target $u_t$ (e.g., $x_t$ or $x_{t-1:t+1}$).
For example, 
one approach for representation learning is autoencoding, that is encoding and then reconstructing the input; in that framework, $u_t$ is the input itself.
We use variational inference (~\citep{kingma2013auto,rezende2014stochastic}) to learn a posterior distribution of $z_t$, \weiran{the introduction of ``variational inference'' is a bit sudden'', again providing the context may help.}
that is $q(z_t|x_{1:T},t)$,
and reconstruct the target $u_t$ based on $z_t$. \par

%The reason that reconstructing $u_t$ can encourage $z_t$ to preserve key information of $u_t$ is explained
%in ~\citep{vincent2008extracting, vincent2010stacked, hjelm2018learning} --
%accurately reconstructing $u_t$ given $z_t$ amounts to maximizing a lower bound on the mutual information between $U$ and $Z$
%(where $u_t$ and $z_t$ are samples of variable $U$ and variable $Z$).
%The authors of ~\citep{alemi2016deep} further show that
%maximizing Evidence Lower Bound (ELBO) of Variational Auto Encoder (with the weight of Kullback–Leibler divergence term being $\beta$) 
%amounts to maximizing a lower bound on the mutual information between variables $U$ and $Z$ w.r.t. some tradeoff. 
%Our experiments in Chapter ~\ref{cha:feedforward} indicate that VAE outperforms its non-variational counterparts in learning compressed codes for speech recognition tasks, 
%presumably due to the regularization effect of the Kullback–Leibler divergence (KL divergence) term. \par
Possible choices of $u_t$ are listed in Table ~\ref{tab:reconstruction-target}.
When choosing $u_t$ to be exactly $x_t$,
we are encouraging $z_t$ to capture very local information --
key information needed to reconstruct a single frame $x_t$, while requesting $z_t$ to reconstruct a (weighted) window as shown in the table 
would encourage $z_t$ to capture contextual information. 
\par

\begin{table}[t]
 \begin{center}
  \begin{tabular}{|c|}
   \hline
   $u_t=x_t$ \\
   \hline
   $u_t=x_{t+1}$ \\
   \hline
   $u_t=x_{t-1}$ \\
   \hline
   $u_t= \left[ x_{t-K:t+K} \right]$ \\
   \hline
   $u_t= \frac{1}{2K+1} \left\{\sum_{j=t-K}^{t+K} x_j \right\}$ \\
   \hline
   $u_t = \left[ \alpha_j x_j \right]_{t-K \leq j \leq t+K} $ s.t. $\sum_{j=t-K}^{t+K} \alpha_j =1 $ and $\alpha_j \geq 0 $ for each $j$ \\
   \hline
   $u_t = x_j $ with probability $p_j$, and $\sum_{j=1}^T p_j = 1$ \\
   \hline
  \end{tabular}
  \caption{Summary of different choices of reconstruction targets $u_t$.}
  \label{tab:reconstruction-target}
 \end{center}
\end{table}

Experimental study suggests that our variational representation learning approaches following this paradigm help learn better-performing representations for a few tasks (e.g., speech recognition, named entity recognition, and text chunking) 
when we jointly optimize the unsupervised and supervised objectives.
%In the unsupervised learning scenario,
%if we use a stacked bidirectional long short term memory network (LSTM ~\citep{hochreiter1997long}) as the encoder to summarize the input sequence,
%and parameterize $z_t$ as a function of the $t^{\text{th}}$ hidden state of the stacked bidirectional LSTM,
%the performance of speech recognizers with the learned representation sequence $z$ as input does not improve over speech recognizers with $x_{1:T}$ as input.
%However, a stacked feedforward neural network taking contiguous frames (e.g., $x_{t-K:t+K}$) as inputs can help learn acoustic representations that could help improve the performance of speech recognition.
One of our findings is that, while variational reconstruction-based methods work well for learning unsupervised feedforward representations, they are less successful for recurrent representations.  For this reason we study several other approaches for recurrent representation learning.
We hypothesize that when a recurrent neural network is used, the full context of the input utterance is available and thus the reconstruction task becomes much easier, thus preventing learning meaningful representations. 
In addition to studying different reconstruction targets, we study a variety of other aspects of variational representation learning models, and propose improvements to them in several learning settings.
\par

Expanding upon our hypothesis and also the success of unsupervised pre-training algorithms across different domains ~\citep{schneider2019wav2vec, devlin2018bert, baevski2020vqwav2vec, baevski2020wav2vec}, we further experiment with some recent unsupervised learning techniques and their extensions.
These methods either try to learn the future contents based on the seen part of the input sequence or learn masked content that is not part of the input to the encoder.
Compared with a bidirectional encoder that has seen the entire sequence and tries to reconstruct all time steps, the learning tasks of these methods are more difficult.
We investigated contrastive predictive coding (CPC, ~\citep{oord2018representation}), masked reconstruction ~\citep{wang2020unsupervised}, masked reconstruction combined with CPC and multi-view masked reconstruction. 
We found that all these techniques improve over baseline recognizers when the size of the unlabeled training data is much larger than the size of labeled training data. \par

\section{Contributions}
\label{sec:intro_contributions}

We summarize our contributions:
%----------------------------------------------------------------------------------------
%    We then talk about the contributions
%----------------------------------------------------------------------------------------
\begin{itemize}
\item[1] 
\textbf{We show variational inference can learn useful representations for several downstream tasks}: 
We systematically study reconstruction-based sequence representation learning and show it is helpful for several sequence labeling and sequence transduction tasks in speech and NLP.
In Chapter ~\ref{cha:feedforward}, 
we show that speech recognition can benefit from the representations learned by VAE (with a feedforward encoder) when the VAE is trained using a much larger unlabeled dataset.
%To the best of our knowledge,
%we are the first to experimentally show that VAEs learn representations that significantly improve speech recognition
%and outperform auto-encoding approaches that do not use variational inference.
In Chapter ~\ref{cha:recurrent}, we extend variational representation learning to recurrent models, and to a multitask learning setting, and show that 
%show that it is possible to use the representations learned by a recurrent variational model to improve the performance of the recognizer of a sequence prediction task through multitask learning.
%Specifically, we show 
variational representation learning can consistently improve the accuracy of speech recognition, named entity recognition, and chunking tasks.

\item[2] 
We contribute several technical improvements to variational sequential models:
%We further enhance variational sequential representation learning in multitask learning scenarios in Chapter ~\ref{cha:recurrent}.
\begin{itemize}
\item[1] \textbf{Bidirectional encoders}: 
Unlike other variational sequential models developed prior to our work ~\citep{chung2015recurrent, fraccaro2016sequential} that focus on generation tasks and typically use unidirectional recurrent layers as their encoders,
\textbf{we propose to use variational bidirectional recurrent layers as encoder to learn better-performing representations for downstream tasks of interest}.
\item[2] \textbf{Learning better representations using heuristic priors}: 
We show that using heuristic priors, e.g., using a posterior of $z_t$ learned in earlier epochs as a prior in later epochs, can provide more informative guidance and assist both feedforward and recurrent encoders in learning better representations for downstream discriminative tasks.
\item[3] \textbf{Factoring label-related and reconstruction-specific information}: 
We factorize the discriminative task-specific and reconstruction task-specific information by introducing one auxiliary latent variable used only for reconstruction tasks. We show such a design encourages the main latent variable to capture more discriminative information w.r.t. the supervised tasks.
\item[4] \textbf{Reconstructing contiguous time steps}: Our vanilla model reconstructs all time steps of an input sequence independently given hidden states of the bidirectional recurrent encoder. 
We explore reconstructing more context-aware targets, like a few contiguous weighted frames, rather than a single frame $x_t$ given the bidirectional recurrent encoder hidden state $h_t$.
\item[5] \textbf{Sharing only low-level representations}: We find that when the supervised loss and the reconstruction loss favor different types of representations, using an encoder fully shared by both supervised and unsupervised tasks has negative impact on the performance of the supervised tasks;
thus, it is preferred to share only the low-level representation layers among the different tasks, and still have private layers for each task.
%it is not good to use a big encoder fully shared by the supervised and unsupervised tasks.
%Instead, it is preferred only to share the low-level representation layers and have private layers for the two parts.
\end{itemize}

\item[3] 
We contribute new techniques for variational representation learning models in a multi-view learning setting (shown in Chapter ~\ref{cha:multiview}). 
Specifically, \textbf{We extend variational canonical correlation analysis} (VCCA ~\citep{wang2016deep}), 
an unsupervised multi-view representation learning method, 
with more informative sample-specific priors.
Our experimental study indicates that the extension of VCCA can learn better representations in terms of speech recognition.
We also explore two specific problems in multi-view learning --- \textbf{cross-domain multi-view learning} and \textbf{label embedding}.
We study transferring the second modality information in the source domain to a target domain without its second modality for cross-domain multi-view learning.
For label embedding, 
we directly treat the label as a second view and use the structural information hidden in the label to assist the representation learning.

\item[4]
We study masked reconstruction for self-supervised learning in Chapter ~\ref{cha:semi-and-pre-training}, and contribute a new approach, \textbf{multi-view masked reconstruction}.
%We study masked reconstruction for self-supervised learning in Chapter ~\ref{cha:semi-and-pre-training}, and contribute \textbf{multi-view masked reconstruction}.
%We find that it is difficult to learn representations that would benefit downstream tasks using the same variational recurrent model we used in Chapter ~\ref{cha:recurrent}. 
%We describe some hypotheses on this observation, and then move towards unsupervised representation learning via predicting unseen context rather than reconstructing what we have seen.
%We experiment with CPC and masked reconstructions, and find both approaches learn helpful information for the downstream speech recognition task.
%However, we find that simply combining CPC with masked reconstruction does not further improve upon the masked reconstruction approach, but multi-view masked reconstruction does consistently (though not significantly) outperform masked reconstruction for speech recognition tasks.
%In multi-view masked reconstruction, we encourage the consistency between the representations of two corrupted inputs (with different masks applied). More specifically, besides performing masked reconstruction, the contextual representation of each corrupted input is also enforced to be more predictive to the latent representations of the other view via using InfoNCE loss.
We also find using a linear adaptation layer is one simple but powerful technique to address the potential domain mismatch between unlabeled datasets (used for pre-training) and labeled datasets.

\end{itemize}

%----------------------------------------------------------------------------------------
%  Introduce the structure of the thesis
%----------------------------------------------------------------------------------------

The remainder of the thesis is structured as follows.
Chapter ~\ref{cha:related} provides a quick summary of work on representation learning based on generative modeling, 
discusses a mutual information maximization view of representation learning and lists related work on representation learning for speech processing.
Chapter ~\ref{cha:feedforward} and ~\ref{cha:multiview} describe our feedforward models for single-view and multi-view representation learning work, respectively.
We then describe our reconstruction-based representation learning using RNN encoders in multitask learning scenarios in Chapter ~\ref{cha:recurrent}, and in unsupervised and semi-supervised setting in Chapter ~\ref{cha:semi-and-pre-training}.
In Chapter ~\ref{cha:semi-and-pre-training}, we demonstrate the difficulty of performing unsupervised sequential representation learning via maximizing ELBO (especially when a RNN encoder is used). We show that more recent pre-training techniques like CPC and masked reconstruction can learn representations more useful for speech recognition.
We extend the current masked reconstruction pre-training approach in multi-view-learning scenario and show it improves upon masked reconstruction for speech recognition tasks.
In Chapter ~\ref{cha:future}, we conclude the thesis and propose a few future research directions.

%----------------------------------------------------------------------------------------
%	Related Work
%       1. (Variations of) auto-encoders
%       2. VAEs and its variations, and their applications to ``representation learning''
%       3. Why VAE is useful for representation learning? What is the problem of VAE?
%       List some studies towards better understanding VAE.
%       4. Sequential extensions of VAE (e.g. VRNNs)
%       5. ELMo, Open-GPT and Bert.
%       6. A line of works that can be unified under the umbrella of MI (e.g. MINE, CPC and DIM)
%.      7. More recen
%----------------------------------------------------------------------------------------
\chapter{Background and Related Work}
\label{cha:related}

%----------------------------------------------------------------------------------------
% Related Work
% 1. One common way to learn representation
%---------------------------------------------------------------------------------------

\weiran{I could understand the flow of thoughts, but I feel the writing/organization can be more ``logical''. One possibility is to break this section into a few subsections with the overal theme being VAE and Mutual information maximization. Your current order is mostly fine: starting with AE, VAE, then mutual information view, then contrastive learning CPC which maximizes mutual information. May also help to hint which of these approaches you will be pursuing along the way.}

Using large amount of data is vital for the success of representation learning.
One way to utilize a large amount of (unlabeled) data $\mathcal{X}$ is to learn a mapping for each $x \in \mathcal{X}$ to a distribution $q_{\phi}(z|x)$ that captures the key information in $x$ in latent variable $z$.
Samples of $ q_{\phi}(z|x)$ or $\mathbb{E}_{q_{\phi}(z|x)} (z)$ can be used as a representation for $x$.
For example, when assuming the posterior $q$ to be a Gaussian distribution, the mean typically would be used as the representation of $x$ for downstream tasks if only one sample can be used.
The mapping can be learned by maximizing the objective:
\weiran{why this objective?? which more general principle is this derived from? consider reorganizing the order of equations? also consider adding a regularization term.}
\begin{equation}
\mathbb{E}_{ q_{\phi}(z|x) } \left\{ \log{ p_{\theta}(x|z) }  \right\}
\label{eqn:learn-representation}
\end{equation}

%According to literature, 
$q$ can be a Bernoulli distribution ~\citep{Bengio2013EstimatingOP}, 
a categorical distribution ~\citep{jang2016categorical, maddison2016concrete, van2017neural}, 
a Gaussian distribution ~\citep{kingma2013auto}, 
a Gaussian mixture ~\citep{makhzani2015adversarial}, 
a Markov chain ~\citep{Salimans2015MarkovCM} or other flexible multimodal distributions ~\citep{rezende2015variational}.
The Gaussian posterior is used most widely due to its simplicity. \par
%\weiran{Provide some examples here for the ``distribution'' here. E.g., if the distribution is Gaussian, $q$ would parameterize the mean and variance.}

%----------------------------------------------------------------------------------------
% Related Work
% 2. AE and DAE
%---------------------------------------------------------------------------------------
When $q_{\phi}(z|x) = \delta \left( z=f_{\phi}(x) \right) $ (where $\delta$ denotes the Dirac-delta distribution and $f_{\phi}$ is a non-linear mapping, e.g. multiple layers of feedforward neural networks, parameterized by $\phi$), the mapping becomes deterministic.
For example, traditional autoencoders (AE) ~\citep{hinton2006reducing} learn a deterministic mapping for each sample $x$ and reconstruct $x$ given the mapping.
By compressing $x$ to a low-dimensional code while still being able to accurately reconstruct $x$, an AE learns a lossy representation that retains enough information to recover $x$.
However, in ~\citep{vincent2008extracting,vincent2010stacked}, the authors point out that merely retaining sufficient information for reconstructing $x$ is not sufficient to learn a robust representation that can generalize well.
AE/DAE and their variations (e.g. sparse AE ~\citep{ng2011sparse}, contractive AE ~\citep{rifai2011contractive}, higher order contractive AE ~\citep{rifai2011higher} and stacked AE ~\citep{vincent2010stacked, masci2011stacked})
have been widely applied in domains like recommendation systems ~\citep{sedhain2015autorec}, speech processing ~\citep{lu2013speech,feng2014speech} and natural language processing ~\citep{liou2014autoencoder}. \par

%----------------------------------------------------------------------------------------
% Related Work
% 3. VAE
%---------------------------------------------------------------------------------------

\begin{figure*}[htbp]
  \centering
  \includegraphics[width=0.9\textwidth]{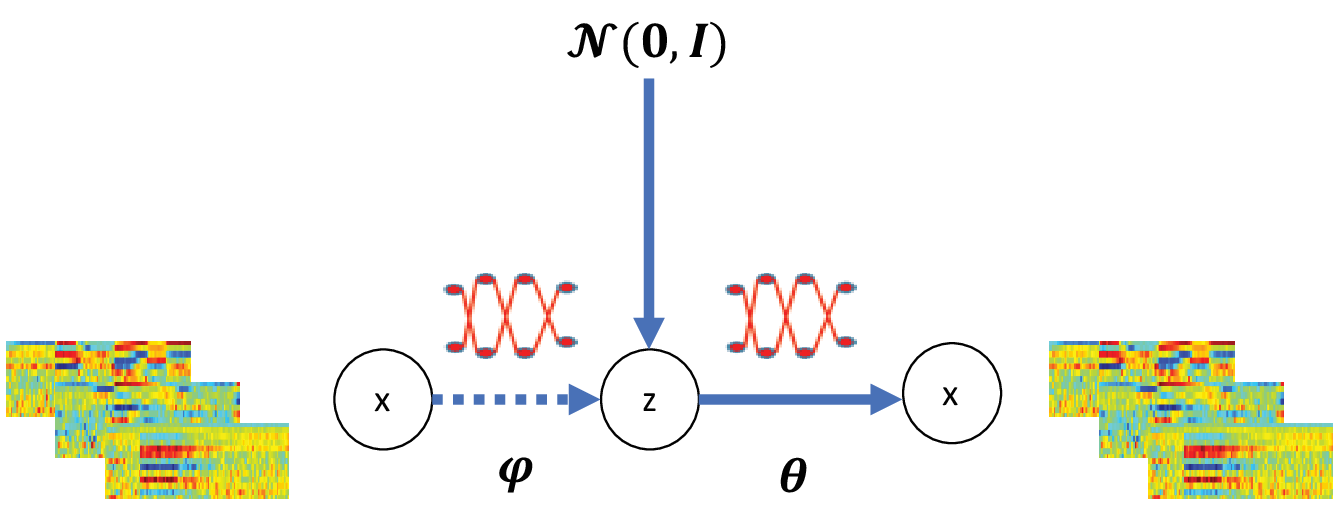}
  \caption{Illustration of a variational autoencoder (VAE),
with the input being acoustic measurements.
A VAE consists of an inference network $q_{\phi}(z|x)$ and a generation
network $p_{\theta}(x|z)$.
We typically use the mean of $q_{\phi}(z|x)$ as a representation of $x$ for downstream tasks. }
\label{fig:vae}
\end{figure*}

We can also consider representation learning from a Bayesian perspective.
Consider the joint density of the latent variable $z$ and observation $x \in \mathcal{X}$:

\begin{equation}
p_{\theta}(x,z) = p_{\theta}(z)p_{\theta}(x|z)
\label{eqn:joint-density}
\end{equation}
with $p_{\theta}(z)=\mathcal{N}(0,I)$ being a common choice, and whose marginal distribution is:

\begin{equation}
p_{\theta}(x) = \int_z p_{\theta}(x,z) dz
\label{eqn:marginal}
\end{equation}

The representation $p_{\theta}(z|x)$ can be calculated via $p_{\theta}(z|x)= \frac{p_{\theta}(z) p_{\theta}(x|z)}{p_{\theta}(x)}$,
which is typically intractable 
\footnote{There is a line of research designing flow-based generation networks, which can implicitly obtain $p_{\theta}(z|x)$ via a sequence of invertible transformations, e.g. \citep{dinh2014nice,dinh2016density,kingma2018glow}, such that the posterior is actually tractable. However, the representation needs to have the same dimension as the high-dimensional input.}
as calculating Equation ~\eqref{eqn:marginal} is intractable when a deep neural network is used.

A variational autoencoder (VAE ~\citep{kingma2013auto} and ~\citep{rezende2014stochastic}, 
as shown in Figure ~\ref{fig:vae}) approximates the intractable posterior using an inference network parameterized by $\phi$.
VAEs convert the posterior approximation problem into an optimization problem. \par

More specifically, a VAE learns inference network parameters $\phi$ and generative model parameters $\theta$ jointly by maximizing the Evidence Lower Bound (ELBO) (for derivation, please see Equation ~\eqref{eqn:elbo-derive} in the Appendix):
\begin{eqnarray}
\mathbb{E}_{q_{\phi}(z|x)} \big\{ \log p_{\theta}(x|z) \big\} - \KL{q_{\phi}(z|x)}{p_{\theta}(z)}
\label{eqn:elbo}
\end{eqnarray}

Equation ~\eqref{eqn:elbo} intends to maintain a small divergence between posterior and prior,
while reconstructing well the observations (using samples of posterior).
The posterior distribution is typically diagonal Gaussian in order to accelerate computation (i.e., the computation on each dimension decorrelates with others).
To approximate the expectation of the log likelihood with respect to the approximate posterior in Equation ~\eqref{eqn:elbo},
Monte Carlo sampling is used.
Sampling from $q_{\phi}(z|x)$ is performed via the ``reparameterization trick'',
that is, drawing samples $\delta$ from $\mathcal{N}(0,I)$ and then computing samples of the posterior as $\mu(x) + \delta \odot \sigma(x)$,
allowing the gradient w.r.t. $(\theta, \phi)$ to be computed easily by automatic differentiation.
\par

%----------------------------------------------------------------------------------------
% Related Work
% 4. Interpretation of VAE
%---------------------------------------------------------------------------------------

There have been many studies aiming to better understand ELBO and to improve VAE ~\citep{he2019lagging, razavi2019preventing, van2017neural, razavi2019generating, shu2018amortized, rainforth2018tighter,alemi2017fixing,zhao2017towards,serban2016piecewise,hoffman2016elbo}. 
~\citep{zhao2017towards, razavi2019preventing, he2019lagging} focus on alleviating the ``posterior collapse'' issue of VAE (i.e., one major issue during the VAE training which would be discussed in later chapters),
%\weiran{very briefly describe what is posterior collapsing? and mention this is discussed in later chapters?}, 
~\citep{van2017neural,razavi2019generating} circumvent the posterior collapse issue by quantizing the learned representations, 
~\citep{hoffman2016elbo} demonstrates that a good prior would further improve ELBO and ~\citep{serban2016piecewise} designs a multimodal prior to match the multiple modes in the data,
while ~\citep{shu2018amortized,rainforth2018tighter,alemi2017fixing} describe the potential issue of maximizing ELBO and show that tighter ELBO does not guarantee better generation and generalization. \par

%----------------------------------------------------------------------------------------
% Related Work
% 5. Unifying perspective
%---------------------------------------------------------------------------------------

All AEs, DAEs and VAEs learn representations by reconstructing the input.
~\citep{vincent2008extracting,vincent2010stacked,hjelm2018learning} motivate 
why we can learn representations by reconstructing the input samples
by demonstrating that  reconstructing an input sample $x$ based on a learned representation $z$ amounts to maximizing one lower bound of mutual information $I(X ; Z)$.
In ~\citep{alemi2016deep}, 
the authors show that the ELBO of a VAE (Equation ~\eqref{eqn:elbo}) is also a lower bound of the below regularized mutual information
\begin{equation}
I(X;Z) - \beta I(Z;i)
\label{eqn:bottleneck}
\end{equation}
when $\beta=1$, where $i$ represents the identity of sample $x$ and $\beta$ is nonnegative.
Note, ``identity" is not the ``label" of $x$. It can simply be the index of $x$ in the training sample list.
%\weiran{Is ``identity'' the best description? It may confuse with the ``label'' of $x$. Try to improve this explanation.}
Thus, all these of ``autoencoding'' approaches can be unified under the umbrella of mutual information, while VAE learns information with (a tunable level of) regularization.
In fact, recent empirical study ~\citep{bowman2015generating} shows that putting a monotonically increasing weight ($\beta$) on the KL divergence term of Equation ~\eqref{eqn:elbo} can in practice encourage VAEs to learn high-level information such as the style, topic, and synthetic features of a sentence.
%\weiran{explain ``better'', I think this reference aims at learning more disentangled representation} 
\par

%----------------------------------------------------------------------------------------
% Related Work
% 6. VRNNs
%---------------------------------------------------------------------------------------

VAEs have also been extended to better model sequence data in a number of ways. 
~\citep{fabius2014variational} learn a single representation for the entire sequence, 
while ~\citep{hsu2017unsupervised} learn both a whole-sequence representation and a set of representations for pre-defined segments of the given sequence. 
For many tasks, 
such as the ones we consider here, 
it is desirable to represent a length-$T$ input sequence $x_{1:T}$ with a corresponding length-$T$ latent sequence $z_{1:T}$ so as to fit directly into typical recurrent network-based prediction models.
Several recent approaches fit this criterion ~\citep{Krishnan2015DeepKF, archer2015black, chung2015recurrent, fraccaro2016sequential, goyal2017z, chen2018variational}.
Learning is again done by maximizing an ELBO, 
with the main differences among approaches being the specific forms of the prior $p(z_{1:T})$ (typically parameterized so as to capture dynamics in the latent space), 
the generation distribution $p_{\theta}(x_{1:T}|z_{1:T})$, 
and the approximate posterior $q_{\phi}(z_{1:T}|x_{1:T})$. 
For example, direct recurrent connections between stochastic variables~\citep{fraccaro2016sequential}, 
or indirect recurrent connections, 
%\weiran{$h_t$ needs to explained to avoid confusion, perhaps with a figure}
e.g.~$h_{t-1} \rightarrow z_{t-1} \rightarrow h_{t} \rightarrow z_{t}$~\citep{chung2015recurrent, goyal2017z} where $h_t$ is the $t^{\text{th}}$ hidden state, 
are often introduced in $p(z_{1:T})$ to model the dependence between neighboring latent variables.  
While this is more powerful than a simpler prior for the purpose of generation, 
it poses challenges for designing the approximate posteriors due to the dependencies among $z_t$'s. 
On the other hand, given $z_{1:T}$, the generation model is often fully factorized into $T$ independent terms: $p_{\theta}(x_{1:T}|z_{1:T}) = \prod_{t=1}^T p_{\theta}(x_t|z_t)$. 
Most prior work on variational recurrent models has focused on generation quality and likelihood evaluation, or more recently,  some variational models focus on learning disentangled static and dynamic representations, including ~\citep{hsu2017unsupervised, Bai2021ContrastivelyDS, Li2018DisentangledSA}.
It is not clear, however, that the learned representations are useful for downstream tasks. \par

%----------------------------------------------------------------------------------------
% Related Work
% 7. Open GPT/EMLO/Pre-trained language model
%---------------------------------------------------------------------------------------
%\weiran{Instead of saying ``specific to NLP'', may draw a bit connection between NLP and speech since both deal with sequence data. Otherwise ``specific'' sounds like irrelevant work that does not apply to speech.}
Pre-training general language models are also a form for learning representations.
%GPT-1/2
~\citep{radford2018improving, radford2019language} show that natural language understanding (comprising a wide range of very diverse tasks like document classification, named entity recognition, semantic similarity assessment and question answering) can be greatly improved via generative pre-training of a language model on unlabeled text from a diverse corpus. 
%ELMO
Building upon unidirectional language models,
~\citep{peters2018deep} trains both deep forward and backward models to learn forward and backward language models respectively,
and remembers the contextual embedding, the concatenation of forward and backward representations, of each word. 
The contextual embedding forms the input for downstream tasks. 
The model is called Embeddings from Language Models (or ELMo), and the authors show that these bidirectional contextual representations can significantly improve state-of-the-art results across a few challenging NLP problems. \par
%APC
Unlike in natural language processing where the input is discrete, 
the input for speech/audio processing is typically continuous.
The authors of ~\citep{Chung2019AnUA} propose an unsupervised autoregressive model called autoregressive predictive coding (APC) for learning a speech representation via predicting the spectrum of a future frame rather than the wave form. 
This idea is largely motivated by the aforementioned large-scale pre-training methods.
%Decora
~\citep{Ling2020DeepCA} also pursues APC style training,
but it learns bidirectional contextual representation by training both forward and backward APC,
then uses both pre-trained forward and backward layers to initialize bidirectional encoders for downstream tasks.
\par

%----------------------------------------------------------------------------------------
% Related Work
% 8. Language modeling vs bidirectional model and BERT
%---------------------------------------------------------------------------------------

However, all the aforementioned pre-training models and variational sequential models are either unidirectional or ELMo style ~\citep{peters2018deep} bidirectional (e.g. forward and backward encoders are trained separately).
~\citep{devlin2018bert} proposes a bidirectional model for learning masked language models named ``BERT'',
where the input is the sentence with randomly selected words masked,
and the deep bidirectional model tries to predict the masked words. \par

To apply BERT to a continuous signals like speech, 
~\citep{wang2020unsupervised} have designed masked reconstruction bidirectional encoders, where the continuous speech signal is masked in both time and frequency domain and then the masked region is reconstructed;
~\citep{baevski2020vqwav2vec} quantize the representations to a sequence of discrete codes and then further use masked language modeling for learning.
Our VAE-based sequential representation learning approach is also natural to deep bidirectional encoding, which is similar to BERT-style pre-training.
It is different from BERT in that the input to our model is not corrupted,
and we learn the contextual representation by reconstructing different forms of context (as listed in Table ~\ref{tab:reconstruction-target}). \par

%----------------------------------------------------------------------------------------
% Related Work
% 9. Other more recent works like MINE, CPC and DIM
% Need to mention wav2vec, vq-wav2vec
%---------------------------------------------------------------------------------------

There is another line of contextual representation learning approaches that are 
under the umbrella of maximizing mutual information,
but are not reconstruction-based.
There is a long history for this kind of algorithms, e.g., the infomax principle ~\citep{linsker1988self} that maximizes the average shannon mutual information between input and output,
and infomax based independent component analysis (ICA) approaches ~\citep{bell1995information, nadal1999sensory,hyvarinen1999nonlinear,almeida2004linear}.
However, most of these approaches are difficult to adapt to deep learning, 
as deep neural estimation of mutual information is not easy.
Recently, ~\citep{belghazi2018mine} proposes a deep neural framework to effectively calculate the mutual information of continuous variables.
The resulting estimation is strongly consistent with the true MI, 
and can be easily combined with a bidirectional architecture. \par

Parallel to ~\citep{belghazi2018mine}, 
contrastive predictive coding (CPC, ~\citep{oord2018representation}) is proposed.
This approach uses probabilistic contrastive loss to induce contextual representations that are predictive of future time steps in latent space.
This amounts to maximizing a lower bound on mutual information between the contextual representation and future time steps.
%\weiran{I think ~\citet{oord2018representation} is published/submitted to arxiv before \citet{hjelm2018learning}, and their motivations are quite similar.}
Motivated by ~\citep{belghazi2018mine, oord2018representation}, deep infomax (DIM ~\citep{hjelm2018learning}) is also proposed and shows consistently good performance on different data sets and different tasks in the vision domain.
In fact, DIM shares some motivations and computations with CPC,
although there are some design and implementation differences.
DIM also explores MI estimators based on different divergences,
including a KL divergence based estimator (~\citep{belghazi2018mine}), 
Jensen-Shannon estimator (~\citep{Nowozin2016fGANTG}),
and noisy-contrastive estimator (used in CPC as a bound on MI).\par

Wav2vec ~\citep{schneider2019wav2vec} is another recently proposed work whose motivation and computations are similar to those of CPC.
Compared with CPC,
wav2vec more thoroughly shows that a simple multi-layer pre-trained CNN can improve upon a strong character-based log-mel filterbank baseline by a large margin.
Further, the authors combined wav2vec with BERT,
resulting in the aforementioned vq-wav2vec ~\citep{baevski2020vqwav2vec}. 
\weiran{since it is aforementioned, can you mention wav2vec a }
\par

%wav2vec and vq-wav2vec

%From this perspective, 
%the main loss and major architecture of deep infomax turn out to be very similar to that of CPC.
%However, by 
%Prior to infomax,
%contrastive predictive coding (Cpc, ~\citep{oord2018representation}) has been 
%and is also a special case of deep infomax, but with very different motivation.

%----------------------------------------------------------------------------------------
% Related Work
% 10. Transformers and state of the art algorithms
% 1. Transformer
% 2. wav2vec 2
% 3. Decora 2
% 4. GPT 3
%---------------------------------------------------------------------------------------
%Transformer
%Pre-training general language model is also heavily used to learn representations.
Besides the aforementioned works,\weiran{I would say transformer is an important progress on neural architecture for sequence modeling. However, you are not using it in the thesis, you could mention why.}
one of the most important advancements in representation learning is the transformer model ~\citep{vaswani2017attention},
an attention-based CNN architecture that can be parallelized more effectively than previous architectures when processing sequence data.
Almost all the state-of-the-art sequence representation learning models are based on transformer,
such as the latest implementation of BERT, 
wav2vec 2.0 ~\citep{baevski2020wav2vec} (the latest version of \citep{schneider2019wav2vec}),
DeCoAR 2.0 ~\citep{Ling2020DeCoAR2D} (the latest version of \citep{Ling2020DeepCA})
and GPT 3 ~\citep{Brown2020LanguageMA} (the latest version of \citep{radford2018improving}).

\chapter{Feedforward Models for Representation Learning}
\label{cha:feedforward}

\begin{figure*}[htbp]
  \centering
  \includegraphics[width=0.9\textwidth]{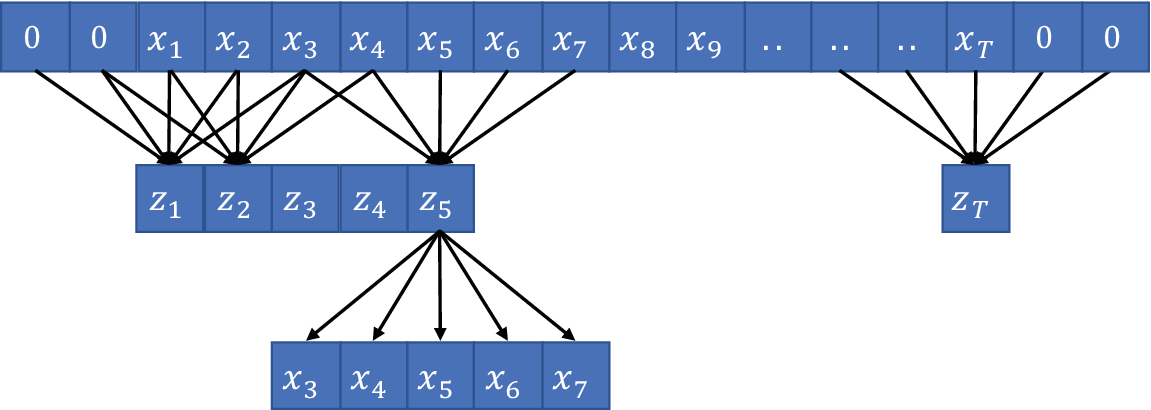}
  \caption{Illustration of feedforward representation learning with context window size $W=5$.
For every $5$ consecutive time steps, we use a feedforward neural network to infer a vector ($z$ in the figure) as a contextual representation of the central time step,
and reconstruct the $5$ consecutive time steps provided the representation.}
  \label{fig:feedforward}
\end{figure*}

In this chapter, we use feedforward neural networks to learn representations for a given sequence $x_{1:T}$. 
We consider learning a contextual representation of each $x_t$ 
by compressing $x_t$ and its neighborhood, 
i.e. a window $\left\{ x_{t-K},\cdots,x_t,\cdots,x_{t+K} \right\}$ centered at $t$,
and reconstructing the window based on the compressed code.
The compressed code would be the desired representation of $x_t$.
The learning procedure is illustrated in Figure ~\ref{fig:feedforward}. \par

~\citep{alain2014regularized} show that regularized autoencoders ~\citep{vincent2008extracting,vincent2010stacked,makhzani2015adversarial} can implicitly learn data-generating density given a large number of samples and enough model capacity.
We are interested in whether autoencoder-type generative models,
in particular variational autoencoders (VAE) ~\citep{kingma2013auto, rezende2014stochastic, doersch2016tutorial}, 
can learn useful representations for downstream sequence prediction tasks like speech recognition,
which has not yet been fully explored. \par

As discussed in ~\citep{alemi2016deep} and also in Chapter ~\ref{cha:related} of this thesis,
Equation ~\eqref{eqn:elbo} with weight $\beta$ on the KL divergence $\KL{q_{\phi}(z|x)}{p_{\theta}(z)}$ term amounts to a lower bound of $I(X;Z)-\beta I(Z;i)$.
Compared with other reconstruction-based approaches that amounts to a lower bound of $I(X;Z)$ (e.g., deterministic autoencoders like AE and denoising AE) ,
tuning $\beta$ allows us to more flexibly regularize the representation learning process, which is a key advantage of VAE-type model over other pure reconstruction-based models. \par
%Further experimental shown in this chapter, and theoretical analysis also meet our expectation that VAE works better than other counterparts under sufficient tuning. \par

In the literature, VAEs have been widely applied in different tasks and settings,
such as controlled generation ~\citep{hu2017toward}, 
unsupervised disentangled representation learning ~\citep{hsu2017unsupervised} 
and semi-supervised learning ~\citep{kingma2014semi}.
However, to the best of our knowledge, 
there has been no systematic study to compare and contrast deterministic autoencoder-type models for unsupervised representation learning.

In my study,
I show that VAE learns representations that benefit a central task in the speech domain -- speech recognition.
My empirical study shows that VAE outperforms other reconstruction-based approaches I have investigated in terms of learning useful representations for speech recognition.
Compared with other deterministic AE approaches, I hypothesize that that the KL divergence term of VAE plays a crucial role as a regularizer towards learning better representations.
I perform ablation studies to support our hypothesis.
I also study the effect of the ``context window size'' ($W=2K+1$) on the quality of the learned representations. 
My empirical study suggests that larger window size typically helps representation learning;
however, the performance deteriorates when the context window is too large.
I further investigate how the amount of unlabeled data affects learned representations.
I show that a sufficiently large amount of unlabeled data is crucial
for autoencoder-type generative representation learning to benefit downstream tasks,
presumably because learning a good enough data-generating distribution requires many samples.
This finding aligns with many more recent works spanning the
NLP ~\citep{devlin2018bert,yang2019xlnet}, vision ~\citep{hjelm2018learning} and speech domains ~\citep{oord2018representation,schneider2019wav2vec},
which also have shown that unsupervised pre-training utilizing large amounts of unlabeled data can enhance downstream tasks.

I also further explore the ``zero extra unlabeled data" scenario, where the supervised tasks and the reconstruction tasks are trained jointly.
I found that autoencoder-type generative models still benefit the supervised tasks in such kind of scenario.
%problem ``Can autoencoder-type generative models benefit supervised tasks if we don't have extra unlabeled data?''.
%We show that the answer is yes if supervised tasks and generative models can be trained jointly.
In this scenario, I train VAE/AE jointly with a speech recognizer,
where the VAE/AE extracts contextual representations that are subsequently fed to a downstream speech recognizer,
as illustrated in Figure ~\ref{fig:feedforward-recognizer}.
I observe clear improvement of such multitask model (speech recognition+reconstruction)
compared with a baseline speech recognizer. \par

\section{Datasets}
\label{sec:data}

In this section, I introduce the three speech corpora that I use throughout this thesis:
1) University of Wisconsin (UW) X-ray Microbeam (XRMB) ~\citep{westbury1994x}, 
2) TIMIT ~\citep{garofolo1993darpa} and 
3) Wall Street Journal (WSJ) ~\citep{paul1992design}. \par

XRMB consists of both acoustic and articulatory measurements, 
but I only use the acoustic measurements throughout this chapter. 
The acoustic features are a $39D$ vector, 
consisting of $13D$ Mel-frequency cepstral coefficients (MFCCs), 
$\Delta$ (first order temporal derivatives) and $\Delta\Delta$ (second order temporal derivatives). 
Specific to feedforward representation learning,
I concatenate a $15$-frame window centered at each frame to incorporate context information. 
I use the first $1500$ utterances as the training set for unsupervised representation learning, 
and use the following $236$ utterances as a dev set for model selection (e.g. to pick the epoch with the best ELBO for VAE, or best reconstruction error for non-variational models). 
I leave the remaining $621$ unseen utterances (corresponding to $12$ speakers) to train speech recognizers. \par

For TIMIT, I follow the standard train/dev/test split, 
which consists of $3696$, $400$ and $192$ utterances respectively. 
The features are log scale filterbank coefficients with $\Delta$ and $\Delta\Delta$, 
a total of $120$ dimensions. 
Similarly to the setting of XRMB, I use window size $15$ for the purpose of incorporating contextual information.
Regarding representation learning,
I use the $3696$ training set utterances to train generative models,
and the $400$ dev set utterances for early stopping. 
All speech recognizers are trained/tuned/tested on the $3696/400/192$ utterances respectively. 
Thus, when performing generative pre-training on TMIMIT, we do not use extra unlabeled data. \par

The full training set of WSJ (typically referred to as ``SI284'') consists of 81 hours of speech.
There is a $15$ hour subset of SI284, which is referred to as ``SI84''. 
``dev93'' and ``eval92'' are used as the development and test sets respectively. 
I use $40$ mel-scale filterbank coefficients with $\Delta$ and $\Delta\Delta$, as well as energy, for a total of $123D$ features for each frame. 
Same as in other related work ~\citep{kim2017joint, bahdanau2016end}, 
I use $32$ distinct character labels in speech recognition experiments, 
including $26$ characters, apostrophe, period, dash, space, noise and a special blank for connectionist temporal classification (CTC).  \par

\section{Generative Acoustic Feature Learning}
\label{sec:15w-exp}

In this section,
I experiment on the model illustrated in Figure ~\ref{fig:feedforward}, a simple adaptation of autoencoder-type models for sequence data.
I try both autoencoders (AE) ~\citep{poultney2007efficient, bengio2007greedy}, denoising autoencoders (DAE, with Bernoulli noise or Gaussian noise) ~\citep{vincent2008extracting, vincent2010stacked} and variational autoencoders (VAE).
I describe these models (and other models for ablation study) in Table ~\ref{tab:model-description}. 
My experimental results indicate that autoencoders, especially VAE, are able to learn useful acoustic features that boost downstream speech recognition. \par

\begin{table} [htbp]
\centering
\begin{tabular}{| c | c | c | }
  \hline
  Methods & Description & Loss \\
  \hline \hline

  VAE & 
\begin{tabular}{@{}c@{}}
Inference network produces a\\
$d-$dimensional Gaussian posterior\\
$\mathcal{N}(\mu, \diag{\sigma^2})$; \\
$\delta \sim \mathcal{N}(0,I_d)$; \\
$\beta \geq 0$;  \end{tabular}
& \begin{tabular}{@{}c@{}}
$ \frac{\norm{x-F_{\theta}(\mu+\delta\odot\sigma)}_2^2}{2}$ \\
$+ \beta \KL{q_{\phi}(z|.)}{p_{\theta}(z)}$ \end{tabular} \\
  \hline

  AE & 
\begin{tabular}{@{}c@{}}
Encoder produces \\
a $d-$dimensional vector $\mu$ \end{tabular}
& $ \norm{x-F_{\theta}(\mu)}_2^2 $ \\
  \hline

\begin{tabular}{@{}c@{}}
  DAE \\
  with Bernoulli noise \end{tabular} & 
\begin{tabular}{@{}c@{}}
Assume $x$ is $k-$dimensional; \\
$\delta \in \{0,1\}^k $, \\
with $\mathbb{E}(\delta)=\{p\}^k$; \\
Encoder takes $x \odot \delta$ as input, \\
and produces a $d-$dimensional vector $\mu$
\end{tabular}
& $ \norm{x-F_{\theta}(\mu)}_2^2 $ \\
  \hline

\begin{tabular}{@{}c@{}}
  DAE \\
with Gaussian noise  \end{tabular}& 
\begin{tabular}{@{}c@{}}
Assume $x$ is $k-$dimensional; \\
$\delta ~\sim \mathcal{N}(\mathbf{1}_k,\gamma^2 I_k)$;\\
Encoder takes $x \odot \delta$ as input, \\
and produces a $d-$dimensional vector $\mu$
\end{tabular}
& $ \norm{x-F_{\theta}(\mu)}_2^2 $ \\
  \hline

  NAE & 
\begin{tabular}{@{}c@{}}
Inference network produces \\
a $d-$dimensional Gaussian posterior \\
$\mathcal{N}(\mu, \diag{\sigma^2})$; \\
$\delta \sim \mathcal{N}(0,I_d)$; \\
$\beta \geq 0$;  \end{tabular}& 
\begin{tabular}{@{}c@{}}
$ \frac{\norm{x-F_{\theta}(\mu+\delta\odot\sigma)}_2^2}{2}$ \\
$+ \beta \norm{\mu}_2^2 $ \end{tabular} \\
  \hline

\begin{tabular}{@{}c@{}}
  AE \\
  with Bernoulli dropout \\
  ~\citep{srivastava2014dropout} \end{tabular} 
& 
\begin{tabular}{@{}c@{}}
Encoder produces \\
a $d-$dimensional vector $\mu$; \\
$\delta \in \{0,1\}^d $, \\
with $\mathbb{E}(\delta)=\{p\}^d$; \\
$\hat{\mu}= \mu \odot \delta $
\end{tabular}
& $ \norm{x-F_{\theta}(\hat{\mu})}_2^2 $ \\
  \hline

\begin{tabular}{@{}c@{}}
  AE \\
  with Gaussian dropout \\
  ~\citep{wang2013fast} \end{tabular} 
&
\begin{tabular}{@{}c@{}}
Encoder produces \\
a $d-$dimensional vector $\mu$; \\
$\delta ~\sim \mathcal{N}(\mathbf{1}_d,\gamma^2 I_d)$;\\
$\hat{\mu}= \mu \odot \delta $
\end{tabular}
& $ \norm{x-F_{\theta}(\hat{\mu})}_2^2 $ \\
  \hline
\end{tabular}
\caption{Description of models used for unsupervised acoustic feature learning experiments. 
  $F_{\theta}(.)$ takes the output of an encoder/inference network as its input, and reconstructs the input.
  Here, $\mu$ in the table corresponds to $z$ in Figure ~\ref{fig:feedforward}, while $x$ in the table corresponds to reconstruction target in Figure ~\ref{fig:feedforward}. For example, the reconstruction target of $\mu=z_5$ is $x_{3:7}$.}
\label{tab:model-description}
\end{table}

%We do experiments, which consists of two steps, on XRMB.
My experiments on XRMB consist of two steps.
I first learn an acoustic feature transformation (training autoencoders) on unlabeled data, 
and then freeze the feature extraction network (the encoder)
and train a recognizer on top of the learned representations (mean value inferred by encoder). \par

The encoder I use to learn the acoustic feature transformation is a three-layer ReLU ~\citep{maas2013rectifier} network 
followed by a linear layer to generate the bottleneck representation (or posterior mean for VAE). 
For VAE, we need an additional linear layer to generate the logarithm of the diagonal covariance matrix during training.
The decoder is also a three-layer ReLU network that takes the bottleneck representation or samples from the VAE posterior as input, 
followed by a linear layer to reconstruct the input. 
Throughout this chapter, I use $1500$ units per feedforward layer.
As described in Section ~\ref{sec:data},
the autoencoder training is performed using the first $1500$ utterances and another $236$ utterances are used for early stopping. \par

In the second step, 
I use connectionist temporal classification (CTC) ~\citep{graves2006connectionist} for speech recognition.
I don't use any language models during speech recognizer training and decoding throughout all experiments in this chapter.
The task here is phonetic recognition on XRMB.
Before training/testing recognizers on the reserved $621$ utterances from $12$ speakers, 
I transform each $15-$frame window centered at each frame to a $70-$dimensional representation and use it as input for training/testing recognizers.
The $621$ feature sequences are then divided into $6$ splits,
and the ASR experiments are done in a $6$-fold manner. 
In each fold, 
$8$ speakers are used for ASR training, 
$2$ speakers are used for ASR hyper-parameter tuning and early stopping, 
and the remaining $2$ speakers are used as a test set.
The speech recognizer is a two-layer BiLSTM CTC network, 
with $256$ hidden units per direction. 
I use dropout ~\citep{dahl2013improving, wager2013dropout,srivastava2014dropout} rate $0.2$, 
batch size $1$,
Xavier initialization ~\citep{glorot2010understanding}, 
ADAM optimizer ~\citep{kingma2014adam} with learning rate $0.0005$, 
first momentum $0.9$ and second momentum $0.999$. 
I train each ASR model up to $20$ epochs, 
and pick the best epoch for each ASR model according to the dev set phone error rate (PER). \par

I compare the different approaches in Table ~\ref{tab:compare-framework}.
For each row (an approach with specific hyper parameters) in the table, the $6-$fold averaged phonetic error rate (PER) on the dev set is reported.
For each approach, I only report averaged PER on the test set for the model with the lowest dev set PER.  \par

\begin{table} [htbp]
\centering
\begin{tabular}{| l | r | r |}
  \hline
  Methods & Dev PER ($\%$) & Test PER ($\%$) \\
  \hline \hline
  1. Baseline (MFCCs) & 11.2 & 11.3 \\
  \hline
  2. AE & \textbf{12.1} & 11.9 \\
  \hline
  3.a DAE (Bernoulli=0.2) & \textbf{11.3} & 12.1 \\
  3.b DAE (Bernoulli=0.4) & 11.8 & - \\
  3.c DAE (Bernoulli=0.5) & 12.2 & - \\
  3.d DAE (Gaussian=0.5) & \textbf{12.1} & 12.4 \\
  3.e DAE (Gaussian=0.6546) & 13.7 & - \\
  3.f DAE (Gaussian=1.0) & 12.6 & - \\
  \hline
  4. VAE ($\beta=1.0$) & \textbf{9.3} & \textbf{9.6} \\
  \hline
\end{tabular}
\caption{Comparison of the quality of representations learned by several unsupervised representation learning approaches. 
The downstream task is phonetic recognition. 
All dev and test set results are given as $6$-fold averaged phone error rates (PERs). }
\label{tab:compare-framework}
\end{table}

In Table ~\ref{tab:compare-framework}, 
the baseline (group 1) is a $2-$layer-bidirectional-LSTM CTC recognizer with MFCCs+$\Delta$+$\Delta\Delta$ as input.
For Bernoulli noise, $p$ is selected from among $\{ 0.2,  0.4, 0.5 \}$. \allowbreak
For multiplicative Gaussian noise following the distribution $\mathcal{N}\left( \mathbf{1},\diag{\gamma^2} \right)$, 
I use $\gamma=\{ 0.5, 0.6546, 1.0 \}$.  \allowbreak
I choose $p$ and $\gamma$ such that $\gamma=\sqrt{\frac{p}{1-p}}$ following ~\citep{srivastava2014dropout}.
According to our observation, 
it is difficult for AE to outperform the baseline recognizer, 
but VAE outperforms the baseline recognizer (as well as AE) considerably.
This demonstrates that, having more regularization in learning data-generating density relative to AE, 
VAE benefits from learning on large quantities of unlabeled data.
Similarly, DAE also outperforms AE substantially.
These findings match prior work ~\citep{bengio2013representation,alain2014regularized}.
%In fact, 
%these findings match the assertion by ~\citep{bengio2013representation,alain2014regularized} that 
%``regularization would assist reconstruction-based approach capture variations along manifold of high density when there is productive resistance between them.'' 
Although VAE is the only technique that outperforms the baseline in the table, 
it is possible that lower noise applied to DAE would further improve the performance of DAE-based approaches.

\section{What Affects Generative Representation Learning?}
\label{sec:ablation-study}

In Section ~\ref{sec:15w-exp},
I have shown that modeling data-generating distributions using a large amount of unlabeled speech data can learn encoders capable of extracting acoustic representations that benefit downstream speech recognition.
I have also seen the success of VAE compared to alternatives (e.g. AE and DAE) in Table ~\ref{tab:compare-framework}.
In this section, I conduct an ablation study to understand the variables that impact generative representation learning.
My exploration suggests an appropriate magnitude of regularization, 
sufficient unlabeled samples (relative to the amount of labeled training samples), and use of contextual information
are the three key ingredients for learning useful representations.

%%%
%%% Now regularization

\subsection{Regularization -- the KL term}
\label{subsec:regularization}

\begin{table} [htbp]
\centering
\begin{tabular}{| l | r | r |}
  \hline
  Methods & Dev PER ($\%$) & Test PER ($\%$) \\
  \hline \hline
  1. Baseline (MFCCs) & \textbf{11.2} & \textbf{11.3} \\
  \hline
  2.a AE & 12.1 & 11.9 \\
  2.b AE (layer-wise Bernoulli Dropout 0.2) & 14.1 & - \\
  2.c AE (layer-wise Bernoulli Dropout 0.4) & 13.1 & - \\
  2.d AE with Bernoulli Dropout on Bottleneck (0.2) & \textbf{11.8} & \textbf{12.0} \\
  2.e AE with Bernoulli Dropout on Bottleneck (0.4) & 12.7 & - \\
  2.f AE with Bernoulli Dropout on Bottleneck (0.5) & 13.4 & - \\
  2.g AE with Gaussian Dropout on Bottleneck (0.5) & 12.4 & 12.2 \\
  2.h AE with Gaussian Dropout on Bottleneck (0.6546) & 12.6 & - \\
  2.i AE with Gaussian Dropout on Bottleneck (1.0) & 12.4 & - \\
  \hline
  3.a DAE (Bernoulli=0.2) & \textbf{11.3} & \textbf{12.1} \\
  3.b DAE (Bernoulli=0.4) & 11.8 & - \\
  3.c DAE (Bernoulli=0.5) & 12.2 & - \\
  3.d DAE (Gaussian=0.5) & 12.1 & 12.4 \\
  3.e DAE (Gaussian=0.6546) & 13.7 & - \\
  3.f DAE (Gaussian=1.0) & 12.6 & - \\
  \hline
  4.a NAE with $\beta=0.0$ & 12.0 & - \\
  4.b NAE with $\beta=0.000001$ & 12.9 & - \\
  4.c NAE with $\beta=0.00001$ & 12.6 & - \\
  4.d NAE with $\beta=0.0001$ & 13.4 & - \\
  4.e NAE with $\beta=0.001$ & 12.2 & - \\
  4.f NAE with $\beta=0.01$ & \textbf{11.8} & \textbf{12.2} \\
  4.g NAE with $\beta=0.1$ & 12.6 & - \\
  4.h NAE with $\beta=1.0$ & 13.1 & - \\
  4.i NAE with $\beta=10.0$ & 12.0 & - \\
  \hline
  5.a VAE ($\beta=10.0$) & 12.2 & - \\
  5.b VAE ($\beta=5.0$) & 11.7 & - \\
  5.c VAE ($\beta=2.5$) & \textbf{9.2} & \textbf{10.0} \\
  5.d VAE ($\beta=1.0$) & 9.3 & 9.6 \\
  5.e VAE ($\beta=0.75$) & 10.6 & - \\
  5.f VAE ($\beta=0.5$) & 11.3 & - \\
  5.g VAE ($\beta=0.1$) & 11.9 & - \\
  5.h VAE ($\beta=0.01$) & 12.1 & - \\
  5.i VAE ($\beta=0.001$) & 11.1 & - \\
  5.j VAE ($\beta=0.0001$) & 11.0 & - \\
  5.k VAE ($\beta=0.00001$) & 11.2 & - \\
  5.l VAE ($\beta=0.000001$) & 10.9 & - \\
  \hline
\end{tabular}
\caption{Ablation study on the effect of regularization.
All dev and test set results are given as $6$-fold averaged phone error rates (PERs). }
\label{tab:compare-framework-regularization}
\end{table}

Regularization plays a crucial role in learning. 
One common way to ``regularize'' autoencoders is to add some noise into the model during training.
~\citep{bishop1995training} shows that training with noise is equivalent to Tikhonov regularization.
Specific to autoencoders,
denoising autoencoders (DAE) ~\citep{vincent2008extracting,vincent2010stacked} are proven to be able to learn robust features leveraging large amounts of unlabeled data.
Later, ~\citep{bengio2013representation,alain2014regularized} show 
why DAE can be seen as a type of contractive autoencoders (CAE) ~\citep{rifai2011contractive, rifai2011higher},
and how concentration regularization helps the autoencoders to capture the data-generating distribution. \par

There is already a line of research trying to better understand VAE and use VAE to learn representations that benefit downstream tasks (e.g., ~\citep{zhao2017towards} explores how to make sure latent variables are not ignored by powerful decoder, ~\citep{alemi2017fixing, rainforth2018tighter} both mention that ``tighter'' ELBO which leads to better likelihood estimation does not necessarily guarantee good representations, ~\citep{shu2018amortized} provides a very interesting perspective to look at ELBO as a ``regularized likelihood'' and propose denoising variational autoencoder). 
In this section, I focus on understanding the benefits of VAE from regularization perspective.
Especially, I am interested in three questions:
a) what is the role of the KL divergence term of an ELBO on learning representations
and b) how important is tuning the weight of KL divergence term for representation learning
and c) can we use dropout to improve autoencoding
and what is the connection between dropout and variational inference.
To answer these questions, I compare and contrast a few autoencoder-type models from the perspective of regularization.
The experimental results are summarized in Table ~\ref{tab:compare-framework-regularization}.
I show that encoders trained using different level of regularization produce representations that result in very different performance on downstream tasks like speech recognition
-- evidence of the importance of ``a suitable level of regularization'' in learning good representations. \par
%We focus on understanding the benefits of VAE from regularization perspective,
%and are especially interested in three questions:
%a) what is the role of the KL divergence term of an ELBO on learning representations
%and b) how important is tuning the weight of KL divergence term for representation learning
%and c) can we use dropout to improve autoencoding
%and what is the connection between dropout and variational inference. \par

In VAE, we typically use prior distribution $p_{\theta}(z)=\mathcal{N}(\mathbf{0}, I_d)$.
Then $\KL{q_{\phi}(z|x)}{p_{\theta}(z)}$, the KL divergence term, is

\begin{eqnarray}
\KL{q_{\phi}(z|x)}{\mathcal{N}(\mathbf{0},I_d)} 
&=& \frac{\norm{\mu}_2^2}{2} + \sum_{i=1}^d \left\{ \frac{1}{2}\sigma^2_i-\log{\sigma_i} \right\} - \frac{d}{2}
\label{eqn:kl_identity}
\end{eqnarray}

Here $\mu$ is the mean value of the Gaussian posterior,
and we typically use $\mu$ as the representation of $x$.
$\diag{\sigma^2}$ is the diagonal covariance matrix of the Gaussian posterior. \par

The KL divergence term ~\ref{eqn:kl_identity} contains two parts: 
the $\mathcal{L}_2$ norm on $\mu$ and \allowbreak $\sum_{i=1}^d \left\{ \frac{1}{2}\sigma^2_i-\log{\sigma_i} \right\} $ which controls per-sample-variance in latent space. 
Minimizing $\sum_{i=1}^d \left\{ \frac{1}{2}\sigma^2_i-\log{\sigma_i} \right\}$ actually encourages each $\sigma_i$ to be $1$. 
When larger $\beta$ is used, it means we put very heavy $\mathcal{L}_2$ regularization on $\mu$ to make it compact,
and in the mean time, encourage $\diag{\sigma^2}$ to be closer to the identity matrix $I_d$, and vice versa. \par
% -- that is, we do not expect the representation ($\mu$) to be compact,
%and we learn a more flexible sample-dependent variance. \par
%\weiran{It is better to explicitly write out the regularization term in this case and give a symbol to the coefficient. Also, in the table, ``NAE'' with $\mathcal{L}=xxx$ shall really be the coefficient=xxx.}
The NAE (Eqn ~\eqref{eqn:NAE}, (4.a) in Table ~\ref{tab:compare-framework-regularization}) is the extreme case of VAE when the second part of the KL divergence is ignored.
\begin{eqnarray}
\mathbb{E}_{q_{\phi}(z|x)} \big\{ \log p_{\theta}(x|z) \big\} - \beta \frac{\norm{\mu}_2^2}{2} + 0 \times \{ \sum_{i=1}^d \left\{ \frac{1}{2}\sigma^2_i-\log{\sigma_i} \right\} - \frac{d}{2} \}
%\KL{q_{\phi}(z|x)}{\mathcal{N}(\mathbf{0},I_d)} 
\label{eqn:NAE}
\end{eqnarray}

%This extreme case model results in similar performance in ASR compared to AE and is much worse than VAE and DAE.
%The observation matches our expectation.
I study regularizing the reconstruction by only using $\mathcal{L}_2$ on mean with different $\beta$
while dropping the term $\sum_{i=1}^d \left\{ \frac{1}{2}\sigma^2_i-\log{\sigma_i} \right\}$.
As shown in 4.b-4.i of Table ~\ref{tab:compare-framework-regularization}, 
the resulting learned representation is worse than VAE though it outperforms plain AE when weight of $\mathcal{L}_2$ equals to $0.01$.
I also tune $\beta$ thoroughly for full VAE as shown in 5.a-5.l of Table ~\ref{tab:compare-framework-regularization},
with the best dev set error rate being $9.2$ with $\beta=2.5$. \par
Further comparison between group 4 and group 5 in Table ~\ref{tab:compare-framework-regularization} 
shows how the two components of ~\eqref{eqn:kl_identity} regularize the representation learning.
We first look at 5.a and 4.i. 
When $\beta=10.0$, there is not much flexibility for both $\mu$ and $\sigma$,
and thus the difference between NAE and VAE is very small.
When $0.01 \leq \beta < 10.0$, 
VAE is typically much better than or comparable to NAE.
When $\beta$ becomes much smaller (e.g. $\leq 0.001$),
the $\mathcal{L}_2$ regularization effect gradually becomes much weaker.
The advantage of VAE presumably arises because that the variance term $\sigma^2$ is still under control and thus provides some level of regularization,
but the variance term on the NAE side is totally free which makes reconstruction in NAE lacking necessary regularization.
All these aforementioned observations suggest the benefit of learning a compact representation ($\mu$)
paired with a suitable sample dependent covariance matrix $\diag{\sigma^2}$. \par

In Table ~\ref{tab:compare-framework-regularization}, 
I also compare VAE with AE with dropout applied.
According to experiments on XRMB (Table ~\ref{tab:compare-framework-regularization}) and TIMIT (Table ~\ref{tab:compare-framework-timit}),
by selecting $\beta$ according to development set performance, 
VAE can outperform DAE and also AE with dropout applied layer-wise or only to the bottleneck layer. 
For details on experiments on TIMIT, 
see Section ~\ref{subsec:data-amount}. \par

Bernoulli dropout is a very popular technique to prevent over-fitting when training neural networks.
Denote the bottleneck feature vector of an AE as $\mu$ given sample $x$.
Bernoulli dropout defines a keep probability $1-p$, where each element of the bottleneck vector $\mu$ is either kept with probability $p$, or set to $0$ (dropped) with probability $p$.
In contrast, Gaussian dropout requires a given noise distribution $\mathcal{N}(1,\gamma^2)$.
Each element of $\mu$ is multiplied by a random value drawn from $\mathcal{N}(1, \gamma^2)$. \par

In Table ~\ref{tab:compare-framework-regularization} and ~\ref{tab:compare-framework-timit}, 
VAE (with $\beta$ selected on dev set) outperforms AE with Bernoulli/Gaussian dropout applied layer-wise or only to the bottleneck representation.
I suspect this is because the reparameterization itself can be viewed as a type of more flexible dropout with a learned per-sample dropout rate. \par

In ~\citep{wang2013fast}, 
the authors show that Gaussian dropout is an approximation of Bernoulli dropout with almost identical regularization effect but converges much faster.
In ~\citep{srivastava2014dropout}, 
the authors experimentally verify that Gaussian dropout outperforms Bernoulli dropout.
Here I show how reparameterization can connect to Gaussian dropout,
which has been discussed in more depth in the papers ~\citep{wang2013fast,gal2016dropout,gal2016theoretically,kingma2015variational}. \par

Assuming $\delta$ is drawn from $\mathcal{N}(\mathbf{1},\gamma^2 I)$ and $\mu$ is the bottleneck feature vector of an AE, we have

\begin{eqnarray}
\mu \odot \delta &=& \mu \odot \mathbf{1} + \mu \odot (\delta-\mathbf{1}) \label{eqn:gp-additive} \\
&=& \mu + \gamma \times \mu \odot \frac{(\delta-\mathbf{1})}{\gamma} \label{eqn:reparameterization}
\end{eqnarray}

Equation \eqref{eqn:gp-additive} shows that Gaussian dropout is equivalent to the additive noise technique used in ~\citep{vincent2010stacked}.
Note that because $\frac{\delta-\mathbf{1}}{\gamma}$ follows $\mathcal{N}(\mathbf{0}, I)$, Equation \eqref{eqn:reparameterization} shows that we can rewrite a Gaussian noise-corrupted bottleneck representation as a sample from a Gaussian diagonal posterior $\mathcal{N}(\mu, \diag{(\gamma \times \mu)^2})$.
From this perspective, a VAE is more flexible than Gaussian dropout because the diagonal standard deviation of the VAE posterior need not be linear correlated with its mean vector.
Also, unlike Gaussian dropout where $\gamma$ is a hyper-parameter for all samples, VAE provides sample-specific ``regularization'' by learning a sample-specific deviation. \par

%%%
%%% Now the amount of data

\subsection{Amount of unlabeled data}
\label{subsec:data-amount}

In this section, I test the different representation learning frameworks listed in Table ~\ref{tab:model-description} on TIMIT.
Training is performed on the $3696$ utterances of the training set, while the $400$ utterances from the dev set are used for hyper-parameter tuning and early stopping.
Both representation learning and recognition training use the same set of utterances. \par

In the ASR training phase, I first use the trained encoders to transform each $15-$frame window centered at each frame into a $70-$dimensional feature vector for all of the $3696/400/192$ utterances. 
I then train a $3-$layer BiLSTM (with subsampling rate $0.5$ in the second and third layers) CTC recognizer using the learned representations.
The size of the hidden layers is $256$ per direction. 
I use dropout rate $0.4$, Xavier initialization, ADAM optimizer with learning rate $0.0005$ and batch size $4$. 
All the recognizers are trained up to $20$ epochs,
and the epoch with best performance on the development set is selected. \par

The experimental results on TIMIT are summarized in Table ~\ref{tab:compare-framework-timit}.
As shown in the table, 
VAE still outperforms other models in terms of PER, 
but none of the pre-trained models outperform the baseline CTC recognizer after fine-tuning. 
My observation that pre-training on $1500$ untranscribed utterances helps ASR models trained on $~200$ transcribed utterances, 
while pre-training on the same $3696$ utterances does not help,
suggests  that it is important to pre-train on large amount of data (relative to the labeled training data) in unsupervised representation learning. \par

To further validate the fact that the size of data used for pre-training is important, 
I conduct experiments on Wall Street Journal (WSJ). 
I use the speech utterances in SI284, but not in SI84, 
to train VAEs using dropout rate $0.2$, different $\beta$,
and dimensionalities selected from among $\{ 90, 120, 150\}$ for latent variables.
The Dev93 set is used for tuning and early stopping in this phase. 
I use window size $7$. \par

In the ASR training phase, 
I use different proportion of SI84 respectively, i.e., $\frac{1}{16}$, $\frac{1}{8}$, $\frac{1}{4}$, $\frac{1}{2}$ and $100\%$ of SI84. 
The motivation here is to further test the importance of the size of the unlabeled dataset relative to the size of the transcribed corpus. 
I use the same setting to train a $3-$layer CTC recognizer as I have done in the TIMIT experiments; 
the only difference is that we are training a character recognizer with $32$ labels (see Section ~\ref{sec:data} for details about the 32 labels). \par

As shown in Table ~\ref{tab:compare-framework-wsj}, 
a speech recognizer trained on the inferred representations can outperform the baseline, when the size of the transcribed dataset is much smaller than the unlabeled speech data. 
For example, when using $\frac{1}{16}$ of SI84 (roughly $1$ hour of speech) or $\frac{1}{8}$ of SI84 (roughly $2$ hours of speech) for speech recognizer training, 
the unlabeled speech data (roughly $65$ hours of speech) can easily help us to improve performance. 
However, 
if we have a larger transcribed dataset (e.g. using $\frac{1}{4}$ of SI84, $4$ hours of speech), the learned representations no longer enhance ASR performance. 
When we have even more labelled utterances (e.g. $50\%$ and $100\%$ of SI84, $7.5$ and $15$ hours of speech respectively), 
the learned representations even slightly hurt performance. \par

The second interesting observation is regarding $\beta$. 
Unlike the observations on XRMB, here $\beta=1$ usually produces useless representations. 
Instead, $\beta=0.1$ or $\beta=0.01$ typically give us the best representations, 
while $\beta=0.001$ seems to provide not enough regularization. \par

\begin{table} [htbp]
\centering
\begin{tabular}{ | l | r | r | r | r | r | r | r | r | r | r |}
\hline  Models & \multicolumn{2}{|c|}{1hr} & \multicolumn{2}{|c|}{2hrs} & \multicolumn{2}{|c|}{4hrs} & \multicolumn{2}{|c|}{8hrs} & \multicolumn{2}{|c|}{$100\%$} \\
  \hline \hline
  & Dev & Test & Dev & Test & Dev & Test & Dev & Test & Dev & Test \\
  \hline \hline
  1. Baseline & 57.2 & 47.3 & 51.2 & 40.1 & \textbf{40.8} & 31.5 & \textbf{33.0} & 26.0 & \textbf{26.1} & 19.5 \\
  \hline
  2.a Z=90, $\beta$=1.0 & 60.5 & - & 56.3 & - & 47.0 & - & 40.5 & - & 34.6 & - \\
  2.b Z=90, $\beta$=0.1 & 56.1 & - & 51.2 & - & 41.8 & - & 35.0 & - & 29.9 & - \\
  2.c Z=90, $\beta$=0.01 & 57.5 & - & 51.9 & - & 42.2 & - & 35.3 & - & 29.5 & -\\
  2.d Z=90, $\beta$=0.001 & 58.9 & - & 52.5 & - & 43.3 & - & 35.1 & - & 29.7 & - \\
  \hline
  3.a Z=120, $\beta$=1.0 & 60.8 & - & 58.1 & - & 48.6 & - & 40.9 & - & 36.2 & - \\
  3.b Z=120, $\beta$=0.1 & 55.3 & - & 51.7 & - & 41.5 & - & 35.5 & - & 29.2 & - \\
  3.c Z=120, $\beta$=0.01 & 56.3 & - & 51.0 & - & \textbf{41.1} & 32.7 & \textbf{34.9} & 28.1 & 29.6 & - \\
  3.d Z=120, $\beta$=0.001 & 55.1 & - & 49.1 & - & 41.7 & - & 35.9 & - & 30.5 & - \\
  \hline
  4.a Z=150, $\beta$=1.0 & 60.0 & - & 57.7 & - & 48.2 & - & 41.4 & - & 36.1 & - \\
  4.b Z=150, $\beta$=0.1 & \textbf{54.8} & 47.2 & \textbf{48.7} & 40.0 & 41.2 & - & 35.2 & - & 29.3 & - \\
  4.c Z=150, $\beta$=0.01 & 56.5 & - & 50.4 & - & 42.7 & - & 35.4 & - & \textbf{29.0} & 21.7 \\
  4.d Z=150, $\beta$=0.001 & 58.1 & - & 52.9 & - & 42.2 & - & 38.3 & - & 31.7 & - \\
  \hline
\end{tabular}
\caption{Comparison of several representation learning approaches on WSJ. 
All numbers are character error rate (CER) on Dev93 and Eval92.
Each row indicates a VAE model using a certain dimensionality of latent variable $z$ and a certain $\beta$. 
Each column corresponds to experiments using a certain portion of SI84 for training the speech recognizer. 
All recognizers are trained up to $20$ epochs with batch size $8$.
%\weiran{give number of hours rather than $1/16$ etc in the header}
}
\label{tab:compare-framework-wsj}
\end{table}

\begin{table} [htbp]
\centering
\begin{tabular}{| l | r |}
  \hline
  Methods & Dev PER ($\%$)  \\
  \hline \hline
  1. Baseline (Log Mel Filter Bank) & 18.7  \\
  \hline
  2.a AE & \textbf{19.3} \\
  2.b AE (layer-wise Bernoulli Dropout 0.2) & 25.6 \\
  2.c AE (layer-wise Bernoulli Dropout 0.4) & 37.4 \\
  \hline
  3.a AE with Bernoulli Dropout on Bottleneck (0.2) & \textbf{22.2} \\
  3.b AE with Bernoulli Dropout on Bottleneck (0.4) & 25.9 \\
  3.c AE with Bernoulli Dropout on Bottleneck (0.5) & 29.0 \\
  3.d AE with Gaussian Dropout on Bottleneck (0.5) & \textbf{19.5} \\
  3.e AE with Gaussian Dropout on Bottleneck (0.6546) & \textbf{19.5} \\
  3.f AE with Gaussian Dropout on Bottleneck (1.0) & 19.6 \\
  \hline
  4.a DAE (Bernoulli=0.2) &  19.3 \\
  4.b DAE (Bernoulli=0.4) &  19.7 \\
  4.c DAE (Bernoulli=0.5) &  \textbf{19.3} \\
  4.d DAE (Gaussian=0.5) & \textbf{19.4} \\
  4.e DAE (Gaussian=0.6546) & 19.3 \\
  4.f DAE (Gaussian=1.0) & 20.2 \\
  \hline
  5.a NAE with $\beta=0.0$ & 19.7 \\
  5.b NAE with $\beta=0.000001$  & 19.3 \\
  5.c NAE with $\beta=0.00001$  & \textbf{19.2} \\
  5.d NAE with $\beta=0.0001$  & 19.5 \\
  5.e NAE with $\beta=0.001$  & 19.3 \\
  5.f NAE with $\beta=0.01$  & 19.5 \\
  5.g NAE with $\beta=0.1$  & 19.9 \\
  5.h NAE with $\beta=1.0$  & 20.4 \\
  5.i NAE with $\beta=10.0$  & 23.5 \\
  5.j 6.a with layer-wise Bernoulli dropout (0.2)  & 24.2 \\
  5.k 6.a with layer-wise Bernoulli dropout (0.4)  & 30.2 \\
  5.l 6.a with $3$ samples from posterior  & 19.5 \\
  5.m 6.a with $5$ samples from posterior  & 19.4 \\
  \hline
  6.a VAE (beta=10.0) & 38.0 \\
  6.b VAE (beta=1.0) & 26.0 \\
  6.c VAE (beta=0.1) & 26.7 \\
  6.d VAE (beta=0.01) & \textbf{19.1} \\
  6.e VAE (beta=0.001) & 19.5 \\
  6.f VAE (beta=0.0001) & 19.1 \\
  6.g VAE (beta=0.00001) & 19.9 \\
  6.h 6.b with layer-wise 0.2 Bernoulli Dropout & 28.8 \\
  6.i 6.b with layer-wise 0.4 Bernoulli Dropout & 73.7 \\
  6.j 6.b with $3$ samples from posterior & 25.2 \\
  6.k 6.b with $5$ samples from posterior & 25.8 \\
  \hline
\end{tabular}
\caption{Comparison of the quality of representations learned by several unsupervised learning approaches on TIMIT.
The dimensionality of the learned bottleneck representation is $70$.
All recognizers are trained up to $20$ epochs with batch size $4$.}
\label{tab:compare-framework-timit}
\end{table}

%%%
%%% Now contextual information

\subsection{Contextual Information}
\label{subsec:window}

In this section, 
I study the effect of the context window size on representation learning. 
%\weiran{This sentence is not necessary because you said early in the section that this section will use ASR as downstream task. } We still take speech recognition as our downstream task for evaluating the quality of the learned representation. \par
In Section ~\ref{sec:15w-exp}, 
I learn representation for each frame of an utterance based on its $15-$frame context. 
Here, I perform generative pre-training of VAE on XRMB, with different window sizes $1$, $3$, $7$, $15$ and $31$,
while all other settings are identical to what I used 
%with identically the same setting (except window size) used to train VAE 
in Section ~\ref{sec:15w-exp}.
The process to train speech recognizers is also identical to that of Section ~\ref{sec:15w-exp}. \par

\begin{table} [htbp]
\centering
\begin{tabular}{| c | c | c | r | r |}
  \hline
  Window Size & $\beta$ & Dropout of VAE & Dev PER ($\%$) & Test PER ($\%$) \\
  \hline \hline
  Baseline (MFCCs) & - & - & 11.2 & 11.3 \\
  \hline
  1 & 0.1 & 0.2 & 10.9 & 11.0 \\
  \hline
  3 & 1.0 & 0.0 & 10.9 & 11.0 \\
  \hline
  7 & 1.0 & 0.0 & 10.4 & 10.8 \\
  \hline
  15 & 2.5 & 0.0 & 9.2 & 10.0 \\
  \hline
  31 & 1.0 & 0.0 & 10.7 & 10.7 \\
  \hline
\end{tabular}
\caption{The effect of window size on learned representations using XRMB. 
The dimensionality of the bottleneck feature is $70$, and all PERs shown in the table are $6-$fold averaged PER. 
For each window size, I try VAEs with different $\beta$ and dropout, but I only report the model that performs best on averaged PER on dev for each given window size.}
\label{tab:window-size}
\end{table}

I expect that larger windows may incorporate more context information and thus help the representation learning. 
However, the observation is complicated. 
Window size $15$ does show better 
performance compared to window size $1$, $3$, and $7$. 
However, the model with window size $31$ seems to work worse than the model with window size $15$. 
According to our observations, 
it is not easy to learn good representations from very high-dimensional inputs.
The content of a larger window is more complex than that of a smaller window.
For example, a $5-$frame window of an acoustic utterance probably consists of frames that all associate with the same phone label,
but a $31-$frame window consists of frames with a few consecutive phones.
On the other hand, we might not have enough samples to learn representations of a very large window of context (e.g. $31-$frame window).
Increasing the window size increases the number of parameters of a feedforward neural network,
but the number of samples remains the same as the case with smaller window size. 
One possibility to learn better performance given larger window size (e.g., $31-$frame) is to treat the $15-$frame representation as a reference, and try to use richer information inside the $31-$frame window to improve upon the reference.
I named this method as ``prior updating" (i.e., replace the vanilla prior by a learned $15-$frame posterior during learning), which would be introduced in next Chapter. \par

\section{Multitask Speech Recognition with Auxiliary Reconstruction Task}
\label{subsec:feedforward-auxiliary}
%\weiran{shall we use ``semi-supervised''?}

\begin{figure*}[htbp]
  \centering
  \includegraphics[width=0.95\textwidth]{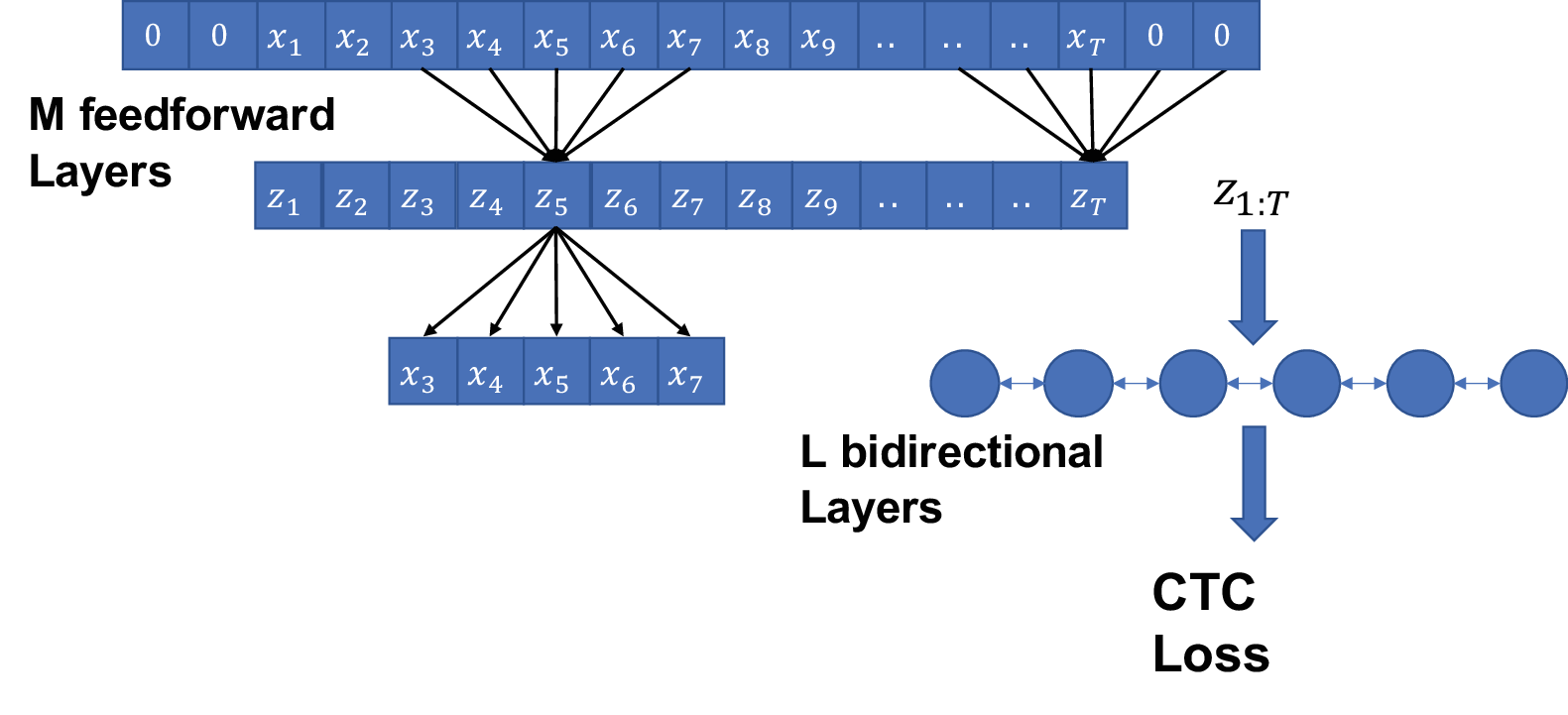}
  \caption{Illustration of CTC-based speech recognition with auxiliary reconstruction task,
with windows size $W=5$.
For every $5$ consecutive time steps,
I use a feedforward neural network to infer a vector as the representation of the central time step,
and reconstruct the $5$ consecutive time steps provided the time-specific representation.
These per-time-step representations also are the input to the speech recognizer branch (I use a CTC recognizer).
In the figure, $M$ indicates the number of feedforward layers,
and $L$ indicates the number of bidirectional recurrent layers used only in supervised tasks.}
  \label{fig:feedforward-recognizer}
\end{figure*}

\begin{table} [htbp]
\centering
\begin{tabular}{| c | c | c | c | r | r | r | r | r |}
  \hline
M & L & Dim(z) & With Reconstruction? & W=1 & W=3 & W=7 & W=15 & W=31 \\
\hline \hline
3 & 0 & 512 & No & 40.7 & 31.2 & 27.9 & 26.6 & \textbf{26.3} \\
3 & 0 & 512 & Yes & 40.4 & 29.3 & 26.3 & \textbf{25.0} & 26.8 \\
\hline \hline
2 & 1 & 512 & No & 24.0 & 20.6 & \textbf{20.3} & 22.9 & 24.4 \\
2 & 1 & 512 & Yes & 24.6 & 20.5 & \textbf{20.2} & 21.7 & 24.7 \\
\hline \hline
1 & 2 & 512 & No & 21.6 & \textbf{19.8} & 20.5 & 21.6 & 23.7 \\
1 & 2 & 512 & Yes & 21.0 & 17.8 & \textbf{17.6} & 18.0 & 20.0 \\
\hline \hline
3 & 0 & 1024 & No & 40.4 & 30.0 & 26.7 & \textbf{25.0} & 26.1 \\
3 & 0 & 1024 & Yes & 40.6 & 30.6 & 27.1 & \textbf{26.3} & 30.0 \\
\hline \hline
2 & 1 & 1024 & No & 23.7 & \textbf{20.2} & 20.5 & 23.3 & 27.0 \\
2 & 1 & 1024 & Yes & 24.9 & 19.8 & \textbf{19.6} & 20.4 & 24.0 \\
\hline \hline
1 & 2 & 1024 & No & 21.9 & \textbf{20.2} & \textbf{20.2} & 22.4 & 25.0 \\
1 & 2 & 1024 & Yes & 21.9 & 17.8 & \textbf{17.4} & 18.2 & 19.8 \\
\hline \hline
\end{tabular}
\caption{Multitask learning on TIMIT. 
All reported numbers are PER on the development set, 400 utterances. 
For different combinations of $M$, $L$ and $Dim(z)$, I compare CTC recognizers on top of $M-$ feedforward layers and $L-$ BiLSTMs as shown in Figure ~\ref{fig:feedforward-recognizer}, with and without using auxiliary reconstruction loss.
For reference, a $3-$layer BiLSTM CTC recognizer achieves dev/test performance $17.3/19.4$, which is the baseline of this table.}
%$L-$layers of BiLSTM which is on top of $M-$ layers of feedforward layers. The }
\label{tab:multitask-feedforward}
\end{table}

I have carefully studied unsupervised generative representation learning in Sections ~\ref{sec:15w-exp} and ~\ref{sec:ablation-study}.
One key observation is that the ratio of
the amount of unlabeled samples relative to 
the amount of labeled samples
has significant effect on the amount of improvement caused by unsupervised pre-training using VAEs.
Only when the ratio is large enough, 
as shown in Table ~\ref{tab:compare-framework-timit},
the unsupervised pre-training via VAE helps with downstream tasks.
%without enough unlabeled data, can generative pre-training potentially fail to generalize well to downstream tasks?
%the generative pre-training can learn feature extractors that generalizes well to the downstream tasks.
%modeling can learn feature extractors that generalize well to unseen development/test set.

In this section, I seek to answer: 
``Can we improve a supervised sequence prediction task (e.g. speech recognition here)
if we do not have, or only have limited amount of unlabeled samples in addition to the labeled training set?"
I study supervised speech recognition with auxiliary reconstruction loss in this section.
Our observation is that a speech recognizer trained jointly with a generative model, (e.g., VAE),
with low-level representations shared by the two tasks,
can enhance the ASR performance
even without using extra unlabeled speech. 
The model architecture I use is explained in Figure ~\ref{fig:feedforward-recognizer}.
I first use $M$ stacked feedforward layers to transform $W$ ($W=5$ in the figure) consecutive frames into a contextual representation $z$,
I then feed the representation sequence $z$ as input to a $L-$layer bidirectional CTC recognizer.
I jointly minimize the CTC loss and maximize the ELBO. The weights on the CTC loss and ELBO are tunable, and sum to one.

I experiment on TIMIT (Section ~\ref{sec:data}) using $\{M=3,L=0\}$, $\{M=2, L=1\}$ and $\{M=1, L=2\}$.
Here, I first evaluate all candidate architectures with depth $3$, and investigate if using more recurrent layers makes the model more powerful.
The experiments are summarized in Table ~\ref{tab:multitask-feedforward}.
%The motivation of trying different combinations of $M$ and $L$ is for better understanding 
As we can see from Table ~\ref{tab:multitask-feedforward}, 
for different choices of $\{M,L\}$, the multitask ASR models typically outperform their corresponding baselines, which in contrast do not have reconstruction loss; The improvement potentially comes from the multitask learning process. 
On one hand, according to existing self-supervised learning works like ~\citep{Pascual2019, Ravanelli2020MultiTaskSL}, representations that work well for multiple tasks can potentially be more robust.
On the other hand, the improvement could be due to the regularization effect (e.g. reparameterization could be connected to Gaussian dropout) from the VAE as discussed earlier in this chapter. 
Another trend we can clearly see from the table is that the more recurrent layers used the lower PER we can achieve.
This makes sense as feedforward layers typically provide limited contextual information compared with recurrent layers.
One interesting observation is that, 
when I use $\{M=1, L=2\}$, the baseline can achieve PER $20.2\%$, which is clearly worse than the PER $17.3\%$ of the $3-$layer BiLSTM CTC recognizer.
However, this gap can be closed by multitask learning (PER $17.4\%$ when using $W=7$) which shows the clear benefit of VAE when jointly trained with the speech recognizer.

\section{Summary}

To summarize this chapter, I have below key contributions and findings:

\begin{itemize}
\item[1] \textbf{VAE can learn representation beneficial to speech recognition tasks}: 
I tried VAE in two scenarios, i.e., unsupervised learning and generative pre-training as an auxiliary task of supervised learning, wherein VAE all helps improve performance.
I also compared VAE and other non-variational autoencoding approaches in terms of unsupervised representation learning.
I found that the representation learned by VAE can significantly outperform its non-variational counterparts in terms of downstream speech recognition tasks.
I also found that using more contextual information and larger training dataset for pre-training are crucial.
\item[2] \textbf{What makes VAE powerful for representation learning}: 
Regularization plays a crucial role in learning. 
One common way to ``regularize'' autoencoders is to add some noise into the model during training.
In VAE, we typically use prior distribution $p_{\theta}(z)=\mathcal{N}(\mathbf{0}, I_d)$.
Then the KL divergence term can be roughly viewed as a $\mathcal{L}_2$ regularization on posterior mean plus another regularization term on variance.
Such regularization term encourages the representation to be compact.
I found by tuning the weight of the KL divergence term to find the proper extent of regularization effect, 
VAE can have superior performance than its non-variational counterparts.
\end{itemize}

%----------------------------------------------------------------------------------------
%	Multiview representation learning
%       1. A quick introduction to multiview representation learning and CCA-based representation
%       learning approaches
%       2. Experiments for feedforward multiview representation learning
%       3. Multiview approaches with BiRNN encoders
%       4. Cross-domain representation learning with multiview data
%----------------------------------------------------------------------------------------
\chapter{Multi-view Representation Learning}
\label{cha:multiview}

%----------------------------------------------------------------------------------------
%       Motivation
%       We also need some references for multi-view representation learningmulti-view
%----------------------------------------------------------------------------------------

In Section ~\ref{cha:feedforward} we showed that generative pre-training (e.g. VAEs and other auto encoders) can learn representations that would benefit downstream supervised tasks like speech recognition.
We also showed that VAEs can improve the performance of target tasks (e.g. speech recognition) when jointly trained with the target tasks.
Such a multitask learning framework is even helpful when we don't have extra unlabeled data.
In this section, we explore using paired-view information for improving the quality of learned representations.
This chapter consists of four sections:
In the first section, we study multi-view representation learning ~\citep{xu2013survey, wang2015deep} and show that variational canonical correlation analysis (VCCA) ~\citep{wang2016deep}
and its extensions learn good acoustic features for speech recognition.
In the second section, we propose a novel method for learning the prior distribution of latent variables for sequence data and show that this technique alleviates the difficulty for learning representations of high-dimensional input.
In the third section, we investigate using the acoustic representation learned in a source domain where we have access to paired-view information to enhance the representation learning in a target domain where we do not have access to paired information.
In the final section, we study ``label embedding'',
where we use the labels as the additional supervised view for learning representations that capture the structure of high dimensional discrete labels.
Strictly speaking,
``label embedding" can also be understood as multi-view representation learning,
as it tries to learn the shared representation between raw input and the labels.
Thus we also put our work on ``label embedding" in this chapter. \par

\section{Multi-view Representation Learning}
\label{sec:multi-view}

With large amount of labeled data,
current supervised learning techniques can learn very powerful predictive models.
However, it can be costly to collect a dataset that consists of a large amount of labeled instances (e.g. ImageNet ~\citep{deng2009imagenet} for computer vision tasks and LibriSpeech ~\citep{panayotov2015librispeech} for ASR).
Thus, representation learning techniques that can reduce sample complexity for supervised learning are important.
In this section, we study representation learning where we assume each sequence has instances from two views during training but only one modality is available at test time,
and the two views share information that are correlated with the downstream task and thus the second view provides weak supervision. \par

There have been a series of works on multi-view representation learning in the deep learning era,
including but not limited to
Heteroscedastic dropout ~\citep{lambert2018deep} which uses the second view to learn the dropout rate;
deep variational canonical correlation analysis (VCCA) ~\citep{wang2016deep},
which is a deep neural network version of probabilistic CCA ~\citep{bach2005probabilistic};
deep canonical correlation analysis (DCCA) ~\citep{andrew2013deep,wang2015deep},
which is a deterministic extension of linear CCA;
and multi-view learning with contrastive loss (CONTRAST) ~\citep{hermann2014multilingual}.
Our contributions are 1) Extending VCCA with ``prior updating", and 2) showing that multi-view variational approaches (e.g. VCCA and its extensions) can be used to learn acoustic representations benefiting downstream speech recognition,
and can outperform a few of their deterministic counterparts (e.g. DCCA and CONTRAST),
as presented in ~\citep{tang2017acoustic, tang2018acoustic}. \par

\subsection{Variational Canonical Correlation Analysis (VCCA)}

\begin{figure*}[t]
 \centering
 \label{fig:vcca-illustration}

\begin{minipage}{1.0\textwidth}
\begin{subfigure}{0.4\textwidth}
  \centering
  \includegraphics[width=0.6\textwidth]{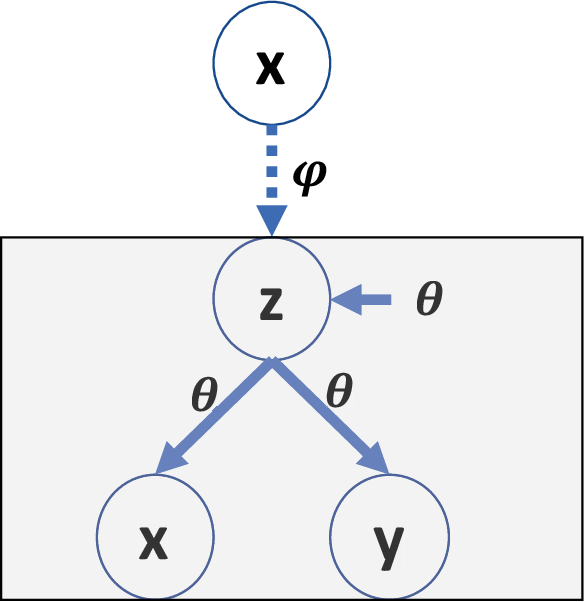}
  \caption{Illustration of deep variational canonical correlation analysis (VCCA).
The part inside the gray box is the graphical model (with solid lines).
The inference model is shown using dashed lines.
$\theta$ and $\phi$ are parameters of the generation and inference networks respectively.}
  \label{fig:vcca}
\end{subfigure}
\qquad
\begin{subfigure}{0.6\textwidth}
  \centering
  \includegraphics[width=0.7\textwidth]{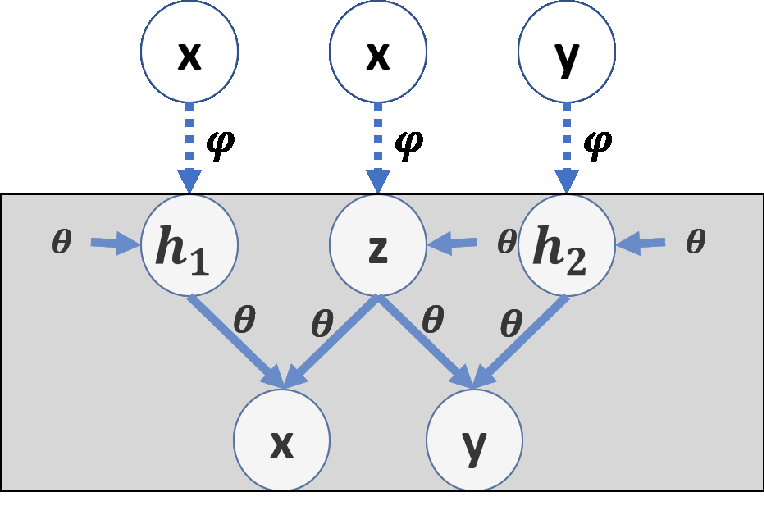}
  \caption{Illustration of deep variational canonical correlation analysis with private variables (VCCAP).
The part inside the gray box is the graphical model (with solid lines).
The inference model is shown using dashed lines.
$\theta$ and $\phi$ are parameters of the generation and inference networks respectively.}
  \label{fig:vccap}
\end{subfigure} \\
\end{minipage}

\begin{minipage}{1.0\textwidth}
\begin{subfigure}{1.0\textwidth}
  \centering
  \includegraphics[width=0.95\textwidth]{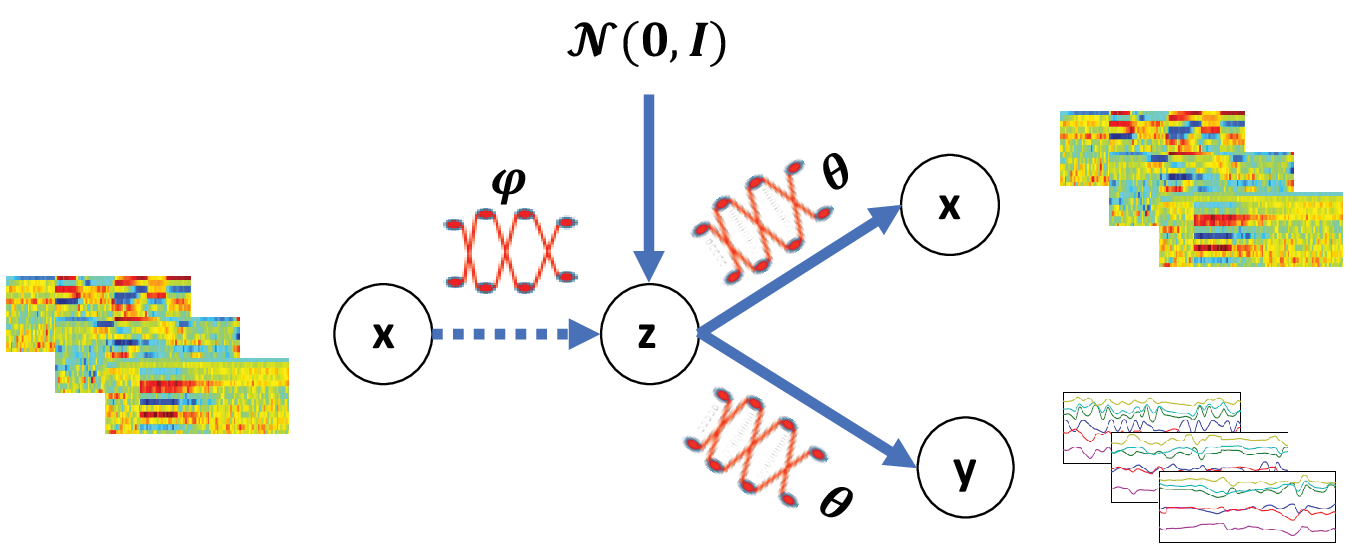}
  \caption{Using VCCA to learn representation for acoustic-articulatory pairs.
Articulation is the process of producing speech signals,
it is a natural parallel view of speech.
The mean value of the posterior of the latent variable $z$ can be used as a feature for downstream tasks.}
  \label{fig:vcca-acoustic}
\end{subfigure}
\end{minipage}

\caption{Multi-view variational representation learning}
\label{subsec:vcca}
\end{figure*}

We first give a very brief introduction to deep variational canonical correlation analysis (VCCA) ~\citep{wang2016deep}, illustrated in Figure ~\ref{fig:vcca}.
Given paired views $x$ and $y$ of the same object,
we expect to learn a latent variable $z$ such that $x$ and $y$ become independent conditioned on $z$. 
We consider a joint density of the latent variable $z$ and observation $(x,y)$ (as described in the graphical model shown in Figure ~\ref{fig:vcca}):
\begin{equation}
p_{\theta}(x,y,z) = p_{\theta}(x,y|z)p_{\theta}(z) = p_{\theta}(x|z)p_{\theta}(y|z)p_{\theta}(z)
\label{eqn:joint-vcca}
\end{equation}

The marginal density for $(x,y)$ we try to maximize is:
\begin{equation}
p_{\theta}(x,y) = \int_z p_{\theta}(x,y,z) dz
\label{eqn:joint-marginal-vcca}
\end{equation}

VCCA uses an inference network parameterized by $\phi$ to infer the approximate posterior $q_{\phi}(z|x)$ of the ground-truth posterior $p_{\theta}(z|x,y)=\frac{p_{\theta}(x,y,z)}{p_{\theta}(x,y)}$.
Similarly to  VAE, VCCA jointly trains the inference network (with parameters $\phi$) and generative model (with parameter $\theta$) by maximizing the Evidence Lower Bound (ELBO) of $\log p_{\theta}(x,y)$ (see Section ~\ref{sec:derive-elbo-vcca} in the Appendix for derivation):
\begin{eqnarray}
\log p_{\theta}(x,y)
&\geq & \mathbb{E}_{q_{\phi}(z|x)} \Bigg\{ \log \left\{ p_{\theta}(x|z) \right\} + \log \left\{ p_{\theta}(y|z) \right \} \Bigg\} \nonumber \\
&-& \KL{q_{\phi}(z|x)}{p_{\theta}(z)}
\label{eqn:vcca-elbo}
\end{eqnarray}

Note that in Equation ~\eqref{eqn:vcca-elbo},
the approximate posterior is inferred using $x$ only;
this is particularly useful when we only have access to the view of $x$ in the downstream tasks.
In general, we can parameterize the posterior based on all observations that are accessible at test time. \par

Though two paired objects share information, 
each object also has object-specific information that is not shared with the other.
Thus, the latent variable $z$ does not only need to encode information shared by $x$ and $y$, but also view-specific information in order to reconstruct both $x$ and $y$.
In order to encourage $z$ to focus more on view-shared information,
VCCA is extended to incorporate latent variables that are specific for each view.
As view-specific latent variables can also encode the view-specific information,
$z$ does not need to encode this information. \par

This method is dubbed VCCA-private (VCCAP, ~\citep{wang2016deep}),
whose graphical model is illustrated in Figure~\ref{fig:vccap}.
Here, the private variables are $h_1$ and $h_2$ for view 1 and view 2 respectively,
and the data likelihood is
\begin{equation}
p_{\theta}(x,y) = \int_{z,h_1,h_2}p_{\theta}(x,y,z,h_1,h_2)dz dh_1 dh_2
\end{equation}
We can factor the joint density according to the graphical model (Figure ~\ref{fig:vccap}):
\begin{eqnarray}
p_{\theta}(x,y,h_1,h_2,z)
&=& p_{\theta}(x,y | h_1,h_2,z) p_{\theta}(z,h_1,h_2) \nonumber \\
& = & p_{\theta}(x|h_1,z) p_{\theta}(y|h_2,z) p_{\theta}(h_1) p_{\theta}(h_2) p_{\theta}(z)
\label{eqn:vccap-generation}
\end{eqnarray}
The inference network defined in Figure ~\ref{fig:vccap} can be written as
\begin{eqnarray}
q_{\phi}(h_1,h_2,z|x,y) = q_{\phi}(h_1|x) q_{\phi}(h_2|y) q_{\phi}(z|x)
\label{eqn:vccap-inference}
\end{eqnarray}

The ELBO of VCCAP is
\begin{eqnarray}
\log p_{\theta}(x,y)
&\geq& \mathbb{E}_{q_{\phi}(z|x)q_{\phi}(h_1|x)} \bigg\{ \log p_{\theta}(x|z,h_1) \bigg\} + \mathbb{E}_{q_{\phi}(z|x)q_{\phi}(h_2|y)} \bigg\{ \log p_{\theta}(y|z,h_2) \bigg\} \nonumber \\
&-& \KL{q_{\phi}(z|x)}{p_{\theta}(z)} \nonumber \\
&-& \KL{q_{\phi}(h_1|x)}{p_{\theta}(h_1)} -\KL{q_{\phi}(h_2|x)}{p_{\theta}(h_2)}
\label{eqn:elbo-vccap}
\end{eqnarray}

$\mathbb{E}_{q_{\phi}(z|x)} \left\{ z \right\} $ is the representation we use in downstream tasks.
As in VCCA, $z$ is inferred only using $x$.
$h_1$ and $h_2$ are view-specific, so $h_1$ is inferred using $x$ only while $h_2$ is inferred using $y$ only.
See Appendix Section ~\ref{sec:derive-elbo-vccap} for the ELBO of VCCAP and its derivation. \par

We also tune the weight ($\beta$) of the KL divergence term(s) for VCCA(P) as we did for VAE. Similarly to VIB ~\citep{alemi2016deep} which provides an interpretation of a motivation for tuning $\beta$ for VAE from an information bottleneck perspective,
VCCA with $\beta \neq 1$ can also be interpreted from the perspective of information bottleneck;
See Appendix Section ~\ref{sec:multi-vib} for derivation. \par

\subsection{Multi-view representation learning experiments}
\label{subsec:multi-view-exp}

In this section, we experimentally compare the quality of learned acoustic features using several multi-view learning approaches.
We focus on how the learned representations can improve speech recognition tasks.
The methods we compare are as follows:
%\weiran{itemize the methods?}
\begin{itemize}
\item[1] Deep canonical correlation analysis (DCCA) ~\citep{andrew2013deep, wang2015deep},
\item[2] Triplet network ~\citep{bertinetto2016fully} with contrastive loss (CONTRAST) ~\citep{hermann2014multilingual,hoffer2015deep,liu2016large},
\item[3] Variational canonical correlation analysis (VCCA) ~\citep{wang2016deep},
\item[4] VCCA-private (VCCAP) ~\citep{wang2016deep}
\item[5] VAE, which performs single-view representation learning. 
\end{itemize}

We use the same XRMB dataset we used in Section ~\ref{sec:data}.
Here we use both acoustic and the paired articulatory measurements for feature learning,
but only use acoustic measurements when training speech recognizers.
Besides this, 
all other experimental setups (e.g., window size, $6-$fold learning procedure) 
for both the multi-view representation learning step and speech recognition step are the same as those described in Section ~\ref{sec:data}.
The articulatory measurements are horizontal/vertical displacement of $8$ pellets attached to several parts of the vocal tract,
which are also concatenated over the same time window as that used for acoustic measurements.
The results are shown in Table ~\ref{tab:multi-view-compare-framework}. 

\begin{table} [htbp]
\centering
\begin{tabular}{| l | r |}
\hline
 Methods & Averaged Test PER ($\%$) \\
\hline \hline
Baseline (MFCCs) & 11.3 \\
\hline
VAE, W=15 & 9.9 \\
\hline
DCCA, W=15 & 11.3 \\
\hline
CONTRAST, W=15 & 10.5 \\
\hline
VCCA, W=15 & 9.2 \\
\hline
VCCAP, W=15 & 8.9 \\
\hline
VCCAP, W=35 & 7.2 \\
\hline
VCCAP, W=71 & 7.5 \\
\hline
VCCAP, W=71+35 & 6.5 \\
\hline
\end{tabular}
\caption{Comparison of different multi-view representation learning methods.
To incorporate context information, the inputs to all feature learning models are $W$-frame windows centered at each frame of each acoustic sequence.
The dimensionality of learned representations is $70$ for all models.
The dimensionality of private variables is $30$ if used.
For each feature learning method, we only report the test set performance of the model with the best $6-$fold averaged dev set PER.
``VCCAP, W=71+35'' refers to a model using ``prior updating'' which is explained in Section ~\ref{sec:pu}
(In short, a model is first trained with a $35-$frame input,
then when training a model with $71-$frame input,
the prior of each $71-$frame window is set to the posterior inferred from the central $35-$frames.)}
\label{tab:multi-view-compare-framework}
\end{table}

According to Table ~\ref{tab:multi-view-compare-framework}, 
when using window size $15$, VCCA(P) outperforms baseline VAE as well as other multi-view methods.
This observation shows the effectiveness of a paired second view in learning better generalized features.
We also observe a similar trend with respect to context window size as observed in Table ~\ref{tab:window-size},
where a larger window size helps to learn better representations,
but learning good representations for very large window sizes is still challenging.
Finally, as VCCAP models both shared and view-specific attributes,
we expect the shared latent variable can better dial in on label-related information and thus enhance speech recognition performance.
To drive view-specific latent variables more focusing on capturing the view-specific information, we use relatively small latent variable sizes for $h_1$ and $h_2$.
As indicated in Table ~\ref{tab:multi-view-compare-framework}, we find that VCCAP works better than VCCA as expected. \par

\section{Prior Updating}
\label{sec:pu}

Here we describe a technique named ``prior updating'' to enhance sequential representation learning. Given a context window $\left\{x_{t-K},\cdots,x_t,\cdots,x_{t+K} \right\}$,
denote the learned posterior of this window as $q(z|x_{t-K:t+K})$.
When learning from a larger context window $\left\{ x_{t-W}, \cdots,x_{t+W} \right\}$ where $W>K$,
we can use $q(z|x_{t-K:t+K})$ instead of $\mathcal{N}(0,I)$ as the prior specific to this $2W+1$ window,
and maximize the sample-specific lower bound
\begin{equation}
\mathbb{E}_{q_{\phi} \left( z|x_{t-W:t+W} \right) } \Big\{ \log{p_{\theta} \left( x_{t-W:t+W}|z \right) } \Big\} - \beta \KL{q_{\phi}\left( z|x_{t-W:t+W}\right) }{ q \left( z|x_{t-K:t+K} \right) }
\label{eqn:sample-specific-elbo}
\end{equation}
Please note, the posterior $q$ without subscript is not being updated in Equation ~\eqref{eqn:sample-specific-elbo}.
The motivations here are twofold:
First, a nearby frame of $x_t$ probably has more cues for predicting $x_t$ than a distant frame.
For example, for speech signals, a nearby frame is more likely within the same phone duration as $x_t$ while a distant frames is not.
Similarly, when generating future video given what we have seen so far, the generated frames are likely to agree with the ground-truth video in the first few frames, but the discrepancy dramatically increases as prediction goes on.
Recent sequential representation learning works,
like contrastive predictive coding (CPC) ~\citep{oord2018representation,schneider2019wav2vec},
also treat nearby future time steps (of time step $t$) as ``positive samples'' and distant time steps as ``negative samples'' when learning the representation for time step $t$.
Thus, when learning representations for time step $t$, it makes sense to more focus on time steps closer to $t$ rather than distant frames. \par

Second, as we can see from Table ~\ref{tab:window-size} and ~\ref{tab:multi-view-compare-framework},
a moderately sized context window works best for representation learning.
When the size of the context window is too small, 
the input contains less contextual information for representation learning.
However, when using a very large context window,
the number of parameters increases considerably, but the number of training samples does not increase.
More parameters would make the optimization more difficult and fewer training samples would make the model more easily overfit.
Thus, as we already have good sample-specific posteriors learned from smaller context windows, 
it makes sense to use these sample-specific posteriors to further guide the learning in a more general context. \par

It is possible to tune the parameter $\beta$ of equation ~\eqref{eqn:sample-specific-elbo} to control the magnitude to which sample-specific priors affect the learned representation.
When using very large $\beta$,
Equation ~\eqref{eqn:sample-specific-elbo} tries to copy the representation already learned using a smaller context window.
Using smaller $\beta$, the learned posterior $q(z|x_{t-K:t+K})$ provides weaker regularization and provides more flexibility to the inference network.
Similarly, we can derive a sample-specific lower bound for VCCA as given by Equation ~\eqref{eqn:sample-specific-elbo-vcca}.
Please see Equation ~\eqref{eqn:elbo-vccap-prior-updating} in Appendix Section ~\ref{subsec:vccap-prior-updating} for the ELBOs for VCCAP when using prior updating.
\begin{eqnarray}
&& \mathbb{E}_{q_{\phi} \left( z|x_{t-W:t+W} \right) } \Big\{ \log{p_{\theta} \left( x_{t-W:t+W}|z \right) } + \log{p_{\theta} \left( y_{t-W:t+W}|z \right) } \Big\}  \nonumber \\
&-& \beta \KL{q_{\phi} \left( z|x_{t-W:t+W} \right) }{q \left( z|x_{t-K:t+K} \right) }
\label{eqn:sample-specific-elbo-vcca}
\end{eqnarray}

As shown In Table ~\ref{tab:multi-view-compare-framework}, VCCAP with $71-$dimensional input actually works worse than VCCAP with only $35-$dimensional input even though VCCAP with $71D$ input sees more contextual information.
We then use VCCAP-35 to infer the posteriors of all $35-\text{frame}$ windows, and use these learned posteriors to further guide the learning of VCCAP-71.
With the help of these sample-specific posteriors, 
VCCAP-71 finally achieves a $6-$fold averaged test set PER of $6.5\%$, 
which outperforms both VCCAP-35 and VCCAP-71 with generic prior. \par

\section{Cross-Domain Multi-View Feature Learning}
\label{sec:cross-domain-unsupervised}

\begin{figure*}[htbp]
  \centering
  \includegraphics[width=1.0\textwidth]{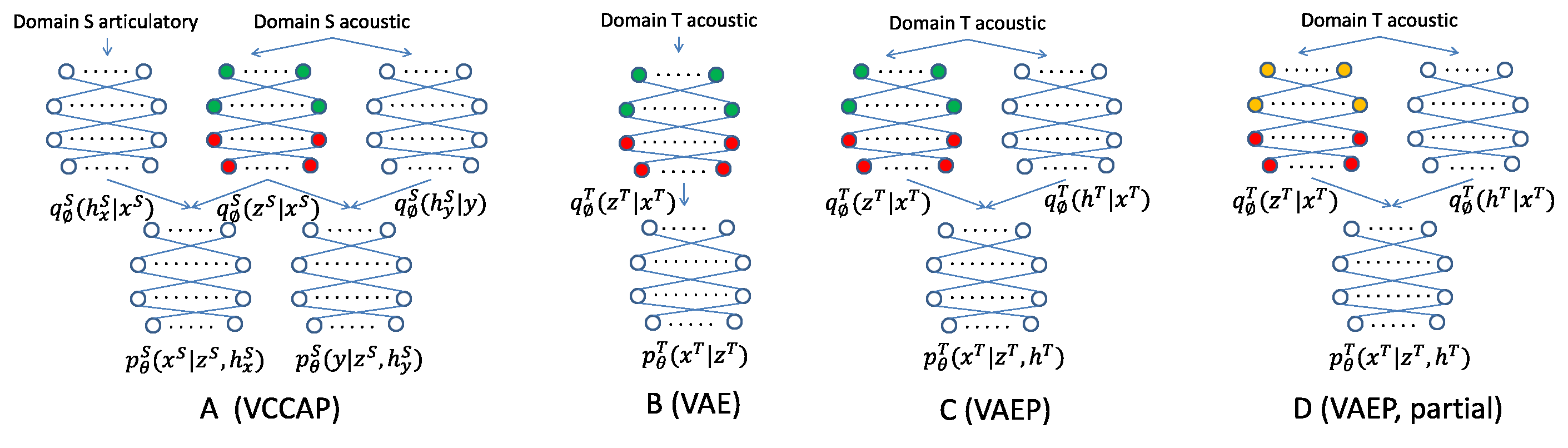}
  \caption{(A): VCCAP model for multi-view data; (B): Variational autoencoder (VAE) for the target-domain acoustics. The projection network is the same as that of VCCAP (indicated by the colors); (C): Like (B), but with additional private variables for the target domain; (D): Like (C), but sharing only part of the projection network (in red) with VCCAP; the other layers (yellow) model domain-specific information. }
  \label{fig:models}
\end{figure*}

Motivated by the observation that paired information is not always available for a given dataset, 
we want to explore the possibility of using cross-domain multi-view information to enhance representation learning in a target dataset.
More specifically, we will consider the learning scenario where we have both source and target domains, 
but we only have paired information in the source domain.
In order to use multi-view information outside the target domain, 
we train an encoder (which learns acoustic features) shared by the source domain and target domain.
We use VCCAP to learn representations shared by both acoustic measurements and the paired view (e.g., articulatory measurement) in the source domain,
and use VAE, whose acoustic encoder is shared with VACCP, to learn acoustic representations in the target domain.
Figure ~\ref{fig:models} describes the few variants we explore in this section. More details are explained in the following sections.
We expect the multi-view representations learned using the source dataset will help with representation learning in the target dataset via a joint training process.

\subsection{Joint modeling of source and target domains}
\label{subsec:joint_model}

\begin{figure*}[htbp]
  \centering
  \includegraphics[width=1.0\textwidth]{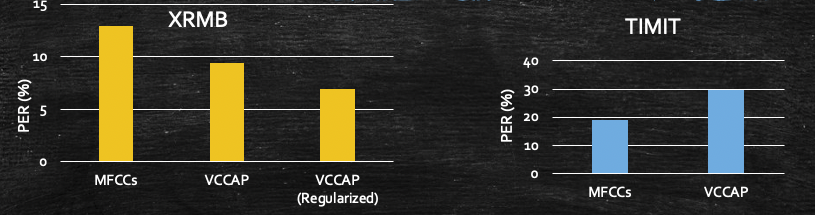}
  \caption{Multi-view representations can help in-domain recognition task but not out-domain recognition task.}
  \label{fig:cross-domain-effect}
\end{figure*}

Using the learned VCCAP network $q_\phi^S(z^S|x^S)$ directly in a target domain does not in general work well if there is significant domain mismatch.
Figure ~\ref{fig:cross-domain-effect} shows a clear example of the effect of domain mismatch.
We train a VCCAP model on XRMB as described in earlier chapters.
The representations inferred by VCCAP on XRMB clearly improve over the baseline on XRMB.
However, the representations inferred by VCCAP on TIMIT clearly hurt the recognition task on TMIT.

To address this domain mismatch effect,
we learn a projection network for the target domain, $q_\phi^T(z^T|x^T)$, that is informed by the source-domain model in various ways. 
One way is to have the two networks fully/partially share parameters, and train the two jointly in a unified model. 
Figures~\ref{fig:models} B,C,D show several options for modeling the target-domain data. 
Architectures B, C, and D can each be combined with A to form three different models that can be viewed as ``weakly supervised'' by the cross-domain articulatory data. 
By ``combining'', here we mean that training is done with a loss that is a linear combination of the multiple relevant losses.

B models the target-domain acoustics with a variational autoencoder (VAE), trained jointly with VCCAP with a shared projection network.
Model C (``VAEP'') is similar to B, but with an additional private variable $h^T$ and corresponding private projection network that is specific to the target domain.
Depending on the degree of domain mismatch, sharing the complete VCCAP network between source and target domains may still be too restrictive. Model D is similar to C, but with only a subset of the VCCAP layers shared.
The hidden layers that are closer to the acoustic input (in yellow) are treated as domain-specific, while the layers closer to the output features (in red) are shared between domains.
The objective function for C and D, for one acoustic frame $x^T$, can be written as:
\begin{eqnarray}
L_{VAEP}(x^T) := \bbE_{\{q_{\phi}^T(z^T|x^T)q_{\phi}^T(h^T|x^T)\}} \Big[ \log\big(p_{\theta}^T(x^T|z^T,h^T) \big) \Big] \nonumber \\
-KL\big(q_{\phi}^T(z^T|x^T)||p(z^T)\big)-KL\big(q_{\phi}^T(h^T|x^T)||p(h^T)\big) 
\label{eqn:vaep}
\end{eqnarray} 
The objective for the combined model on $S$ and $T$ is
\vspace{-.05in}
\begin{equation}
(1-\beta)\bbE(L_{VCCAP}(x^S,y^S)) + \beta \bbE(L_{VAEP}(x^T))
\label{eqn:cross-domain-unsupervised}
\end{equation}
where $\beta>0$ is a hyper-parameter and $p(h^T)$ and $p(z^T)$ are set to $\mathcal{N}(0,I)$. The feature vector used for downstream tasks is the mean of $q_{\phi}^T(z^T|x^T)$. 
In practice, we train all of the models with minibatch gradient descent methods. We use a joint loss for data from both domains by taking each minibatch to include some data drawn independently from each domain; for each domain-specific loss term we use the corresponding subset of the minibatch. \par

One way to address the potential domain mismatch when using the learned features in a new domain is to add explicit domain adaptation layers.
In this approach, the projection network $q_{\phi}^S(z^S|x^S)$ is shared with the target domain.  
However, two additional fully connected layers, one with ReLU~\citep{maas2013rectifier} activation and one linear transformation, are used to transform the target input data before being fed to the VCCAP projection network.  
The output of this composed projection network is the input to the recognizer. 
All training is done end-to-end. This simple model is labeled ``VCCAP + adaptation layers'' in Table ~\ref{tab:timit}. 

\subsection{Joint training of target recognizer and features}
\label{subsec:joint_training}
An alternative to explicit domain adaptation is to adapt implicitly, by keeping the feature projection structure fixed but jointly learning it along with the recognizer.  
As in previous chapters, for recognizers we use bidirectional long short-term memory (LSTM) recurrent neural networks (RNNs)~\citep{hochreiter1997long,schuster1997bidirectional} trained with the connectionist temporal classification (CTC) loss~\citep{graves2006connectionist}, which have recently achieved very good performan in ASR (e.g.,~\citep{zweig2017advances}).
This ``implicit adaptation approach" may be preferable over adding extra layers, which can result in an overparameterized model.
To be more concrete, for the feature learning model we will use VCCAP+VAEP from the previous section, since (as will be shown in Section~\ref{subsec:cross-domain-experiment}) it is the best-performing unsupervised model (although the approach in this section can be used with any of the feature learning losses).  
Denoting one target-domain acoustic utterance $\mathbf{x}^T$ and one frame $x^T$,
the objective function is
\begin{eqnarray}
\alpha \bigg \{(1-\beta)\bbE\left[ L_{VCCAP}(x^{S},y) \right] + \beta \bbE \left[ L_{VAEP}(x^T) \right] \bigg\} \nonumber \\
+ (1-\alpha) \bbE \big[ L_{CTC} \left( \mathcal{F}_{VAEP}(\mathbf{x}^T) \right) \big]
\label{eqn:cross-domain-multitask}
\end{eqnarray}
where $\mathcal{F}_{VAEP}(\mathbf{x}^T)$ is the sequence of means of $q_{\phi}^T(z^T|x_i^T)$ for all frames $i$ in $\mathbf{x}^T$; these are the learned features that are used as input to the target-domain recognizer. Here, $\alpha$ is a tunable tradeoff parameter between the recognizer and feature learning losses.

\subsection{Joint training of source and target recognizers}
\label{subsec:shared_blstm}

Finally, if we have access to labels for both the source and target domains, we may be able to benefit from jointly training recognizers for both domains, without direct use of the learned feature projection network in the target domain. 
In this approach, the source-domain recognizer uses VCCAP-based features fed into an LSTM-CTC recognizer, and the target-domain recognizer uses the original acoustic features fed into another LSTM-CTC recognizer.

We only share the topmost recurrent layer of the two recognizers for the two domains, which are trained jointly.
The idea here is to implicitly use the cross-domain articulatory data by encouraging the two recognizers to agree.  Although source-domain labels are present, the articulatory data may still help as a form of regularizer.
While this may seem like a very weak use of the articulatory data, this approach obtains surprisingly good improvements in phonetic recognition on the target domain (see Section ~\ref{subsec:cross-domain-experiment}).

\subsection{Experiments}
\label{subsec:cross-domain-experiment}

We use three datasets: the U. Wisconsin X-ray microbeam database (XRMB) of simultaneous acoustic and articulatory recordings~\citep{westbury1990x}, TIMIT~\citep{garofolo1993darpa}, and Wall Street Journal (WSJ)~\citep{paul1992design}. 
Details regarding the three datasets and our experimental setups on XRMB and WSJ can be found in Section ~\ref{sec:data}.

Unlike in Chapter ~\ref{cha:feedforward} where we use $120$-$D$ features for TIMIT, in this section, the acoustic input features for TIMIT are also $39$-$D$ MFCCs, but speaker-normalized (via mean and variance normalization).
This is for the purpose of directly applying VCCAP trained on XRMB on TIMIT. \par
Unless specified otherwise, the inputs to both VCCAP and VAE(P) are the acoustic features concatenated over a $15$-frame window centered at each frame.  The inputs to the RNN recognizers are either MFCCs, filterbank features, or learned acoustic feature sequences without windowing.
The final task is phonetic recognition, evaluated using phonetic error rate (PER). XRMB is always used as the source domain.  We consider three source-target domain pairs:
\begin{itemize}
\item[1] XRMB(35) $\rightarrow$ XRMB(12): This setup follows earlier work~\citep{wang2015deep, wang2016deep, tang2017acoustic}.  We split the 47 XRMB speakers into two disjoint sets, consisting of 35 and 12 speakers respectively.  We treat the 35 speakers as the source ``domain'' and the 12 speakers as the target ``domain'', and we do not access the articulatory data for the target speakers.  We perform recognition experiments in a $6$-fold setup on the $12$ target speakers, where in each fold we train on $8$ speakers, tune on $2$, and test on $2$; we then report the average performance over the $6$ test sets.  This can be viewed as a very mild case of cross-domain learning.  As shown in prior work, in this setting we can improve target speaker performance by simply using features learned on the source speakers.  Our experiments in this setting are intended to ensure that our approaches still work in this mild case.
\item[2] XRMB $\rightarrow$ TIMIT: In this setting we use XRMB
as the source domain and TIMIT as the target domain.  One prior paper has explored an application of multi-view feature learning from XRMB to TIMIT, but in a more limited setting with fewer speakers and with shallow (kernel-based) feature learning models~\citep{Arora2013MultiviewCA}.
\item[3] XRMB $\rightarrow$ WSJ: Here we use XRMB
as the source domain and WSJ as the target domain.  Whereas XRMB and TIMIT have similar amounts of data, WSJ is much larger, so we may expect that any external multi-view data will have a smaller effect.  We include both TIMIT and WSJ as target domains, both to test this possibility and more generally to measure applicability across target domains.
\end{itemize}

We implement our models using TensorFlow~\citep{Abadi2016TensorFlowAS}.  
In setting (1), all models are trained to $300$ epochs; that is, all frames in the target dataset are used $300$ times. We 
optimize with Adam~\citep{kingma2014adam} (except for TIMIT recognizers which are optimized with vanilla SGD), dropout~\citep{srivastava2014dropout} at a rate of $0.2-0.4$ (tuned for each setting), 
and a batch size of $200$ frames for both VCCAP and VAE(P).  For recognizer training, we use a batch size of $1$ utterance for XRMB, $2$ for TIMIT and $16$ for WSJ.
For XRMB and TIMIT, we train the RNN recognizers (and multitask models) to $50$ epochs; for WSJ we train to $20$ epochs. 
These training settings (and also trade-off hyperparameters $\alpha$ (Eq.~\ref{eqn:cross-domain-multitask}) and $\beta$ (Eq.~\ref{eqn:cross-domain-unsupervised} and ~\ref{eqn:cross-domain-multitask})) were determined by tuning on the corresponding development sets.
All recognizers are speaker-independent (no speaker adaptation),
and performance is evaluated via the dev/test PER. We use $2$-layer bidirectional LSTM recognizers for all experiments (when we need to train a recognizer).

\begin{table}[t]
  \caption{Phonetic error rates (PER, \%) for XRMB(35)$\rightarrow$XRMB(12):  Averaged PER over 6 folds of RNN CTC recognizers trained on acoustic features learned by different models.}
  \label{tab:xrmb}
  \centering
  \begin{tabular}{|l||r|}
    \hline
 Method & Test \\
	\hline\hline
 1. Recognizer (baseline) & 12.9 \\
        \hline
 2. Recognizer (15) & 17.5 \\
        \hline
 3. VCCAP & 9.4 \\
	\hline
 4. VCCAP + Recognizer & 8.9 \\
        \hline
 5. VCCAP+VAEP + Recognizer & { 7.3} \\
        \hline
 6. VAEP+VAEP + Recognizer & 8.6 \\
  	\hline
 7. VAEP+VAEP & 14.2 \\
  	\hline
 8. VCCAP+VAEP &  7.4 \\
        \hline
  \end{tabular}
\end{table}

\subsubsection{Main results}
\label{subsubsec:corss-domain-main-results}

\begin{figure*}[htbp]
  \centering
  \includegraphics[width=0.95\textwidth]{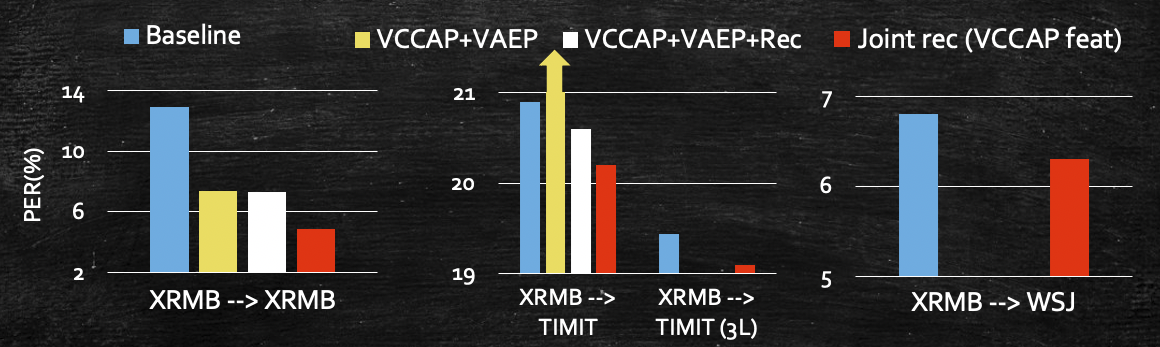}
  \caption{ Overview of experimental results.  ``VCCAP+VAEP" represents the unsupervised model described in Section ~\ref{subsec:joint_model}; ``VCCAP+VAEP+Rec" represents the model with target domain recognizer jointly trained as described in Section ~\ref{subsec:joint_training}; ``Joint rec (VCCAP feat)" represents the model with both source and target domain recognizers jointly trained as described in Section ~\ref{subsec:shared_blstm}.}
  \label{fig:cross_overview}
\end{figure*}

Our recognition results are given in Tables~\ref{tab:xrmb},~\ref{tab:timit} and ~\ref{tab:wsj},
and are summarized and visualized in Figure ~\ref{fig:cross_overview}.
We begin by summarizing our baseline and best final results, and explore more detailed comparisons in the following sections. 
According to Figure ~\ref{fig:cross_overview},
cross-domain multi-view information generally helps in the three scenarios XRMB(35)$\rightarrow$XRMB(12), XRMB $\rightarrow$ TIMIT and XRMB $\rightarrow$ WSJ.
Though VCCAP+VAEP works very well on the ``XRMB $\rightarrow$ XRMB" scenario,
it does not improve upon baselines in the target domain when there is clear domain mismatch.
Supervised fine-tuning seems to be necessary to overcome the domain mismatches. \par

Our best model in the XRMB(35)$\rightarrow$XRMB(12) setting is VCCAP+VAEP jointly trained with a RNN recognizer. The RNN recognizer is identical to the baseline recognizer, except that the input to this recognizer is the mean of $z^T$ rather than MFCCs.  Our best $6$-fold averaged PER is $7.3\%$, while the baseline is $12.9\%$.  
In the XRMB $\rightarrow$ TIMIT and XRMB $\rightarrow$ WSJ settings, our best model is the jointly trained RNN with shared upper layer between the source and target domains, as described in~\ref{subsec:joint_training}.  
For TIMIT our best $2$-layer model yields $1.4/0.6\%$ absolute dev/test set improvements.
The best WSJ result improves by $3.4/2.9\%$ absolute over the baseline.
It is to be expected that the best performance is obtained in the setting with the most supervision.
However, even with less supervision, our best models in Tables~\ref{tab:xrmb},~\ref{tab:timit} and ~\ref{tab:wsj} still outperform the baselines.

\subsubsection{XRMB(35) $\rightarrow$ XRMB(12)}
\label{subsubsec:xrmb}
Table~\ref{tab:xrmb} gives the $6$-fold averaged test set phonetic error rates on XRMB(12). Row $1$ corresponds to the baseline, where the RNN recognizer with MFCC inputs is trained with the CTC loss only. Since our feature learning experiments use windowed $15$-frame features, we also include (row $2$) the performance of a baseline RNN recognizer that uses windowed $15$-frame MFCC inputs to make sure that any improvement is not due simply to the concatenated inputs; in fact, this baseline does much worse. Row $3$ uses acoustic features learned from an unsupervised VCCAP model trained on XRMB(35), reproducing the setting used in prior work~\citep{tang2017acoustic}.\footnote{The results here are improved over the ones in~\citep{tang2017acoustic} due to improved optimization and tuning, and a new TensorFlow version.}
Row $4$ corresponds to an end-to-end version of row $3$.
The improvement from row $3$ to row $4$ shows the benefit of learning the features jointly with the recognizer. Row $5$ adds private target-domain features learned with VAEP to the joint training, which produces the best results.

One possibility is that we are mainly benefiting from the extra acoustic data. To check this hypothesis, we train a recognizer jointly with VAEP for both the target and source domains, using just the acoustic data. Indeed, this model (row $6$) also improves greatly over the baseline, obtaining an average PER of $8.6\%$. However, we still have a sizable gain from $8.6\%$ to $7.3\%$ PER by training on the multi-view data. Row $7$ uses the same model as row $6$ for modeling the acoustic input of both source and target domains, but without joint training with a recognizer. Row $8$ uses the same feature model as row $5$, but also without jointly training with a recognizer. The large gap between rows $8$ and $7$ indicates the significant advantage of using the unlabeled external-domain acoustic-articulatory pairs over extra acoustic inputs from external domains. 
In these XRMB experiments, the source and target ``domains'' are very well matched, and we always use models with shared projection networks across domains.  In the next two subsections, we consider the two settings with much larger domain mismatch, and include experiments with partially shared projection networks.

\subsubsection{XRMB $\rightarrow$ TIMIT}
\label{subsubsec:timit}

\begin{table}[t]
  \caption{PER when the target domain is TIMIT. ``Partial" means the projection networks of the two domains are partially shared (Figure ~\ref{fig:models} A, D).}
  \label{tab:timit}
  \centering
  \begin{tabular}{|l||r|r|}
    \hline
 Method & Dev & Test \\
	\hline\hline

 1. Recognizer (baseline) & 19.8 & 20.8 \\
        \hline
 2. Recognizer (15) & 22.8 & 24.2 \\
        \hline
 3. VCCAP & 29.7 & - \\
        \hline
 4. VCCAP + adaptation layers & 19.0 & - \\
        \hline
 5. VCCAP+VAEP & 25.3 & - \\
        \hline
 6. VCCAP+VAEP + Recognizer  & 19.2 & - \\
        \hline
 7. VCCAP+VAEP, partial & 24.9 &  - \\
        \hline
 8. VCCAP+VAEP, partial + Recognizer & 18.8 &  - \\
        \hline
 9. Two recognizers (acoustic input) & 19.2 & 20.6 \\
        \hline
 10. Two recognizers (VCCAP feature input) & { 18.4} & { 20.2} \\
        \hline

  \end{tabular}
\end{table}

In table ~\ref{tab:timit}, row $1$ is the baseline recognizer, and row $2$ again shows that concatenating acoustic frames does not help.
Row $3$ shows that directly using VCCAP learned on XRMB fails to generalize to TIMIT; this differs from the XRMB(35) $\rightarrow$ XRMB(12) experiments, presumably due to the much larger domain difference here. Row $5$ introduces domain-specific private variables; the improvement over row $3$ shows their benefit. Row $7$ is similar to row $5$ but with a partially shared projection (Section~\ref{subsec:joint_model}).
Rows $4$, $6$, and $8$ use the target domain labels via end-to-end joint training of the features and recognizer. Compared to the XRMB(35) $\rightarrow$ XRMB(12) setting, we obtain a smaller improvement by learning features using XRMB, but there is still an appreciable improvement.
Row $10$ corresponds to the case where we use the training labels for both source and target domains, and train two domain-specific recognizers jointly with a final shared layer (Section~\ref{subsec:shared_blstm}), which produces our best results.  Again, we check whether this improvement could be due solely to having additional acoustic data by training a similar jointly trained pair of recognizers on the acoustic input only in the two domains; the result, shown in row $9$, is worse than row $10$, indicating that only having the cross-domain acoustic data is less helpful than cross-domain acoustic-articulatory pairs.

\subsubsection{XRMB $\rightarrow$ WSJ}
\label{subsubsec:wsj}

In the previous sections we compared a variety of models on XRMB and TIMIT.  Based on the success of the supervised ``two recognizers'' approach in the XRMB $\rightarrow$ TIMIT setting, we only consider this approach for WSJ.
Similarly to 
the TIMIT experiments (rows $9$ and $10$ of Table~\ref{tab:timit}), we train source and target recognizers with the topmost layer shared, using either VCCAP features (Table~\ref{tab:wsj}, row $3$) or acoustic measurements (Table~\ref{tab:wsj}, row $2$) as input to the source-domain recognizer.
Somewhat surprisingly, the external acoustic data from XRMB improves the WSJ recognition performance despite the small relative size of XRMB. However, the additional use of the articulatory data in XRMB is much more helpful than the external acoustic data alone, similarly to the corresponding TIMIT results.

\begin{table}[t]
    \caption{Phonetic error rates for XRMB$\rightarrow$WSJ experiments. The baseline is trained to epoch $20$ while the other two rows only be trained to epoch $15$.}
  \label{tab:wsj}
  \centering
  \begin{tabular}{|l||r|r|}
    \hline
 Method & Dev & Test \\
	\hline\hline
        \hline
 1. Recognizer (baseline) & 10.9 & 8.7 \\
        \hline
 2. Two recognizers (acoustic input) & 9.9 & 8.3 \\
        \hline
 3. Two recognizers (VCCAP feature input) & {7.5} & {5.8} \\
        \hline
  \end{tabular}
\end{table}

\section{Label Embedding for sequence data}
\label{sec:le}

\begin{figure*}[htbp]
  \centering
  \includegraphics[width=0.8\textwidth]{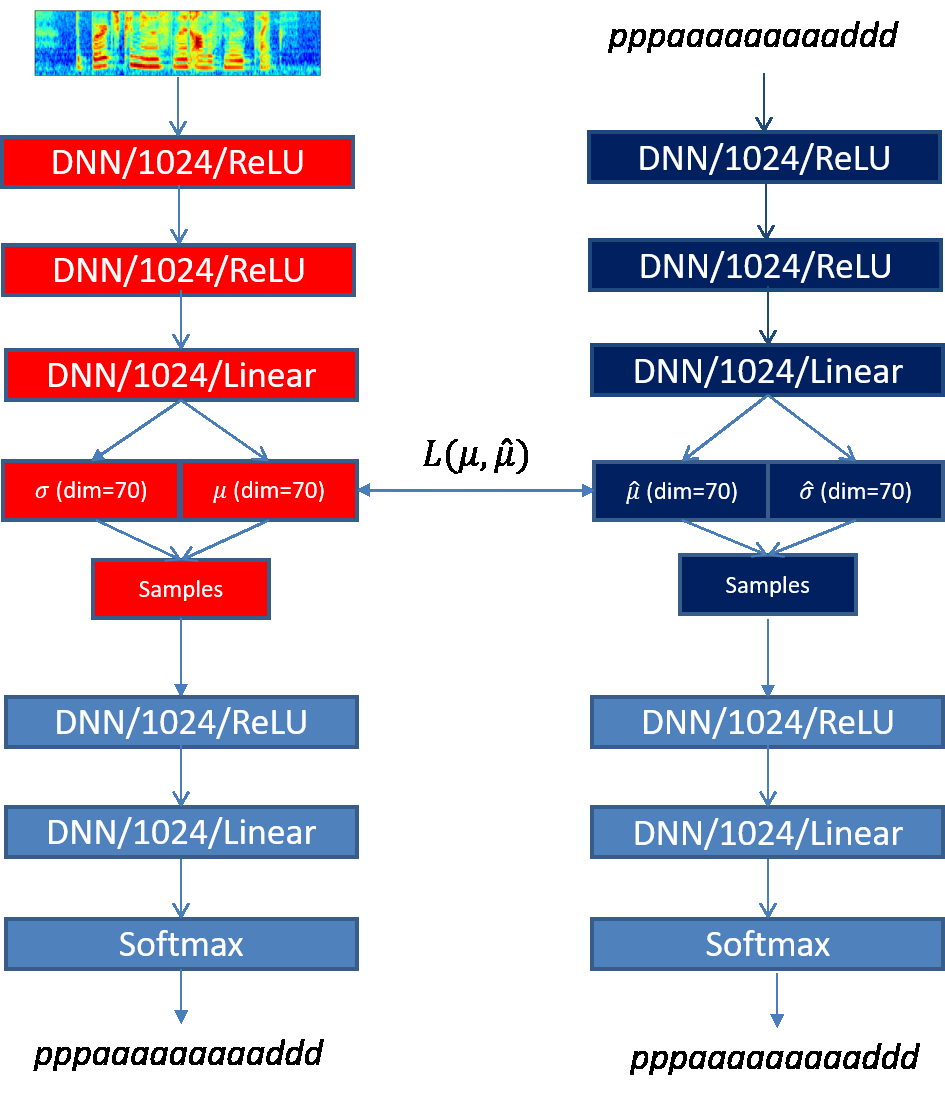}
  \caption{Illustration of our label embedding model.}
  \label{fig:lb}
\end{figure*}

Ground-truth labels themselves might exhibit internal structure (e.g., the frame labels of acoustic sequences typically exhibit segmental structure).
We thus explore regularizing representation learning by encoding label structure and using the compressed representation as a regularizer.
Our method is called ``label embedding" and is illustrated in Figure ~\ref{fig:lb}.
The basic idea is to train two encoders where one encodes the input while the other encodes the per-frame annotation.
In this situation, we expect the internal representations of the two encoders to be similar to each other. \par

Our work is motivated by ~\citep{karita2018semi, karita2019semi, mostajabi2018regularizing}. 
In ~\citep{karita2018semi, karita2019semi}, the authors show that autoencoding both speech and text while using a shared text decoder can enhance the intermediate representation of speech.
They use a domain loss to alleviate the domain difference between the speech and text domain representations.
Compared with ~\citep{karita2018semi, karita2019semi},
our work does not try to use large amount of unpaired text.
Our contribution is that we use similarity loss to push paired acoustic measurements and their frame labels to have similar representations.
We also explore the possible choices of similarity loss. \par

From the perspective of learning shared representations between an input and its paired label,
our work is much closer to ~\citep{mostajabi2018regularizing}.
In ~\citep{mostajabi2018regularizing}, the authors train an encoder over the set of annotations in the first phase, then train the actual encoder using the phase-one encoder as a regularizer. 
Unlike ~\citep{mostajabi2018regularizing} which uses a two-phase approach,
our approach is end-to-end.
Additionally, we focus on extracting temporal structure hidden in the sequential labels of the input sequences,
which is not done in ~\citep{mostajabi2018regularizing}. \par

We experiment on a phone classification task in this section.
In Figure ~\ref{fig:lb},
the input is a $15-$frame acoustic measurement vector with corresponding per-frame label ``pppaaaaaaaaaddd''.
We train two variational autoencoders with independent inference networks, but with a shared decoder.
The inference networks are $3-$layer feedforward neural networks as specified in Figure ~\ref{fig:lb}.
The learned posteriors are $\mathcal{N} \left( \mu,\diag{\sigma^2} \right) $ and $\mathcal{N} \left( \hat{\mu}, \diag{\hat{\sigma}^2} \right)$ respectively.
We use an auxiliary loss, $\mathcal{L}(\mu,\hat{\mu})$, to enforce the similarity between the $\mu$ and $\hat{\mu}$.
We do not force the two diagonal covariance matrices to be similar.
The shared decoder then takes the samples from two posteriors and reconstructs two copies of the label ``pppaaaaaaaaaddd''.
Assuming the label vocabulary size is $L$, the input to the ``label embedding'' branch is a $15$ $L-$dimensional one-hot vector,
while the output of the decoder is a $15\times L$ matrix indicating the probability of each label for each time step.
Given acoustic input $x_{1:15}$ and its label $l_{1:15}$,
and denoting the losses of the two branches as $\mathcal{L}_{\text{acoustic}}$ and $\mathcal{L}_{\text{label}}$ respectively,
the final objective we try to minimize is
\begin{eqnarray}
\mathcal{L}(x_{1:15},y_{1:15}) = && \alpha_1 \mathcal{L}_{\text{acoustic}} \left( y_{1:15} \vert x_{1:15} \right) \nonumber \\
&+& \alpha_2 \mathcal{L}_{\text{similarity}} \left( \mu(x_{1:15}), \hat{\mu}(l_{1:15}) \right) \nonumber \\
&+& (1-\alpha_1-\alpha_2)  \mathcal{L}_{\text{label}}(y_{1:15})
\label{eqn:lb-loss}
\end{eqnarray}
where $\alpha_1 \in [0,1]$, $\alpha_2 \in [0,1]$ and $\alpha_1+\alpha_2 \leq 1$.
There are a number of choices for similarity losses, including but not limited to $\mathcal{L}_1$, $\mathcal{L}_2$, cosine distance, contrastive loss and CCA loss.
For CCA loss, we introduce a Lagrange multiplier to incorporate orthogonal constraints in CCA loss as described in ~\citep{andrew2013deep}.
Appendix Section ~\ref{sec:similarity} provides details of CCA loss and other choices of similarity losses.  \par

We perform a phone classification experiment on two datasets XRMB and TIMIT.
The settings on incorporating context are the same as described in Section ~\ref{sec:15w-exp}.
The data partition for TIMIT is also the same as in Section ~\ref{sec:data}.
For XRMB, we use the $1500/236$ utterances we used for training and tuning in unsupervised training in Section ~\ref{sec:data} as our training and dev set.
We use all of the $12$ speakers we used for the $6-$fold experiments in Section ~\ref{sec:data} as our test set.
Other differences from Section ~\ref{sec:15w-exp} include:
\begin{itemize}
\item[1] There is no unsupervised learning here. We have access to per-frame labels.
\item[2] The task we are interested in is phone classification rather than phonetic recognition.
\item[3] For TIMIT, we use $48$ phone labels in training, and reduce to $39$ phone labels in evaluation, as is often done for TIMIT ~\citep{lee1988large}. The label mapping can be found in ~\citep{lopes2011phoneme}. For XRMB, we directly use $39$ phone labels in training and evaluation.
\end{itemize}

As we are concatenating a $15-$frame window centered at each frame,
we get $15$ predictions for each time step, denoted as a $15 \times L$ matrix $P_t^{15 \times L}$.
Following ~\citep{jaitly2014autoregressive}, we calculate a geometric mean of $P_t^{15 \times L}$.
The geometric mean $p_t^{1 \times L}= \left( p_{t}^{(1)},\cdots,p_{t}^{(L)} \right)$ for each time step is:
\begin{equation}
\text{For} \quad l\in \{1,\cdots,L \}, \quad p_t^{(l)} = \prod_{t'=t-7}^{t'=t+7} \Bigg\{  \left\{P_t^{(t',l)} \right\}^{\frac{1}{15}}  \Bigg\}
\label{eqn:geometric-evaluation}
\end{equation}

The predicted label for $x_t$ is $\argmax_{l}{p_t^{(l)}}$.
We then evaluate the dev set and test set prediction errors and summarize the results in Table ~\ref{tab:label-xrmb} and ~\ref{tab:label-timit}. \par

\begin{table} [htbp]
\centering
\begin{tabular}{| l | r | r |}
\hline
 Methods & Dev Error ($\%$) & Test Error ($\%$)\\
\hline \hline
1. Classifier only, no label embedding branch & 14.2 & 12.9 \\
\hline
2. Classifier plus label embedding branch, no $\mathcal{L}_{\text{similarity}}$ & 14.2 & - \\
\hline
3. $\mathcal{L}_2$ Loss & 14.2 & - \\
\hline
4. Cosine Loss & 14.0 & - \\
\hline
5. Contrastive Loss & 14.0 & - \\
\hline
6. CCA Loss & \textbf{13.5} & \textbf{12.4} \\
\hline
\end{tabular}
\caption{Label embedding experiments on XRMB.
Besides row $1$, all other rows correspond to models with different choices of similarity loss $\mathcal{L}$.}
\label{tab:label-xrmb}
\end{table}

\begin{table} [htbp]
\centering
\begin{tabular}{| l | r | r |}
\hline
 Methods & Dev Error ($\%$) & Test Error ($\%$) \\
\hline \hline
1. Classifier only, no label embedding branch & 21.8 & 23.2 \\
\hline
2. DART in ~\citep{jaitly2014autoregressive}\footnote{We use the frame error rate (FER) shown in the paper, not PER.} & - & 24.2 \\
\hline
3. Classifier plus label embedding branch, no $\mathcal{L}_{\text{similarity}}$ & 21.1 & - \\
\hline
4. $\mathcal{L}_2$ Loss & 21.2 & - \\
\hline
5. Cosine Loss &  21.1 & - \\
\hline
6. Contrastive Loss & \textbf{20.9} & \textbf{22.3} \\
\hline
7. CCA Loss & \textbf{21.0} & - \\
\hline
\end{tabular}
\caption{Label embedding experiments on TIMIT.
Except for row $1$ and $2$, all other rows correspond to models with different choices of similarity loss $L$.}
\label{tab:label-timit}
\end{table}

From the tables we observe that ``label embedding'' does impose some extra regularization in this setting on a phone classification task.
It consistently improves upon the baseline on XRMB and TIMIT.
Comparing the different similarity losses, we found CCA loss works best.
However, as shown in Equation ~\eqref{eqn:contrastive} in Appendix Section ~\ref{sec:similarity}, we do not use most recent contrastive loss variations (e.g. ~\citep{chen2017beyond,he2016multi}), and we only use one negative sample -- a permutation of $l_{1:15}$.
Contrastive loss-based joint models have the potential to work better if using better and more negative samples.
We also study the effect of the trade-off hyper-parameter $\alpha_1$ (when $\alpha_2=\frac{1-\alpha_1}{2}$) in Appendix Section ~\ref{sec:alpha-effect}.

\section{Summary}

To summarize this chapter, I have below key contributions and findings:

\begin{itemize}
\item[1] \textbf{Multi-view representation learning is ubiquitous}: 
I studied multi-view representation learning in three different scenarios. In the first scenario,  different views of the same object (e.g., acoustic and articulatory measurements) are available during training, and one or more views are missing in the test time.
In the second scenario, only one view is available during training; However, we can perform transfer learning to utilize the knowledge acquired from multi-view representation learning on other datasets to enhance single-view representation learning on the target dataset.
In the third scenario (where we only have utterances and the corresponding sequential labels), I show that the sequential label is naturally a second view and thus its structure information can be used to improve representation learning.
\item[2] \textbf{Cross-domain multi-view learning}: 
As mentioned in the scenario 2 in bullet point one, it is still possible to perform multi-view representation learning even when the extra modality is missing. This is also one of my major contributions to the multi-view representation learning.
Specifically, I investigated learning acoustic encoder using extra modality information when we actually don't have access to the paired extra-view information (e.g., articulatory measurements). 
I explored different transfer learning techniques and optimization tricks to exploit the learned acoustic-articulatory mapping in the dataset where acoustic-articulatory pairs are accessible. 
I show that such kind of acoustic-articulatory mapping information can benefit speech recognition tasks (especially on small datasets).
\item[3] \textbf{Better priors help multi-view representation learning in complex scenarios}: 
My another major contribution to multi-view representation learning is on using informative prior to enhance optimization and thus obtain representations that enhance downstream speech recognition.
I found that contextual information helps with multi-view representation learning (e.g., using $15$ consecutive MFCCs is better than $7$). 
However, when the window size is too big, the quality of the learned representations start to decay. 
I showed that, a more informative prior is critical for learning high-quality representation when the size of the input window is large; we also proposed a method to construct informative priors that help with VCCA.
%by using the representation learned using a moderately sized context window (e.g., $31$) as the prior representation to guide representation learning using a larger context window (e.g., $71$),
%the learned representation can better help downstream phone recognition task.
%stably and clearly outperform, in terms of phone recognition, representation of $31$-sized context window and $71$-size context window without using a proper prior.
\end{itemize}

%----------------------------------------------------------------------------------------
%	Recurrent representation learning
%       1. Motivate why we need RNN-based encoder
%       2. Build better recurrent representation learning models
%       3. Experiments to compare different models with/without different tricks
%       for supervised representation learning.
%       4. Experiments to compare different models with/without different tricks
%       for semi-supervised learning.
%----------------------------------------------------------------------------------------
\chapter{Recurrent representation learning with auxiliary loss}
\label{cha:recurrent}

%----------------------------------------------------------------------------------------
%       0. Motivation
%----------------------------------------------------------------------------------------

As discussed in Chapter ~\ref{cha:feedforward}, context information plays a crucial role in learning good representations. 
Feeding a window consisting of consecutive time steps to a feedforward neural network is an intuitive way to introduce context information. 
Under an unsupervised learning setting, we observed that it is challenging to learn a good representation using a feedforward neural network when the input is a large window spanning many time steps.
Though we proposed a two-step learning approach (see Section ~\ref{sec:pu}) to alleviate this difficulty, the feedforward neural network becomes cumbersome when the inputs grow to a higher dimension. 
Unlike feedforward neural networks which require fixed dimensional input, bidirectional recurrent neural networks ~\citep{schuster1997bidirectional} are a more natural choice for modeling variable-length sequence data.
Our experiments in Section ~\ref{subsec:feedforward-auxiliary} also indicate that bidirectional recurrent neural networks are more powerful for exploiting context information to learn representations that would benefit the speech recognition task.
\par

In this chapter, I explore using recurrent neural networks (which are more powerful for modeling sequence input) to learn per-time-step representations of sequence data.
I study recurrent representation learning in the next two chapters under three scenarios: 
1) auxiliary representation learning where I train jointly with two losses, a supervised loss and an auxiliary representation loss (e.g., ELBO or reconstruction loss),
2) semi-supervised representation learning where we use extra unlabeled data and 
3) unsupervised representation learning where we only optimize representation loss (e.g., ELBO or reconstruction loss).
Due to the success of VAE in the feedforward representation learning scenario (shown in our experiments in Chapter ~\ref{cha:feedforward}), 
I use a variational sequential framework in this and the remaining chapters.
I focus on auxiliary representation learning (using multitask training) in this chapter and leave semi-supervised learning and unsupervised learning for the next chapter. \par

There have been a few existing variational sequential models
~\citep{chung2015recurrent, fraccaro2016sequential, goyal2017z, shabanian2017variational}, 
most of which mainly focus on maximizing the sequence likelihood and improving the quality of the generated samples rather than obtaining good representations for downstream tasks.
%Given the trained encoders of these models, there are two potential limitations for using these proposed approaches to improve the performance of downstream discriminative tasks.
The applicability of these approaches for downstream tasks is limited by two obstacles.
One obstacle is the discrepancy between architectures.
Those above-mentioned sequential generative models are typically based on (stacked) unidirectional RNNs,
but bidirectional RNNs are generally more powerful for modeling sequential data.
For example,
~\citep{Graves2005BidirectionalLN} shows that bidirectional LSTM improves upon unidirectional LSTM for phoneme classification and recognition, 
\citep{Arisoy2015BidirectionalRN} shows bidirectional RNNs can benefit automatic speech recognition, 
and \citep{Cui2018DeepBA, Alawneh2020ACO} show that bidirectional RNN can benefit traffic speed prediction and human activity recognition.
Due to this architecture discrepancy, 
it is nontrivial to use the encoders of these generative models to warm-start a bidirectional recognizer (e.g., BiLSTM ASR recognizer).
Similarly, it is also less intuitive to train a shared encoder for both unidirectional generative models
and a bidirectional model (for a supervised task) via multitask learning. \par

The other obstacle is the difficulty of estimating marginal distributions of latent variables. 
When the latent variables over different time steps are not independent of each other,
the marginal distribution of $q_{\phi}(z_t)$ (by marginalizing out all $z$s except $z_t$) is typically a multi-modal distribution without an analytical form. 
This is also true for priors.
To get samples of $z_t$, we need nested sampling during typically costly training. 
To address this and the above considerations,
I propose a variational sequential model built upon stacked bidirectional LSTMs
and tackle the problem of modeling the posterior distributions.
In this chapter, I focus on a multitask learning framework, where the learned representations are fed to downstream classifiers/recognizers, 
and the negative ELBO and discriminative loss are minimized jointly. 
%We test our approach on sequence labeling tasks and TIMIT phone recognition. 
\par

%----------------------------------------------------------------------------------------
%       1. Basic model
%----------------------------------------------------------------------------------------
\section{Basic Model}
\label{sec:basic-model}

\begin{figure*}[htbp]
  \centering
  \includegraphics[width=0.90\textwidth]{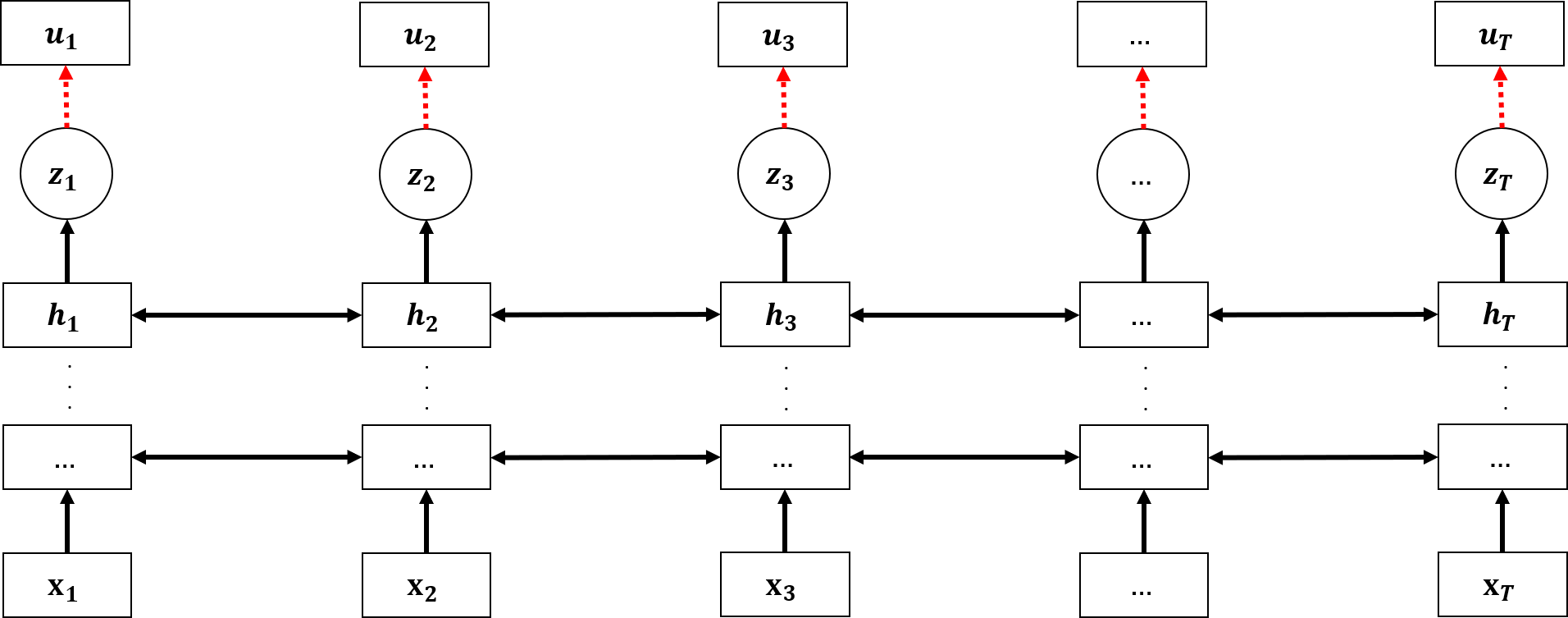}
  \caption{RecRep: Our basic variational sequential representation learning model.
I use red dashed arrows to indicate generation. 
$u_t$ is the ground truth target to be generated. 
For the basic model, $u_t=x_t$; 
however, $u_t$ can be more complex (e.g., weighted consecutive frames) that encourages latent variables to capture more contextual information.}
  \label{fig:basic}
\end{figure*}

I refer to our variational sequential model as the ``Recurrent Representation Learning Model'' (RecRep). 
RecRep is illustrated in Figure ~\ref{fig:basic}. 
RecRep uses stacked bidirectional LSTMs to deterministically encode the input sequence $x_{1:T}$ into $h_{1:T}$. 

Given $h_{1:T}$, 
I parameterize $T$ independent Gaussian distributions $\big\{ q_{\phi}(z_t|h_t) = \mathcal{N} \left( \mu_t, \diag{\sigma^2_t} \right) \big\}_{t \in 1,\cdots,T}$, 
with $\mu_t$ and $\log \sigma^2_t$ being transformations of $h_t$.\footnote{In my implementation, I also add non-linear layers on top of linear layers to increase the complexity of $\sigma_t$.}
The generation model $F_{\theta}(z_t)$ is a multilayer ReLU network
followed by a linear layer parameterizing a Gaussian distribution for reconstruction, or the logits of a multinomial distribution for supervised learning.
I use samples 
\footnote{In actual implementation, we introduce one additional hyparameter $\kappa$ which is in the range $[0,1]$. 
The sample for discriminative model is $\mu_t + \kappa \delta_1 \odot \sigma_t$ while the sample used for reconstruction is $\mu_t + \delta_2 \odot \sigma_t$. 
$\delta_1$ and $\delta_2$ are independently drawn from $\mathcal{N}(0,I)$. 
The motivation here is that the discriminative task may need less uncertainty than generation task.} 
of $z_{1:T}$ as input for training downstream recognizers/classifiers, 
and at test time we use the mean of the posterior distributions $q_{\phi}(z_t|h_t)$ as input to downstream models. 
The average ELBO of RecRep for input $x_{1:T}$ is

\begin{equation}
\text{ELBO} \left( x_{1:T}, u_{1:T} \right) = \frac{1}{T} \sum_{t=1}^T \Bigg\{  \mathbb{E}_{q_{\phi}(z_t|h_t)} \Big\{ \log{p_{\theta}(u_t|z_t)} \Big\}  -\beta \KL{q_{\phi}(z_t|h_t)}{\mathcal{N}(0,I)} \Bigg\}
\label{eqn:recrep-elbo}
\end{equation}

Given $x_{1:T}$ and its corresponding label $l_{1:T}$, we can jointly optimize the ELBO and the supervised loss in a multitask learning framework.
The complete objective we attempt to minimize is

\begin{equation}
\mathcal{L} \left( x_{1:T},l_{1:T} \right) = (1-\alpha) \sum_{t=1}^T \log{p(l_t|z_t)} - \alpha \text{ELBO} \left( x_{1:T}, u_{1:T} \right)
\label{eqn:recrep-joint-loss}
\end{equation}
with $\alpha \in [0,1]$. \par

Note, besides motivating Equation ~\eqref{eqn:recrep-joint-loss} from multitask learning perspective,
it can also be motivated and derived from a constrained VAE perspective as shown in ~\citep{hope2020learning}. 
That is, minimizing ~\eqref{eqn:recrep-joint-loss} is approximately equivalent to maximizing the ELBO with the constraint that the supervised loss ($\sum_{t=1}^T \log{p(l_t|z_t)}$) should to be smaller than a pre-defined threshold $\epsilon$. 
\begin{equation}
\max \text{ELBO} \left( x_{1:T}, u_{1:T} \right), \text{subj. to:} \sum_{t=1}^T \log{p(l_t|z_t)} \leq \epsilon
\label{eqn:constrained-seq-vae}
\end{equation}

%----------------------------------------------------------------------------------------
%       2. Incorporate contextual information 1 (Pyramidal Generation)
%----------------------------------------------------------------------------------------
\section{Pyramidal Model}
\label{sec:pyramid}

\begin{figure*}[htbp]
  \centering
  \includegraphics[width=0.92\textwidth]{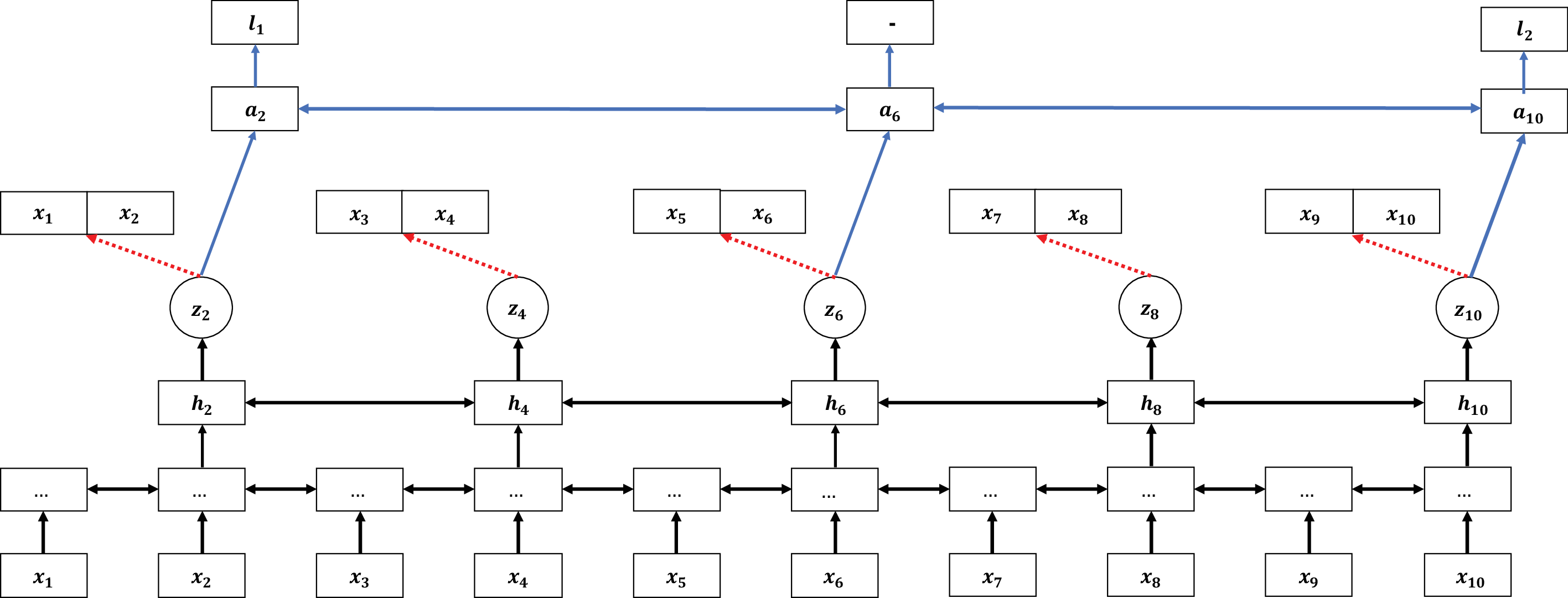}
  \caption{RecRep-Pyramid-MT: A multitask model consisting of a CTC recognizer and our pyramidal variational sequential model. 
The two parts share the first layer, second layer and latent variables. 
The generation part of the model reconstructs the concatenation of $x_{2k-1}$ and $x_{2k}$ given $z_{2k}$.
In this example, the per-time-step label of $x_{1:10}$ is ``$l_1l_1l_1l_1l_1l_1l_1l_2l_2l_2$'', 
and the CTC output is ``$l_1\_l_2$'', where ``$\_$'' is the blank in CTC.}
  \label{fig:pyramid}
\end{figure*}

Per-frame phonetic/character labels of acoustic utterances typically exhibit segmental structure.
Thus reducing time resolution may not hurt the performance of the recognition task
and can significantly accelerate the training and inference ~\citep{chan2016listen}. 
Pyramidal RNNs ~\citep{graves2012hierarchical} reduce the time resolution per layer.
As there are fewer time steps for the latent variable sequence than for the input sequence length
when using pyramidal RNNs,
we may use each latent variable to reconstruct a few neighboring input frames.
Taking Figure ~\ref{fig:pyramid} as an example, 
we would have the pyramidal architecture produce the latent variables $z_2$, $z_4$, $z_6$, $z_8$ and $z_{10}$ 
, where $z_2$ is responsible for reconstructing the concatenation of $x_1$ and $x_2$, $z_4$ for $x_3$ and $x_4$, and so on. \par

Figure~\ref{fig:pyramid} shows a multitask model that is designed for speech recognition, 
where I use CTC loss for training the recognizer.
Unlike the basic model illustrated in Figure ~\ref{fig:basic} where the generative model and the recognizer share all the recurrent layers,
the recognizer shown in Figure~\ref{fig:pyramid} has has an additional recurrent layer (in blue).
The design is motivated because CTC loss and per-frame reconstruction loss favor different sorts of information.
CTC outputs many ``blank'' labels and would encourage the representations from the final layer to be ready to be converted to ``blanks".
Though these consecutive blanks implicitly indicate the duration of one label,
a single ``blank" does not contain the critical information to reconstruct one frame,
and converting large amounts of different frames into blanks may cause reconstructive information to be lost.
In contrast,
our generative model with per-frame reconstruction loss encourages representations of each time step to capture the specific characteristics of that time step, e.g., phonetic information, speaker information, and channel information.
Due to this conflict of interest, 
I leave one private layer to CTC and expect this private layer to transform the representations learned by RecRep-Pyramid to forms better suited for CTC.
\footnote{Please note, we can potentially jointly train with different ASR models (e.g. RNN transducer ~\citep{Graves2012SequenceTW} and encoder-decoder ~\citep{Bahdanau2016EndtoendAL}) other than CTC. The conflicts as mentioned above may be specifically due to jointly training the RecRep and CTC.} 
 \par

The averaged ELBO of RecRep-Pyramid for input $x_{1:T}$, with $u_t = [x_{2k-1},x_{2k}]$, is:
\begin{equation}
\frac{1}{\floor{\frac{T}{2}}} \sum_{k=1}^{\floor{\frac{T}{2}}} \Bigg\{  \mathbb{E}_{q_{\phi} \left( z_{2k} \vert h_{2k} \right)} \Big\{ \log{p_{\theta} \left( [x_{2k-1},x_{2k}] \vert z_{2k} \right) } \Big\}  -\beta \KL{q_{\phi} \left( z_{2k} \vert h_{2k} \right) }{\mathcal{N}(0,I)} \Bigg\}
\label{eqn:recrep-elbo-pyramid}
\end{equation}
and the complete objective given $x_{1:T}$ and its corresponding label $l_{1:M}$ is

\begin{equation}
\mathcal{L} \left( x_{1:T},l_{1:M} \right) = (1-\alpha) \log{ p \left( l_{1:M} \vert \{z_{2k}\}_{k \in 1,\cdots,\floor{\frac{T}{2}}} \right) } - \alpha ELBO \left( x_{1:T} \right)
\label{eqn:recrep-pyramid-joint-loss}
\end{equation}

%----------------------------------------------------------------------------------------
%       3. Improve Pyramidal Generation
%.      -- Self prior updating and auxiliary latent variables
%----------------------------------------------------------------------------------------
\section{Improving RecRep}
\label{sec:strengthen}

I propose two extensions for improving RecRep. 
One extension is ``self prior updating'' which provides more informative priors during training. 
Though it looks similar, this ``self prior updating'' is slightly different from the one introduced in Section ~\ref{sec:pu}. 
Basically, ``prior updating" introduced in Section ~\ref{sec:pu} uses posteriors learned from a smaller context as new priors when learning from a larger context.
However, ``self prior updating" uses posteriors learned from earlier epochs as new priors in later epochs.
For simplicity,
I use ``prior updating" to refer to ``self prior updating" in this chapter.
The second extension introduces a per-frame auxiliary latent variable to capture information unrelated to downstream discriminative tasks that is nevertheless vital for reconstruction. 
Owing to the existence of this auxiliary latent variable, 
the primary latent variable can focus on capturing task-specific discriminative information. \par

\subsection{Self prior updating}
\label{subsec:self-prior-updating}

RecRep (and its variant RecRep-Pyramid) uses $\mathcal{N}(0,I)$ as its per-time-step prior. 
As discussed in ~\citep{chung2015recurrent}, given a sequence of length $T$, 
\weiran{no need to use this notation which is not very standard, you can use words:}
 $\{ \mathcal{N}(0,I) \}^T$ may be too restrictive for sequence representation learning as this prior ignores dependencies between time steps.
In this section, I propose a simple yet powerful approach called ``self prior updating'' (``prior updating" for short), where the prior of each time step is iteratively replaced by the inferred posterior using an encoder with weights of the previous (or even earlier) epoch. 
For example, in the $c^{\text{th}}$ epoch of the optimization process, 
I use $\left\{ q_{{\phi}^{(c-1)}}(z_t|h_t) \right\}_{1\leq t \leq T}$, the posteriors inferred using the model weights of $(c-1)^{\text{th}}$ epoch, as a sequence prior for maximizing the modified ELBO in Equation ~\eqref{eqn:centroid}. 
Note, $\phi^{(c-1)}$ will not be updated during the $c^{\text{th}}$ epoch. 
\begin{equation}
\sum_{t=1}^T \Big\{ \mathbb{E}_{q_{{\phi}}(z_t|h_t)} \big\{ \log{ p_{ {\theta} }(x_t|h_t) } \big\} - \beta \KL{q_{{\phi}}(z_t|h_t)}{q_{{\phi}^{(c-1)}}(z_t|h_t)} \Big\}
\label{eqn:centroid}
\end{equation}

There are two potential benefits of ``prior updating" if we assume $q_{{\phi}^{(c-1)}}$ is already an informative representation.
On one hand, as the prior of time step $t$ is a function of hidden state $h_t$, 
such a position-dependent prior contains more contextual information than the generic prior $\mathcal{N}(0,I)$.
The modified ELBO (Equation ~\eqref{eqn:centroid}) thus has an intuitive interpretation as re-encoding each latent variable $z_t$ w.r.t more informative prior information (searching $q^{(c)}$ near $q^{(c-1)}$ while optimizing the expected log likelihood). 
\weiran{The above intuitions are OK. But the reasoning below is a bit contrived. While 5.3.2 may be true, your actual KL term (measuring difference between current and previous $q(z_t|h_t)$) is easy to optimize since that term is zero if you do not move away from previous model.}
On the other hand,
as we are pushing the posterior of every time step of each sequence to a unique prior (different from $\mathcal{N}(0,I)$),
the averaged KL divergence 
\begin{equation}
\mathbb{E}_{x} \Bigg\{ \mathbb{E}_{t} \Big\{ \KL{q_{{\phi}}(z_t|h_t)}{\mathcal{N}(0,I)} \Big\} \Bigg\}
\end{equation}
is less prone to approach 0. 
In other words, the RecRep with ``prior updating" is more resistant to the posterior collapse described in ~\citep{bowman2015generating}. 
I discuss ``prior updating" from the perspectives of re-encoding latent variables and preventing posterior collapse in more detail in the following sections.

\subsubsection{Re-encoding w.r.t. more informative prior information}
\label{subsubsec:prior}
\weiran{this subsection is still a bit contrived.}

%\weiran{while I understand what this equation means mathematically, I am not sure how crude that approximation is, and what are doing seems to be far from this. I can see the intuition of aggregated posterior here, but the connection does not seem strong enough. I suggest dropping this subsection.}
%\weiran{this seems new, but you did not really explore this empirically, right?}

Given the inference network parameters $\phi^{(c-1)}$,
for each sequence $x \in \mathcal{D}$,
we can infer hidden states ($h_{1:T}$) and posteriors $q_{{\phi}^{(c-1)}}(z_t|h_t)$ for $1\leq t \leq T$.
Interpreting these hidden states as ``samples",
we can estimate a new prior distribution $p'(z)$ using

\begin{eqnarray}
p'(z) = \mathbb{E}_{x} \Bigg\{ \mathbb{E}_{1 \leq t \leq T} \Big\{ q_{{\phi}^{(c-1)}}(z_t|h_t)  \Big\} \Bigg\}
\label{eqn:new_prior}
\end{eqnarray}
$p'(z)$ is a Gaussian mixture as we typically parameterize each $q_{{\phi}^{(c-1)}}(z_t|h_t)$ as Diagonal Gaussian distribution.
The ``prior updating" idea introduced earlier in this chapter actually implements the below approximation of the KL divergence between posteriors and the prior $p'(z)$:

\begin{eqnarray}
&& \mathbb{E}_{x \in \mathcal{D}, t } \Big\{ \KL{q_{{\phi}}(z_t|h_t)}{p'(z)} \Big \} \nonumber \\
&=& \mathbb{E}_{x \in \mathcal{D}, t } \Bigg\{ \mathbb{E}_{q_{{\phi}}(z_t|h_t)}{\log{\frac{q_{{\phi}}(z_t|h_t)}{p'(z)}}} \Bigg \} \nonumber \\
&\approx& \mathbb{E}_{x \in \mathcal{D}, t } \Bigg\{ \mathbb{E}_{q_{{\phi}}(z_t|h_t)}{\log{\frac{q_{{\phi}}(z_t|h_t)}{q_{{\phi}^{(c-1)}}(z_t|h_t)}}} \Bigg \} \label{eqn:approx_mixture} \\
&=& \mathbb{E}_{x \in \mathcal{D}, t } \Big\{ \KL{q_{{\phi}}(z_t|h_t)}{q_{{\phi}^{(c-1)}}(z_t|h_t)} \Big \} \nonumber 
\end{eqnarray}

Equation ~\eqref{eqn:approx_mixture} assumes that for the samples $z \sim q_{{\phi}^{(c)}}(z_t|h_t)$,
$p'(z) \approx q_{{\phi}^{(c-1)}}(z_t|h_t)$.
In other words,
they likely all belong to one Gaussian component of the mixture prior $p'(z)$, which is $q_{{\phi}^{(c-1)}}(z_t|h_t)$.
This ``one Gaussian component" assumption is somewhat too strong.
%For example, the posteriors of neighboring time steps, e.g. $q_{{\phi}^{(c-1)}}(z_{t-1}|h_{t-1})$, $q_{{\phi}^{(c-1)}}(z_t|h_t)$ and $q_{{\phi}^{(c-1)}}(z_{t+1}|h_{t+1})$ can be quite similar to each other as neighboring frames presumably share the same phonetic labels .
%Two frames from two different utterances (e.g., the same sentence spoken by one speaker twice) can also have similar posteriors.
However, 
I still use $q_{{\phi}^{(c-1)}}(z_t|h_t)$ throughout this thesis %to represent these similar posteriors,
and I find such a simple proxy works well in practice. 
Although not investigated within this thesis, it is possible to derive a more accurate approximation than Equation ~\eqref{eqn:approx_mixture},
which is described in the appendix (Section ~\ref{subsec:appendix_multiple}). \par

\paragraph{Hyper-parameter Tuning}
\label{par:prior-hyper-parameter}

Regarding ``prior updating", I also emphasize the following:
%there are some other bullet points related to optimization that we need to pay attention to:
\begin{itemize}
\item[1] \textbf{Starting Epoch}: In the first few epochs, the learned inference network (e.g. $\phi^{(c-1)}$) is not optimized to do high-quality inference. 
Thus the learned posteriors may not be informative enough to serve as ``priors" in the following optimization process. To avoid this pitfall, I only start ``prior updating" after a few epochs.
\item[2] \textbf{Saving the best epoch}: If we have some metrics to evaluate the quality of $\phi^{(c-1)}$, e.g., the ELBO in the validation set or the performance of downstream tasks (e.g., speech recognition) when using $\phi^{(c-1)}$ to generate features, 
it is possible to only perform ``prior updating" when improved metrics on the validation set was observed.
\item[3] \textbf{Updating frequency}: Although we can always use the latest learned posteriors as new priors to guide further optimization, I observe that it may not work the best. 
In practice, we tune the frequency that priors are updated (e.g., every five epochs using the latest inference network) in practice.
\item[4] \textbf{$\beta$}: The $\beta$ parameter controls the flexibility of the optimization. 
To allow more flexibility to search for good priors in the early stage,
one should avoid using extremely large $\beta$.
However, as the optimization goes on, one can gradually increase $\beta$ to stabilize the optimization. For simplicity, in this thesis, I don't use a schedule for $\beta$; instead, I use a tuned fixed value.
\item[5] \textbf{Regarding proxy priors}: I describe other solutions for approximating $p'(z)$ in Appendix Section ~\ref{subsec:appendix_multiple}.
%Regarding that part, the window size $2K+1$ (Equation ~\eqref{eqn:context-prior}) for approximating the prior and the number of finite steps $K$ if flow prior is used can both be tuned. 
%\textit{In fact, it is even possible to combine both flow prior and Equation ~\eqref{eqn:context-prior} to obtain both global and local regularization during optimization.}
\end{itemize}

\subsubsection{Posterior Collapse}
\label{subsubsec:collapse}

In ~\citep{bowman2015generating}, the researchers identify an issue called ``posterior collapse" for variational autoencoders.
This describes the situation when the model relies solely on the powerful decoder (e.g., when the output of the decoder becomes almost independent of the latent variable), the latent variables stop encoding crucial information of the seen input.
The KL divergence between posterior and prior then approaches $0$, and the VAE thus fails to learn meaningful representations. 
This subsection describes the techniques I use (including ``prior updating") to prevent posterior collapse. 
\weiran{if ``prior updating'' is a technique for preventing collapsing, then you could move this subsubsection before ``prior updating'', and here you say ``other techniques for preventing posterior collapsing besides prior updating''.}
\par

% shallow generation network
It is generally understood that a powerful generation network does not rely strongly on high-quality posteriors to reconstruct the input. 
For example, when training Generative Adversarial Networks (GANs) ~\citep{Goodfellow2014GenerativeAN}, a powerful generation network can transform a sample from $\mathcal{N}(0, I)$ to a high-quality image/speech. 
Thus, to enforce the posterior distribution to carry more useful information, I use a relatively shallow feedforward neural network as our reconstruction network. 
This is viable as our goal is more to extract useful features than to generate high quality samples.
\par

% complex target
Besides using a shallow generation network, we can also use a more complex reconstruction target.
As mentioned earlier in Chapter ~\ref{cha:introduction}, 
the reconstruction target for each time step $t$ can also be a more complex target like a (weighted) context window.
To better reconstruct such a target, 
a relatively shallow reconstruction network would have to rely more on hints from informative representations, which might also help prevent the posterior collapse issue. \par

% prior updating prevents posterior collapse
\weiran{These are already covered. Consider re-organize things as mentioned in previous comment.}
%Besides tuning the hyper-parameter $\beta$ (weight of KL divergence term), 
%our prior updating approach actually also prevents the posterior collapse.
%In fact, our prior updating approach reduces the sensitivity of posterior collapse to the choice of beta.
In fact, our prior updating approach also prevents the posterior collapse.
Assume that after first $c-1$ epochs, 
the average distance between a posterior (e.g. $q_{\phi^{(c-1)}}(z_t|h_t)$, with input sequence $x$ randomly selected, and $h_t$ being the $t^{\text{th}}$ hidden state) and the prior $\mathcal{N}(0,I)$ is $\Delta$,
that is,
\begin{equation}
\mathbb{E}_{x\sim \mathcal{D}, t \sim \mathcal{U}(1,|x|)} \Bigg\{ \KL{ q_{ \phi^{(c-1)}}(z_t|h_t) } { \mathcal{N}(0,I)} \Bigg\} = \Delta
\end{equation}

As the optimization reduces the reconstruction error and the divergence between $q_{\phi^{(c)}(.)}$ and $q_{\phi^{(c-1)}(.)}$,
there is no clear driving force pushing $\mathbb{E}{\KL{ q_{ \phi^{(c)}(.) } } { \mathcal{N}(0,I)} }$ towards $0$.
%Actually, if $\beta$ is relatively large, 
When the optimization converges, the $q_{ \phi^{(c)}}(z_t|h_t)$ would be almost the same as $q_{ \phi^{(c-1)}}(z_t|h_t)$, thus the average distance between the posterior (after $c$ epochs) and the vanilla prior $\mathcal{N}(0,I)$ would still be around $\Delta$.
\par

% tuning beta
%In this thesis, we also experiment with generic prior while carefully tuning the hyper-parameter $\beta$, the weight on the KL divergence term to avoid posterior collapse.
%Small $\beta$ is less demanding in pushing posteriors to be similar to the prior and thus helps prevent posterior collapse when using $p(z)=\mathcal{N}(0, I)$. \par

\subsection{Auxiliary latent variables}
\label{subsec:auxiliary}

\begin{figure*}[htbp]
 \centering
 \includegraphics[width=0.76\textwidth]{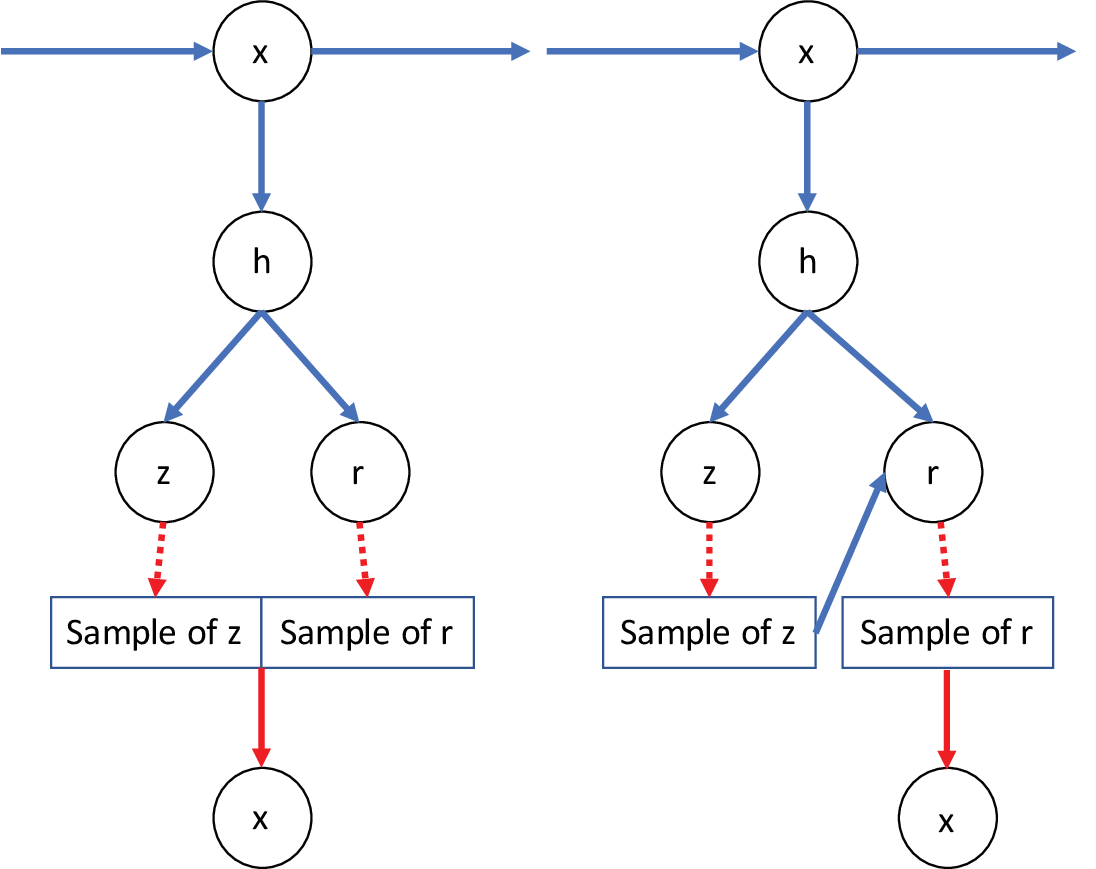}
 \caption{Illustration of two architectures I investigated to introduce auxiliary latent variables. 
 Red dashed lines indicate drawing samples and red solid lines indicate reconstruction. 
 The figure shows one time step of the whole sequential model.
 Discriminative loss (ignored in this figure) only takes $z$ as input.}
 \label{fig:hierarchy}
\end{figure*}

Discriminative tasks typically do not need precisely the same set of attributes as reconstruction tasks.
For example, to reconstruct an acoustic unit,
the latent representation (input to the generation network) needs to encode linguistic information, acoustic channel information, speaker identity, etc.
However, 
suppose we want to train a robust speech recognizer (which is speaker/background-independent), in that case,
the acoustic channel and speaker identity are variables that need to be ``filtered out" by the recognizer.
Motivated by this observation,
when RecRep is jointly trained with a discriminative loss,
I introduce one more set of latent variables $\left\{r_t \right\}_{t \in 1,\cdots,T}$ (as shown in Figure ~\ref{fig:hierarchy}) to capture information unrelated to the discriminative task but important for reconstructing the input. 
\par

Using disentangled latent variables for learning representations is not new.
In ~\citep{hsu2017unsupervised, Hsu2019HierarchicalGM}, the authors use a global latent variable to capture information like speaker identity,
and in ~\citep{Narayanaswamy2017LearningDR}, the authors propose to learn disentangled representations encoding different attributes into a subset of variables. 
The authors employ a general graphical model structure in the encoder and decoder, allowing them to train partially-specified models with strong assumptions on a subset of interpretable variables.
In ~\citep{Denton2017UnsupervisedLO}, a new model for learning disentangled representations from video is proposed in which each frame is factorized into a stationary part and a temporally varying component. \par

Our proposed auxiliary latent variable approach also attempts to learn two latent variables for each frame of the sequence, 
and thus is most similar to ~\citep{Denton2017UnsupervisedLO}. 
I explore two possible architectures, as shown in Figure ~\ref{fig:hierarchy}. \par

In the first architecture, $r_t$ is independent of $z_t$ given hidden state $h_t$. 
This kind of architecture is inspired by multi-space VAE ~\citep{zhou2017multi} and VCCAP ~\citep{wang2016deep},
although the two models are not designed for learning sequential representations.
Similar to $z_t$, I also parameterize $r_t$ as a diagonal Gaussian and use the concatenation of the samples of $r_t$ and $z_t$ to reconstruct $x_t$.  
The ELBO we seek to maximize is shown in Equation ~\eqref{eqn:recrep-multispace-elbo} for this architecture with two independent latent variables. For simplicity, I ignore the discriminative loss and the per-utterance ELBO is also written without updated priors. 
\begin{eqnarray}
\frac{1}{T} \sum_{t=1}^T \Bigg\{ &&\mathbb{E}_{q_{\phi}(z_t|h_t)q_{\phi}(r_t|h_t)} \Big\{ \log{p_{\theta}(x_t|z_t,r_t)} \Big\} \nonumber \\
&-& \beta \left( \KL{q_{\phi}(z_t|h_t)}{\mathcal{N}(0,I)} + \KL{q_{\phi}(r_t|h_t)}{\mathcal{N}(0,I)} \right) \Bigg\}
\label{eqn:recrep-multispace-elbo}
\end{eqnarray}

In the second architecture, $r_t$ is designed to depend on $z_t$, and I only use $r_t$ for reconstructing the input $x_t$. 
I expect both architectures can help $z_t$ focus on information only related to discriminative tasks, 
with $r_t$ being conditionally supervised by the label information in the second architecture. 
The ELBO of this architecture is shown in the following Equation ~\eqref{eqn:recrep-hierarchical-elbo-2}
\begin{eqnarray}
\frac{1}{T} \sum_{t=1}^T \Bigg\{ && \mathbb{E}_{q_{\phi}(r_t|h_t,z_t)q_{\phi}(z_t|h_t)} \Big\{ \log{p_{\theta}(x_t|r_t)} \Big\} \nonumber \\
&-&\beta \left( \KL{q_{\phi}(z_t|h_t)}{\mathcal{N}(0,I)} + \KL{q_{\phi}(r_t|h_t,z_t)}{\mathcal{N}(0,I)} \right) \Bigg\}
\label{eqn:recrep-hierarchical-elbo-2}
\end{eqnarray}

\subsection{Experiments}
\label{subsec:recurrent-exp}

I report our experimental results on two sequence labeling tasks, named entity recognition (NER) ~\citep{sang2003introduction} and text chunking ~\citep{sang2000introduction}, 
and also one sequence transduction task -- TIMIT phone recognition. 
I use RecRep for NER and text chunking and RecRep-Pyramid for phone recognition on TIMIT.
I show that the proposed ``prior updating" and ``auxiliary latent variables" do help RecRep(-Pyramid) outperform baseline models on multitask learning. \par

\subsubsection{CoNLL 2003 named entity recognition and coNLL 2000 chunking}
\label{subsubsec:recurrent-labeling}

\begin{table}[t]
 \begin{center}
  \begin{tabular}{|l|r|r|r|r|r|r|}
   \hline
   \textbf{Model} & \textbf{NER DEV} & \textbf{NER TEST} & \textbf{Chunking DEV} & \textbf{Chunking TEST} \\
   \hline \hline
   1.Baseline & 93.3 & 89.3 & 94.1 & 93.1 \\
   2.StocCon-MT & 92.4 & - & 92.8 & - \\
   3.StocCon-MT+P & 93.2 & - & 94.0 & -\\
   4.RecRep-MT & 92.5 & - & 93.5 & - \\
   5.RecRep-MT+F & 92.6 & - & 93.5 & - \\
   6.RecRep-MT+H & 92.8 & - & 93.6 & - \\
   7.RecRep-MT+P & 93.6 & - & 94.1 & - \\
   8.RecRep-MT+H+P & 93.7 & 89.8 & 94.5 & 93.7 \\
   \hline
  \end{tabular}
  \caption{F1 score of NER on CoNLL 2003 and Text chunking on CoNLL 2000. 
``Baseline'' = Two layer bidirectional GRU recognizer without any representation learning loss. 
``MT" indicates multitask learning.
``+F'' indicates using auxiliary latent variable $r_t$ with $r_t$ independent of $z_t$ given $h_t$, 
``+H'' indicates using $r_t$ where $r_t$ depends on $z_t$. 
``+P'' indicates using ``prior updating''.}
  \label{tab:ner+chunking}
 \end{center}
\end{table}

For named entity recognition and text chunking, I follow the setting of ~\citep{peters2017semi}. 
The input is a pre-trained word embedding (GloVe $100-$dimensional embedding ~\citep{pennington2014glove}) concatenated with the output of a character RNN embedding. 
The baseline prediction model is a $2-$layer bidirectional GRU RNN ~\citep{cho2014properties,chung2015gated}.
I also compare with another baseline sequential representation learning framework, StocCon, for multitask learning (StocCon-MT).
StocCon is the basic model (Figure ~\ref{fig:basic}) with a unidirectional recurrent connection between each pair of two neighboring latent variables; 
that is, $z_t$ is parameterized using the concatenation of $h_t$ and samples of $z_{t-1}$.
$p_{\theta}(x|z)$ is a multinomial distribution rather than Gaussian due to the discrete nature of inputs,
as in Equation (6) and (7) in ~\citep{miao2016neural}. 
The generation network $F_{\theta}(z)$ tries to reconstruct 
the (pre-trained) embedding of the target word. 
The hidden state units for all recurrent layers are $100$ per direction,
and I use unit size $100$ for $z_t$ and $25$ for $r_t$.
Based on Table ~\ref{tab:ner+chunking}, I observe the following:

\begin{itemize}
\item[1] RecRep-MT itself does not outperform the well-trained strong baseline. However, RecRep-MT assists with prior updating, and hierarchical latent variables show clear improvement over the baseline for both NER and text chunking.

\item[2] We can see StocCon-MT generally fails to outperform RecRep-MT, although StocCon-MT has a higher model capacity (e.g., the extra stochastic recurrent connections for StocCon compared with RecRep-MT).
As mentioned at the beginning of this chapter, this may be because it is challenging to estimate marginal distributions of latent variables when the latent variables over different time steps are not independent, requiring nested sampling during typically costly training.
However, it is possible that a more sophisticated sampling and inference strategy can improve StocCon-MT (e.g., using the sampling and inference solutions proposed in ~\citep{fraccaro2016sequential}).  

\item[3] We can see that our proposed ``prior updating" clearly helps both StocCon-MT and RecRep-MT, which matches our expectation that more informative priors can further improve the quality of the learned representations.

\item[4] RecRep-MT+H marginally improves over RecRep-MT while RecRep-MT+F performs similarly to RecRep-MT. 
Unlike with flattened latent variables (``+F"), when using hierarchical latent variables (e.g., ``+H"), the primary latent variable does not need to store too much information related to reconstruction. However, the reconstruction now depends on the discriminative information we have seen. Such conditional reconstruction could be a better way to encourage disentangling between reconstruction-specific information and discriminative task-related information. 
\end{itemize}
\par

\subsubsection{Phone recognition on TIMIT}
\label{subsubsec:recurrent-asr}

For phone recognition on the TIMIT data set, 
I use the same data processing and train/dev/test split as in ~\citep{tang2017end}. 
The baseline recognizer is a $3-$layer stacked bidirectional LSTM ~\citep{hochreiter1997long} network with pyramidal subsampling as shown in Figure ~\ref{fig:pyramid}. 
I use $256$ hidden units per direction for all LSTM layers throughout this chapter, unless specifically mentioned. \par

\begin{table}[htbp]
 \begin{center}
  \begin{tabular}{|l|r|r|}
   \hline
   \textbf{Model} & \textbf{DEV} & \textbf{TEST} \\
   \hline \hline
   1.Baseline & 17.3 & 19.4 \\
   \hline
   2.StocCon-Pyramid-MT & 17.6 & - \\
   \hline
   3.RecRep-Pyramid-MT & 17.2 & - \\
   4.RecRep-Pyramid-MT+F & 16.8 & 18.3 \\
   5.RecRep-Pyramid-MT+H & 16.8 & 18.6 \\
   \hline
   6.RecRep-Pyramid-MT+P & 17.0 & - \\
   7.RecRep-Pyramid-MT+F+P & 16.8 & - \\
   8.RecRep-Pyramid-MT+H+P & 16.7 & 18.3 \\
   \hline
  \end{tabular}
  \caption{TIMIT phonetic error rates (\%). ``Baseline'' = CTC recognizer without representation learning loss.}
  \label{tab:timit-multitask}
 \end{center}
\end{table}

Table ~\ref{tab:timit-multitask} shows the same trend as Table ~\ref{tab:ner+chunking}. 
StocCon struggles to produce representations that improve over baselines.
I suspect this is because of the issue mentioned earlier in this section, namely the difficulty in learning a posterior $q_{\phi} \left( z_{1:T}|h_{1:T} \right) $ and marginals $\left\{ q_{\phi}(z_t|h_t) \right\}_{t \in 1,\cdots,T}$.

Our RecRep-Pyramid-MT models (without stochastic recurrent connections, and directly learning the marginals)
outperform StocCon-MT and consistently improves upon the baselines when using prior updating or (hierarchical) auxiliary latent variables. 
In the remaining part of this subsection, I do two ablation studies.
One ablation study is to understand if prior updating can help to improve the learned representations in a stable fashion. 
The other ablation study is to understand the benefit of using RecRep-Pyramid in multitask learning as a regularizer.
I leave more detailed hyper-parameter tuning information for the Appendix, in Section ~\ref{sec:tune-on-timit}.\par

\begin{table} [htbp]
\centering
\begin{tabular}{| c | c | c | c | c | r | r |}
 \hline
 $Dim(z)$ & $\alpha$ & $\beta$ & $\kappa$ & dropout & PER ($\%$)  & PER ($\%$) with prior updating \\
 \hline \hline
 150 & 0.5 & 0.01 & 0.1 & 0.4 & 18.2 & 17.2 \\
 \hline
 150 & 0.5 & 0.01 & 1.0 & 0.4 & 18.2 & 17.2 \\
 \hline
 150 & 0.5 & 0.01 & 0.0 & 0.4 & 17.9 & 17.2 \\
 \hline
 250 & 0.5 & 0.01 & 0.1 & 0.4 & 18.1 & 17.0 \\
 \hline
 150 & 0.6 & 0.01 & 0.1 & 0.4 & 18.0 & 17.3 \\
 \hline
 150 & 0.4 & 0.01 & 0.1 & 0.4 & 18.0 & 17.5 \\
 \hline
 150 & 0.3 & 0.01 & 0.1 & 0.4 & 18.0 & 17.4 \\
 \hline
 150 & 0.2 & 0.01 & 0.1 & 0.4 & 18.2 & 17.4 \\
 \hline
 150 & 0.5 & 0.001 & 1.0 & 0.4 & 17.9 & 17.4 \\
 \hline
 150 & 0.5 & 0.01 & 0.0 & 0.4 & 17.8 & 17.1 \\
 \hline
 150 & 0.5 & 0.01 & 1.0 & 0.5 & 17.2 & 17.2 \\
 \hline
\end{tabular}
\caption{Dev set performance for ``prior updating".
I select the best $11$ RecRep-Prymida-MT models we have trained. 
I apply prior updating to re-train $11$ models using the same hyperparameters.
The last model shown in this table uses $\mathcal{L}_2$ weight $10^{-7}$, and all other models do not use $\mathcal{L}_2$ regularization.
$\alpha$ is the weight on the ELBO as shown in Equation ~\eqref{eqn:recrep-joint-loss},
and $\kappa$ (introduced in Section ~\ref{sec:basic-model}) is the hpyerparameter I introduced to control how much uncertainty is used for discriminative task.}
\label{tab:prior-updating-timit}
\end{table}

\begin{figure*}[t]

\begin{minipage}{1.0\textwidth}
\begin{subfigure}{1.0\textwidth}
 \centering
 \includegraphics[width=0.9\textwidth]{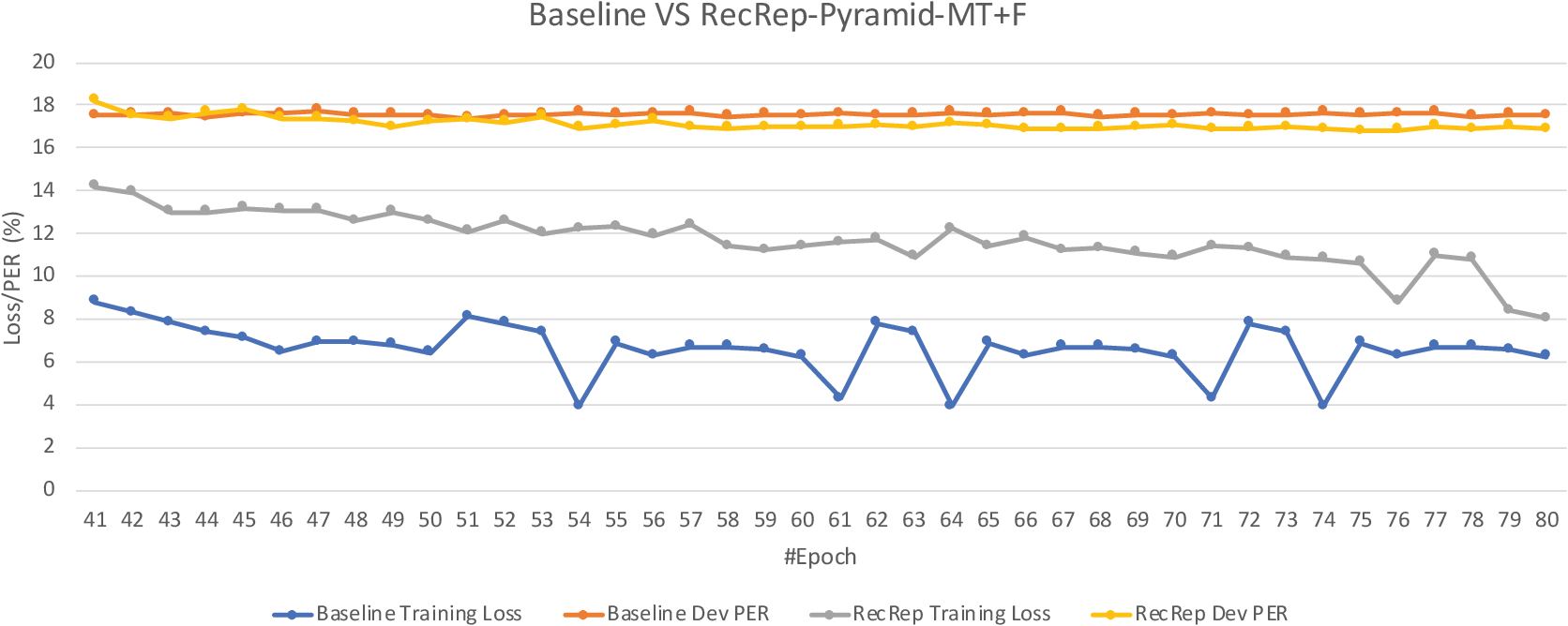}
 \caption{Comparison of dev PER and training loss.}
 \label{fig:timit-cmp1}
\end{subfigure}
\end{minipage}

\begin{minipage}{1.0\textwidth}
\begin{subfigure}{1.0\textwidth}
 \centering
 \includegraphics[width=0.9\textwidth]{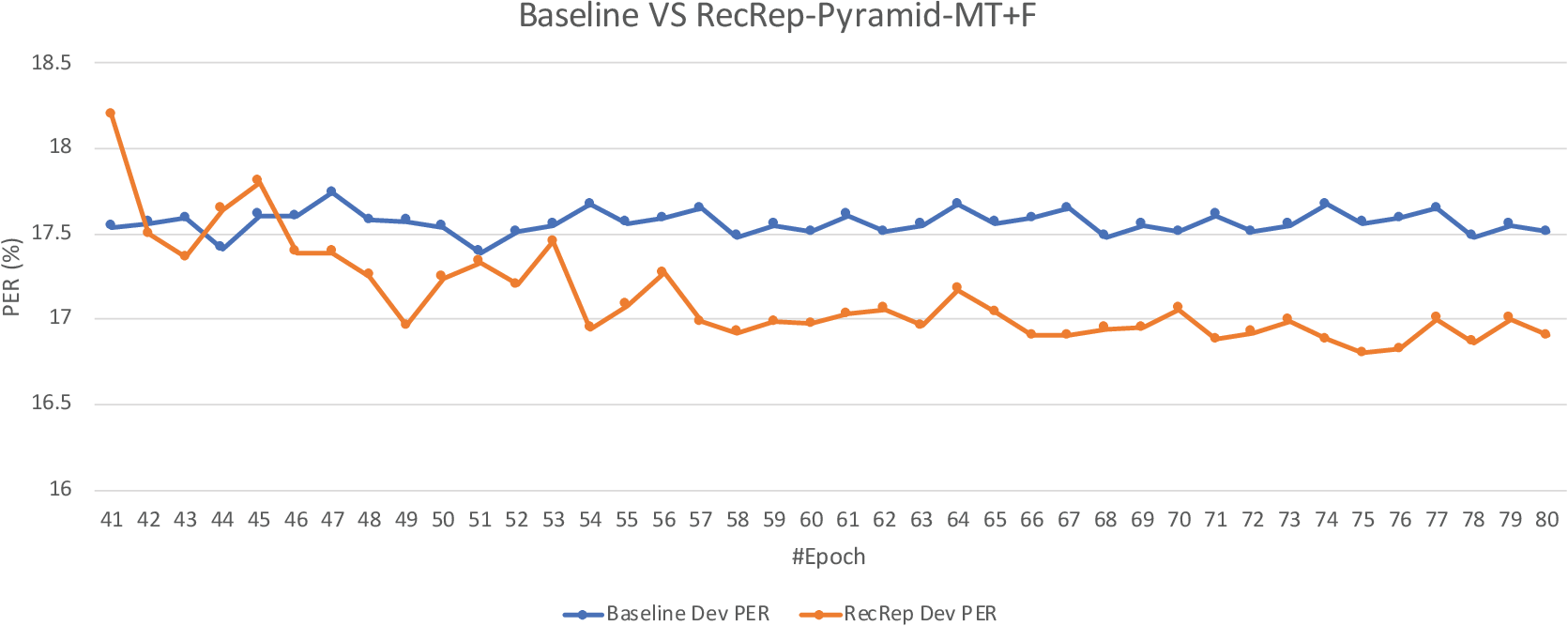}
 \caption{Comparison of dev PER.}
 \label{fig:timit-cmp2}
\end{subfigure}
\caption{Comparison of CTC and RecRep+F regularized CTC on training loss and dev PER.}
\end{minipage}

\end{figure*}

\paragraph{Does prior updating help?}
I describe our ablation study on the prior updating. 
I show that the models with prior updating consistently improves upon corresponding models without prior updating,
and in most cases, clearly outperform the corresponding models.
I select the best $11$ RecRep-Pyramid-MT models (all trained up to $80$ epochs),
and re-train another $11$ models using the same hyperparameters as the $11$ RecRepPyramid-MT models with prior updating applied.
The prior updating starts from the $10$th epoch with a prior updating frequency of $10$. 
All other experimental settings are identical to those used for RecRep-Pyramid-MT.
As shown in Table ~\ref{tab:prior-updating-timit},
prior updating reduces the phone error rate on the dev set significantly.
The consistent benefits across different hyper-parameters as shown in Table ~\ref{tab:prior-updating-timit},
and the benefits we have seen earlier in this section indicate that this simple technique helps improve representation learning.

\paragraph{RecRep-Pyramid as a regularizer}
I also observed the ``regularization effect'' of the RecRep-Pyramid in our multitask learning framework. 
Compared with the CTC baseline recognizer, RecRep-Pyrmamid regularized CTC (e.g., RecRep-Pyramid-MT) has a worse CTC training loss but outperforms the baseline in the dev set,
in terms of both CTC loss and phone error rate. \par

In Figure ~\ref{fig:timit-cmp1}, 
I show the trend of training loss and dev PER of CTC and ``RecRep-Pyramid+F". 
The figures show that CTC regularized by the generative model has higher training loss and converges more slowly. 
According to Figure ~\ref{fig:timit-cmp2}, the dev PER of CTC stabilizes around epoch $40$, but in the generative model-regularized CTC the dev PER continues to decrease.. 
The CTC regularized by ``RecRep-Pyramid+F" starts to outperform the baseline around epoch $50$.

\section{RecRep-Pyramid with window reconstruction}
\label{sec:window-loss}

As mentioned in the introduction, and also in the caption of Figure ~\ref{fig:basic},
the reconstruction target of time step $t$ can be $x_t$ or a (weighted) window centered around frame $t$.
In this section, I experiment with reconstructing a $W-$frame window.
Compared with reconstructing a single frame $x_t$ as in RecRep or reconstructing the concatenation of two consecutive time steps as in RecRep-Pyramid (shown in Figure ~\ref{fig:pyramid}),
reconstructing a $W-$frame window encourages the per-time-step representation
to capture more context information hidden in the $W$ time steps.

Denoting the parameters for inference and generation networks $\phi$ and $\theta$ respectively,
according to ELBO, we are encouraging the posterior distribution $q_{\phi}(z_t|h_t)$ to approximate the distribution $p_{\theta}(z_t|u_t)$.
As $z_t$ will not be used for the reconstruction of targets other than $u_t$,
if $u_t$ is a single frame,
the ELBO simply encourages $z_t$ to capture local information of $x_t$;
if $u_t$ is a context window,
maximizing ELBO encourages $q_{\phi}(z_t|h_t)$ to be similar to $p_{\theta}(z_t|u_t)$ and learn the context information from $u_t$. \par

I compare window-reconstruction-based RecRep-Pyramid-MT
with baseline and vanilla RecRep-Pyramid-MT (reconstructing a concatenation of two consecutive time steps).
I use WSJ for this experiment (with the same setup described in ~\ref{sec:data}).
All recognizers are trained using $\frac{1}{2}$ of SI84 of WSJ,
and I report the character error rate (CER) on dev93. \par

\begin{table} [htbp]
\centering
\begin{tabular}{| c | r | r | r | r| r | r | r | r |}
 \hline
 & $\alpha=0.0$ & $W=2$ & $W=3$ & $W=5$ & $W=7$ & $ W=9$ \\
 \hline \hline
 Dev CER ($\%$) & 27.6 & 28.1 & 27.5 & 26.8 & 26.9 & 27.5 \\
 \hline
 \begin{tabular}{@{}c@{}}
 Reconstruction Loss \\
@ best epoch \\
\end{tabular}
& - & 159.7 & 157.0 & 157.4 & 155.4 & 155.5 \\
 \hline
\end{tabular}
\caption{Comparison of window-reconstruction-based RecRep-Pyramid-MT, baseline and vanilla RecRep-Pyramid-MT.
$\alpha$ is the weight on the averaged ELBO term, when $\alpha=0.0$, the RecRep-Pyramid-MT becomes baseline.
Except the first column (Baseline), I use $\alpha=0.3$ for all the other columns shown in the table.
$W=2$ is the vanilla RecRep-Pyramid-MT.
Reconstruction loss here refers to averaged per-time-step log-likelihood normalized by window size $W$.
The latent variable size is $200$.}
\label{tab:recrep-win}
\end{table}

%% We observe that RecRep-Pyramid can not improve baseline
%% We show the benefits of learning the context information, making the learned representations smoother.
According to earlier experiments described in this chapter (e.g., Table ~\ref{tab:timit-multitask} and Table ~\ref{tab:ner+chunking}),
RecRep-Pyramid-MT and RecRep-MT are either only comparable to, or fail to outperform the baseline.
I observe the same phenomenon on the dev set of WSJ (e.g., the column of $W=2$ is worse than the column of $\alpha=0.0$).
However, according to Table ~\ref{tab:recrep-win},
I observe that RecRep-Pyramid-MT with window reconstruction outperforms the basic RecRep-Pyramid-MT.
I find an absolute CER reduction of $0.6$ to $1.3$ given different window sizes.
Also, the window-based reconstruction helps the CTC achieve (slightly) lower CER in the dev set than baseline CTC, and the benefit is consistent across all window sizes I have tried.
I observe that moderate window sizes $5$ and $7$ are the most helpful for character recognition. 
Collectively, these observations suggest the benefit of reconstructing a complex unit over a single frame. \par

As mentioned earlier in Section ~\ref{sec:pyramid},
I hypothesize that as CTC loss and per-frame reconstruction loss favor different types of representations,
RecRep-Pyramid-MT could potentially struggle between favoring the two losses.
According to Table ~\ref{tab:recrep-win},
I observe that reconstructing a window somewhat alleviates the conflict between reconstruction loss and CTC loss.
I observe that while window reconstruction typically helps RecRep-Pyramid-MT achieve lower dev set CER,
it also achieves lower per-frame reconstruction loss. 
I suspect that this is because window reconstruction losses induce smoother temporal representations, which are better suited for CTC.
CTC tends to output many continuous ``blanks" (see the example in the caption of Figure ~\ref{fig:pyramid}) and thus encourages higher-level LSTM layer representations to be smoother.
When reconstructing a window $[x_{t-W,t+W}]$ centered at $t$ for each time step $t$, 
the latent variable encodes less the per--frame variance as the reconstruction targets of two neighboring time steps becomes even more similar to each other. \par
%the reconstruction targets of two neighboring time steps becomes even more similar to each other and the per-frame variance is less encoded in the latent variable. \par

\section{Shallow Recurrent Representation Learning}
\label{sec:reducing-conflict}

We hypothesized that CTC loss and per-frame reconstruction loss favor different types of representations. 
Thus, I experiment with a model variant where the speech recognizer (CTC) and variational sequential model only share low-level features while the two branches have more private layers to learn task-specific representations.
I expect such kind of shallow model would achieve better performance.
I would also compare with autoencoding with another learning approach -- autoregressive prediction (the FB model described later) in this section.
\par

To make the comparison with a three-layer BiLSTM CTC recognizer fair, 
I design our RecRep-Pyramid-MT in such a way that the speech recognizer and variational sequential model share two BiLSTM layers, and the speech recognizer only has one private BiLSTM layer,
as shown in Figure ~\ref{fig:pyramid}.
Prior to this section, all experiments for multitask speech recognition in this chapter have followed this default setup.
This section explores the model architecture where the speech recognizer (CTC) and variational sequential model only share one BiLSTM layer while the speech recognizer has two private BiLSTM layers.
As CTC loss and per-frame reconstruction loss favor different types of representations,
I expect using more speech-recognition-specific layers will help reveal this conflict and improve speech recognition performance. \par

\begin{figure*}[htbp]
 \centering
 \includegraphics[width=0.8\textwidth]{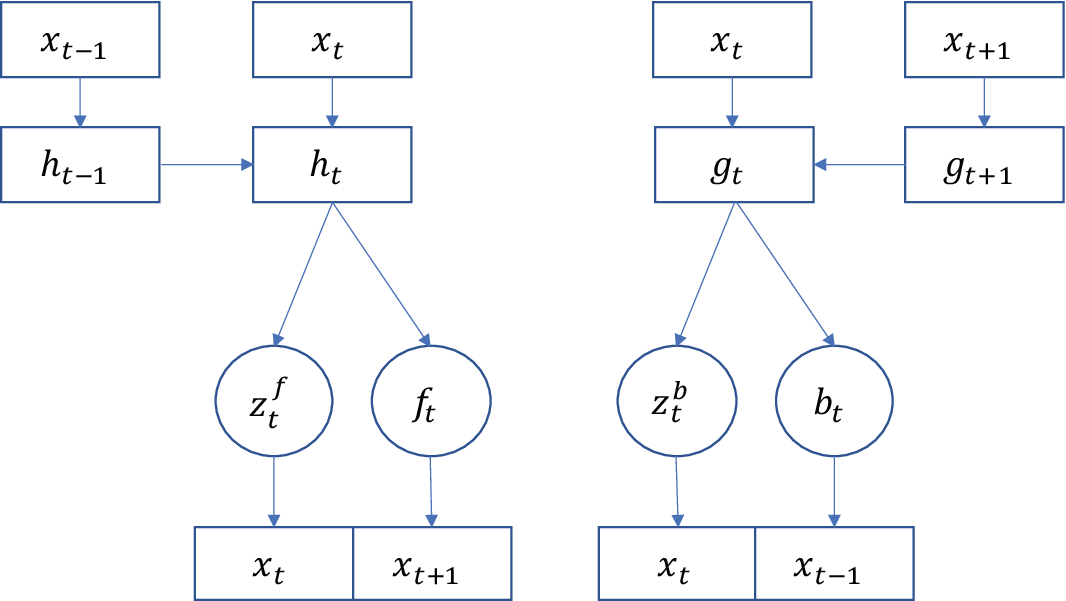}
 \caption{Forward and backward variational recurrent model.}
 \label{fig:single-layer}
\end{figure*}

I investigate two types of shallower variational sequential models.
The first one is a one-layer RecRep-Pyramid, 
which is called RecRep-Pyramid-1L.
Its corresponding multitask learning version is denoted RecRep-Pyramid-MT-1L.
For RecRep-Pyramid-MT-1L,
the representation learning layer is a single bidirectional recurrent layer;
the latent representations on top of this layer are then fed to feedforward neural networks for reconstruction tasks and provided to two-layer bidirectional LSTMs for the speech recognition task.
The second shallow variational sequential model I use is illustrated in Figure ~\ref{fig:single-layer}.
I refer to this model as ``forward-backward model" (FB for short).
Its corresponding multitask version is named FB-MT. \par

For the FB model, 
I use a one-layer unidirectional LSTM as our forward encoder.
Similar to our approach in Section ~\ref{subsec:auxiliary},
I also learn two latent variables $z^f_t$ and $f_t$ for each time step $t$.
Here, both $z^f_t$ and $f_t$ are functions of the hidden state $h_t$ (the summarization of $x_{\leq t}$);
$f_t$ is used to predict the next time step input $x_{t+1}$
while $z^f_t$ is used to reconstruct the current time step input $x_t$.
Similarly, 
I also have one one-layer unidirectional LSTM as our backward encoder,
and use the hidden state $g_t$ (the summarization of $x_{\geq t}$)
to parameterize $b_t$ which is then used to predict $x_{t-1}$. \par

Unlike how RecRep reconstructs current frame $x_t$, where $x_t$ and its neighbors have been observed by the encoder,
the FB model, 
which tries to predict the next and previous unseen frames,
can be understood as learning a legitmate spectrogram language model $p(x_t|x_{1:t-1})$ (or $p(x_t|x_{t+1:T})$ for the backward case).
\par

As the forward summarization of $x_{\leq t}$ and backward summarization of $x_{\geq t}$ both contain the information relevant to reconstruct $x_t$,
I also introduce two more latent variables, $z^f_t$ and $z^b_t$ respectively,
which are used to reconstruct the current time step $x_t$. 
Thus, for the FB model,
there could be three variations: 1) The model only predicts unseen next/previous frames, 
2) the model uses only reconstruction loss similar to RecRep, 
and 3) the model includes both reconstruction loss and loss for predicting unseen frame. \par

I jointly train a forward encoder, backward encoder, and stacked bidirectional CTC recognizer.
The forward encoder (single layer) and the bottommost forward layer of the CTC recognizer share the same parameters,
and similarly for the backward encoder and the backward layer of the CTC recognizer share the parameters.
The samples of the four latent variables ($f_t$, $b_t$, $z^f_{t}$ and $z^b_{t}$) are concatenated and fed to subsequent private layers of the CTC recognizer.
The samples are concatenated as $\left[ f_t ; \frac{z^f_{t}+z^b_{t}}{2} ; b_t \right]$. \par

I compare RecRep-Pyramid-MT, RecRep-Pyramid-MT-1L, FB-MT (all three variations), and the baseline (three-layer BiLSTM CTC recognizer with hidden state units of $256$ per direction).
I use the performance of character speech recognizers to evaluate the quality of the learned representations.
All recognizers are trained using $\frac{1}{2}$ of SI84, and all experimental setup details are the same as in Section ~\ref{sec:data}.

\begin{table} [htbp]
\centering
\begin{tabular}{| c | r | r |}
 \hline
 Models & Dev CER ($\%$) & Test CER ($\%$) \\
 \hline \hline
  \begin{tabular}{@{}c@{}}
  1. Baseline (3-layer BiLSTM)  \\
  $256$ hidden units \\
  per direction \\
  \end{tabular}
& 27.1 & 19.9 \\
 \hline
 \begin{tabular}{@{}c@{}}
  2. RecRep-Pyramid-MT \\
  $Dim(z_t)=200$ \\
  $\alpha=0.3, \beta=0.001, \kappa=0.0$\\
  \end{tabular}
& 28.1 & - \\
 \hline
 \begin{tabular}{@{}c@{}}
  3. RecRep-Pyramid-MT-1L \\
  $Dim(z_t)=200$ \\
  $\alpha=0.3, \beta=0.001, \kappa=0.0$ \\
  \end{tabular}
& \textbf{25.8} & - \\
 \hline
 \begin{tabular}{@{}c@{}}
  4. RecRep-Pyramid-MT-1L \\
  $Dim(z_t)=200$ \\
  $\alpha=0.3, \beta=0.001, \kappa=0.01$ \\
  \end{tabular}
& \textbf{25.4} & \textbf{18.6} \\
 \hline
 \begin{tabular}{@{}c@{}}
  5. FB-MT \\
  $Dim(z^f_t)=0$ and $Dim(z^b_t)=0$ \\
  $Dim(f_t)=185$ and $Dim(b_t)=185$\\
  $\alpha=0.3, \beta=0.001, \kappa=0.1$ \\
  \end{tabular}
& \textbf{25.8} & - \\
 \hline
 \begin{tabular}{@{}c@{}}
  6. FB-MT \\
  $Dim(z^f_t)=0$ and $Dim(z^b_t)=0$ \\
  $Dim(f_t)=150$ and $Dim(b_t)=150$ \\
  $\alpha=0.3, \beta=0.001, \kappa=0.1$
  \end{tabular}
& 26.0 & - \\
 \hline
 \begin{tabular}{@{}c@{}}
  7. Another baseline \\
  Model of row $6$ with \\
  $\alpha=0.0, \beta=0.0, \kappa=0.0$
  \end{tabular}
& 26.8 & - \\
 \hline
 \begin{tabular}{@{}c@{}}
  8. FB-MT \\
  $Dim(z^f_t)=0$ and $Dim(z^b_t)=0$ \\
  $Dim(f_t)=150$ and $Dim(b_t)=150$ \\
  $\alpha=0.3, \beta=0.0, \kappa=0.0$ \\
  \end{tabular}
& 27.0 & - \\
 \hline
 \begin{tabular}{@{}c@{}}
 9. FB-MT \\
 $Dim(z^f_t)=370$ and $Dim(z^b_t)=370$ \\
 $Dim(f_t)=0$ and $Dim(b_t)=0$ \\
 $\alpha=1e-3, \beta=0.01, \kappa=0.0$ \\
 \end{tabular}
 & 26.5 & - \\
 \hline
 \begin{tabular}{@{}c@{}}
 10. FB-MT \\
 $Dim(z^f_t)=70$ and $Dim(z^b_t)=70$ \\
 $Dim(f_t)=150$ and $Dim(b_t)=150$ \\
 $\alpha=0.3, \beta=0.01, \kappa=0.0$ \\
 \end{tabular}
 & 26.8 & - \\
 \hline
\end{tabular}
\caption{
Comparison of Baseline, FB-MT, and RecRep-Pyramid-MT-1L.
$\alpha$ is the weight of the averaged ELBO (see Equation ~\eqref{eqn:recrep-joint-loss}),
$\beta$ is the weight on the KL divergence term,
and $\kappa$ affects the magnitude of jitter from the mean (see Footnote $2$ of this Chapter).
All speech recognizers are trained to convergence (e.g., no improvement on dev set error rate observed).
}
\label{tab:one-layer-model}
\end{table}

\begin{figure*}[htbp]
 \centering
 \includegraphics[width=0.9\textwidth]{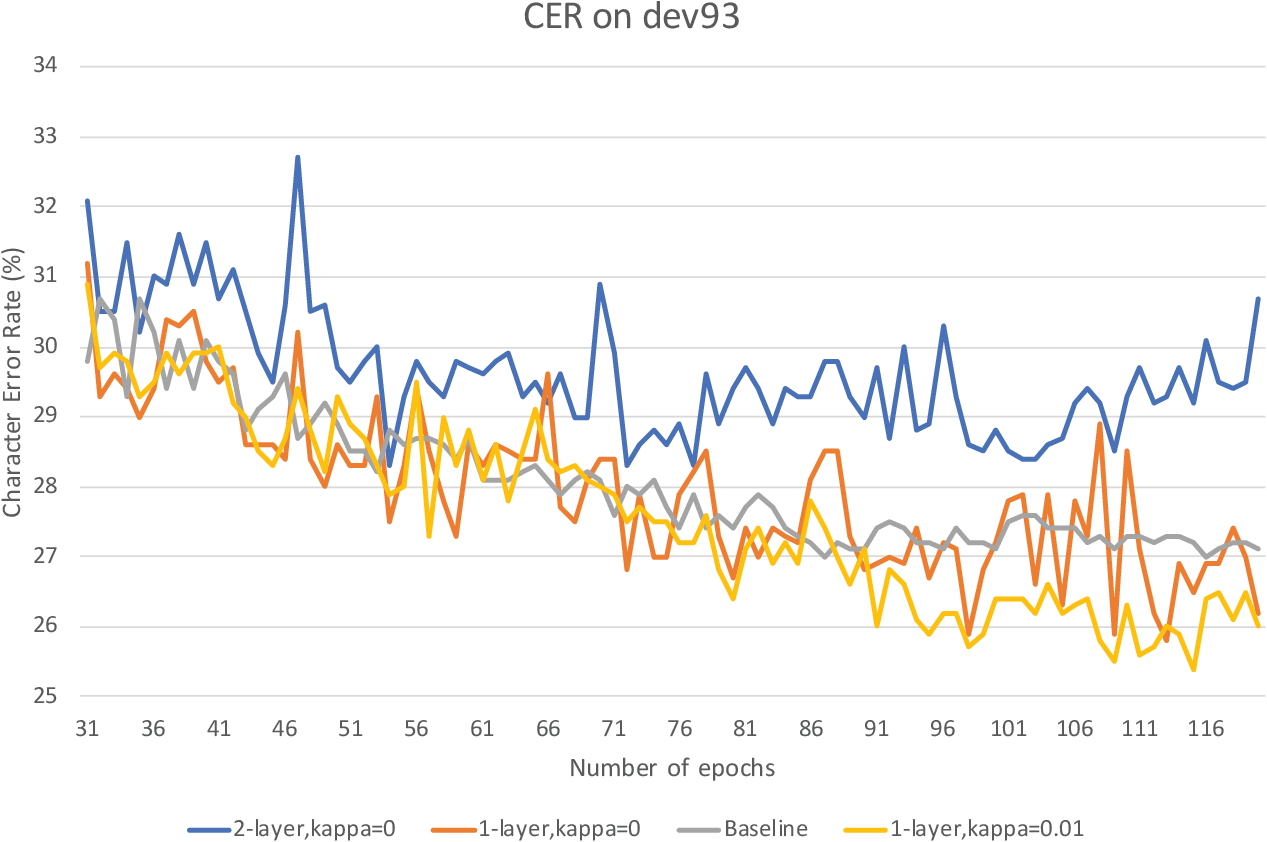}
 \caption{Comparison of Baseline, RecRep-Pyramid-MT and RecRep-Pyramid-MT-1L in terms of error rate on dev93. 
 The hyperparameter $\kappa$ is explained in Footnote $2$ of this Chapter. 
 $\kappa$ is used to bound the variance of samples used for the speech recognition task. 
 The smaller $\kappa$ is, the smaller variance.}
 \label{fig:wsj-layer-cmp}
\end{figure*}

From Table ~\ref{tab:one-layer-model},
I observe that both RecRep-Pyramid-MT-1L (Row $3,4$)
and FB-MT (Row $5-6,9-10$)
outperform the baseline (three-layer BiLSTM CTC recognizer) and basic RecRep-Pyramid-MT (CTC and recurrent sequential model sharing two bidirectional layers).
Please note,
this finding does not necessarily indicate that
shallow encoders (e.g., one-layer BiLSTM) learn better speech representations than deeper encoders (e.g., two-layer BiLSTM).
I hypothesize that RecRep-Pyramid-MT-1L has better performance than RecRep-Pyramid-MT because RecRep-Pyramid-MT-1L has more speech-recognition-specific recurrent layers and thus has a greater capacity to alleviate the conflict between CTC loss and reconstruction loss.
If I also use two private recurrent layers for RecRep-Pyramid-MT, it could presumably achieve better speech recognition performance than RecRep-Pyramid-MT-1L. \par

According to Table ~\ref{tab:one-layer-model},
RecRep-Pyramid-MT-1L performs best among all models,
but the gap between the best RecRep-Pyramid-MT-1L model (row $4$) and the best FB-MT model (row $5$) is not significant (e.g., CER $25.4$ vs $25.8$),
and they all significantly outperform the baseline model (row $1$) and basic RecRep-Pyramid-MT (row $2$). 
Note that I use latent variables of dimension $200$ for basic RecRep-Pyramid-MT and for RecRep-Pyramid-MT-1L.
I use latent variables of larger units for FB-MT; that is, $185$ for both $f_t$ and $b_t$.
In recent work ~\citep{Chung2019AnUA}, 
the authors try to predict additional unseen frames beyond the next unseen frame to encourage the encoder to learn more context information. 
Their model learns speech representations that benefit a few tasks such as phone classification, speaker identification, and speech recognition.
I suspect that modifying the FB-MT loss to predict additional frames in this way could also make FB-MT more powerful.
\par

Compared with the model of Row $5$, Row $6$ uses $f_t$ and $b_t$ of smaller units,
and performs slightly worse.
Row $7$ is another baseline model without recurrent representation learning.
Compared to the baseline (Row $1$), 
Row $7$ has a linear dimension reduction layer between the first two recurrent layers.
This model (Row $7$) is designed to match the speech recognizer part of Row $6$ exactly.
We can see that Row $6$ has lower CER than Row $7$ and Row $8$,
where Row $8$ is the deterministic counterpart of Row $6$.
The observation that Row $6$ is better than Row $8$ echoes
our findings in Table~\ref{tab:compare-framework} that VAEs outperform AEs.
Another interesting observation is that FB-MT with both losses predicting the future and reconstructing the current time step leads to
worse performance than FB-MT with only losses predicting the future or reconstructing the current time step. \par

I also plot the dev set error rate curve (Figure ~\ref{fig:wsj-layer-cmp}) to compare the baseline, basic RecRep-Pyramid-MT, and RecRep-Pyramid-MT-1L models.
The plot clearly shows the advantage of RecRep-Pyramid-MT-1L in terms of dev set error rate.
According to row $3$ and $4$ in Table ~\ref{tab:one-layer-model} and this plot,
a non-zero $\kappa$ seems helpful for stabilizing the validation set character error rate and improving the model's performance.
In Appendix Section ~\ref{sec:one-for-multitask},
I describe my experiments on TIMIT which also uses one-layer BiLSTM for representation learning in multitask learning.
The experimental results are consistent with
Figure ~\ref{fig:wsj-layer-cmp} and Table ~\ref{tab:one-layer-model}. \par

\section{Summary}

To summarize this chapter, I have below key contributions and findings:

\begin{itemize}
\item[1] \textbf{Multitask recurrent representation learning}: 
I have proposed a variational bidirectional encoder framework,
which directly learns posterior distributions that decouple over frames when conditioned on bidirectional RNN outputs (e.g., $h_t$'s).
Our bidirectional encoder can be easily jointly trained with downstream tasks across NLP and speech domains.
I have proposed to factor the per-time-step information into
a discriminative task relative component and a reconstruction-specific component.
I have also proposed to update the per-time-step priors dynamically, 
discussed the resulting benefits and limitations in depth, and propose potential extensions in the Appendix (Section ~\ref{sec:approx}) to more accurately estimate new priors.
Our experimental results show that the proposed multitask representation learning framework and its extensions work well for tasks that include speech recognition, named entity recognition, and chunking. \par
\item[2] \textbf{Different reconstruction targets}: I have also performed an ablation study to investigate the effects of different reconstruction targets and different encoder architectures.
Regarding the reconstruction target $u_t$, our experimental results suggest that reconstructing a window (of proper size) rather than reconstructing a single frame works better (i.e., achieving lower CER).
Unexpectedly, I find that the benefit from reconstructing a window is not big enough in multitask learning setting. \par
\item[3] \textbf{Different tasks favor different representations}: 
I hypothesized that CTC loss and per-frame reconstruction loss favor different types of representations. 
Thus, I experiment with a model variant where the speech recognizer (CTC) and variational sequential model only share low-level features while the two branches have more private layers to learn task-specific representations.
The ablation studies indicate that, when the depth of encoder is fixed, sharing fewer layers between the generative model and the speech recognizer in multitask learning works better than sharing more layers.
Please note that this observation does not necessarily mean deeper generative models is less useful than shallower generative moels.
I expect deeper generative models to learn more useful speech representations in an unsupervised representation learning scenarios than shallower generative models due to the higher model capacity.
Even if we restrict ourselves to a auxiliary/multitask learning setting,
there is no evidence that RecRep-Pyramid-MT with one variational recurrent layer (shared by CTC and RecRep) and $4$ CTC-specific layers would work better than RecRep-Pyramid-MT with two variational recurrent layers (shared by CTC and RecRep) and $3$ CTC-specific layers.
As mentioned before,
minimizing reconstruction loss and minimizing CTC loss may require very different final layer representations, and I attribute the success of RecRep-Pyramid-MT-1L and FB-MT to better handling of this conflict.
\end{itemize}

%----------------------------------------------------------------------------------------
%	Recurrent representation learning
%       1. Unsupervised pre-training failed
%       2. Pre-training based methods
%       3. Pre-training with VAE
%            a. In the input
%            b. In the latent space
%----------------------------------------------------------------------------------------
\chapter{Recurrent Representation Learning Using Unlabeled Data}
\label{cha:semi-and-pre-training}

%----------------------------------------------------------------------------------------
%       0. Motivation
%----------------------------------------------------------------------------------------

Recently, there have been many works attempting to learn speech representations under unsupervised setting, 
such as ~\citep{Dunbar2019TheZR, Chorowski2019UnsupervisedSR, Kamper2017ASF, Harwath2016UnsupervisedLO, Baevski2021UnsupervisedSR}.
These works learn representations without using label information and thus can utilize a large amount of unlabeled data.
In Chapter ~\ref{cha:feedforward} and Chapter ~\ref{cha:multiview}, I studied unsupervised representation learning and showed learned speech representations can benefit speech recognition tasks.
However, the feedforward encoder I used cannot easily take a complete utterance as input (e.g., we need to pre-segment the input utterance into overlapping pieces and then feed these pieces to the encoder separately), 
and thus fails to encode long-range dependency in the learned representations.
In Chapter ~\ref{cha:recurrent}, I studied using RecRep and its variants for auxiliary/multitask representation learning.
I found that an auto-encoding style learning objective, when jointly trained with the loss of the discriminative task, consistently improves the performance of a few sequence prediction tasks.
However, I have not yet explored in-depth learning speech representations utilizing a large amount of unlabeled data, with an encoder taking the complete utterance as input.
This chapter studies semi-supervised and unsupervised representation learning in this context, in particular focusing on evaluating how the speech recognition task benefits from the learned representations. \par

I first study vanilla RecRep-Pyramid in both semi-supervised and unsupervised learning settings.
Our experiments show that it is challenging to significantly improve upon the baseline using RecRep-Pyramid-MT in a semi-supervised learning scenario.
I also find that pre-training with RecRep-Pyramid is even less helpful than pre-training with feedforward VAE (taking a segment of a complete utterance as input) for speech recognition tasks.
I discuss the reasons that RecRep-Pyramid may fail to learn high-quality acoustic representations in the semi-supervised and unsupervised feature learning settings.
One of our hypotheses is that RecRep-Pyramid does not actively encourage encoders to learn representations that encode contextual information -- the learning objective of basic RecRep-Pyramid-MT only encourages the encoder to encode the temporal information rather than the context information.
On the other hand, as the encoder has seen the whole utterance, ``reconstructing what we have seen" facilitates training but not generalization given a powerful bidirectional encoder. \par

Unlike RecRep, which sees the complete utterance via bidirectional encoder before reconstructing the utterance,
recent pre-training approaches usually predict the unseen content of the input utterance.
For example, there are a few recent approaches for unsupervised representation learning ~\citep{oord2018representation, hjelm2018learning} based on the idea of maximizing a lower bound on or an approximation of mutual information (MI) between the representation of the seen context and future-time-step inputs.
Wav2vec ~\citep{schneider2019wav2vec} is one recent work that shows unsupervised pre-training can improve ASR trained on smaller labeled data sets,
where the authors use a loss similar to InfoNCE loss which is used in ~\citep{oord2018representation}.
The authors of wav2vec ~\citep{schneider2019wav2vec} further extend wav2vec to vq-wav2vec ~\citep{baevski2020vqwav2vec},
which learns to discretize the continuous wav2vec representations and use the discretized representations as input to BERT. 
Though simply quantizing the wav2vec representations does not show any advantage in speech recognition tasks over wav2vec,
the wav2vec+BERT pre-training does provide a better initialization to the acoustic model than wav2vec. 
Parallel to this line of MI-related research, there is also an unsupervised autoregressive model (e.g., APC, ~\citep{Chung2019AnUA}) that predicts the spectrum of future frames
, which is motivated by pre-trained language models in the NLP domain. \par

Though those methods obtain impressive results, they are not directly applicable to pre-training bidirectional encoders.
~\citep{Ling2020DeepCA, Ling2020DeCoAR2D} extend APC by combining ELMo style bidirectionality and the reconstruction objective of APC. The authors named the proposed method as ``deep contextualized acoustic representations (DeCoAR)".
Another approach to pre-train bidirectional encoder is called ``masked reconstruction"  ~\citep{wang2020unsupervised}.
Unlike the ELMo style bidirectionality of DeCoAR, masked reconstruction uses a bidirectional encoder to take inputs with temporal slices and spectrum domain slices masked and predicts the unseen content based on upstream and downstream information.
Though all use a BERT-Style training, vq-wav2vec is based on discretized representations while masked reconstruction directly reconstructs the missing part of the surface feature based on the observed remaining.
%Unlike the ELMo style bidirectionality of DeCoAR, and also different from vq-wav2vec, which uses Bert-Style training based on discretized representations,
%Masked reconstruction ~\citep{wang2020unsupervised} uses a Bert-Style training taking continuous signal as input.
%Masked reconstruction uses a bidirectional encoder to take inputs with temporal slices and spectrum domain slices masked
%and predicts the unseen content based on upstream and downstream information.
Unlike ~\citep{wang2020unsupervised},
wav2vec 2.0 ~\citep{baevski2020wav2vec} masks the speech input in the latent space.
Via contrastive loss, the model encourages the context representation centered over a mask to be similar to the correct quantized speech representations.
The authors claim that models pre-trained by wav2vec 2.0, after fine-tuning on transcribed speech, can outperform the state-of-the-art semi-supervised methods. \par

Due to the recent success of pre-training methods for learning acoustic representations,
I also investigate InfoNCE loss and masked reconstruction in this chapter for unsupervised representation learning.
I first explore using Contrastive Predictive Coding (CPC ~\citep{oord2018representation}) for acoustic feature learning and confirm that CPC can outperform VAEs and RecRep-Pyramid for the speech recognition task.
I explored ``predicting the future" (the FB model) in Section ~\ref{sec:reducing-conflict} in multitask learning scenario and observe no clear difference in terms of ASR performance between representations learned by ``predicting the future" and "reconstructing what we have seen" in multitask learning scenario.
%Our simple ``predicting the future" strategy only predicts the surface feature $x_t$ given the autoregressive representation of $x_{1:t-1}$;
The CPC differs from our simple ``predicting the future" strategy in that it enforces the autoregressive representation to be more predictable to a longer duration of unseen future context, and uses a loss other than reconstruction loss.
Although I find no clear difference in terms of ASR performance between representations learned by ``predicting the future" and "reconstructing what we have seen" in multitask learning scenario,
I do observe the clear benefits of CPC over RecRep-Pyramid (``reconstructing what we have seen") in unsupervised representation learning. \par
Though it is possible to employ techniques I used in Chapter ~\ref{cha:recurrent} to improve RecRep-Pyramid in unsupervised learning settings,
due to our observation that CPC significantly outperforms basic RecRep-Pyramid,
I have decided to focus on the learning paradigm of ``predicting unseen content" for unsupervised representation learning.
I also investigate a combination of masked reconstruction (~\citep{wang2020unsupervised}) and CPC,
but find that this hybrid approach does not outperform the masked reconstruction learning method.
I also explore improving upon masked reconstruction by using a more difficult learning objective, incorporating multi-view learning and a simple domain adaptation trick,
and find that all of these tricks improve representations for speech recognition tasks.

%----------------------------------------------------------------------------------------
%       1. Data and setup
%----------------------------------------------------------------------------------------

\section{Data and setup}
\label{sec:pre-training-data}

In this chapter, I focus on improving phone-based and character-based connectionist temporal classification (~\citep{graves2006connectionist}, CTC) systems.
I use the WSJ data set for semi-supervised representation learning (Section ~\ref{sec:recurrent-semi-supervised}) and unsupervised representation learning (Section ~\ref{sec:recurrent-unsupervised}),
and follow the same data processing and train/dev/test partitions as described in Section ~\ref{sec:data}.
Speech utterances belonging to SI284 are used for unsupervised representation learning and are also used as unlabeled data in semi-supervised learning.
(Part of) SI84 is used to train character-based or phone-based CTC systems
with dev93 and eval92 used as dev/test sets. \par

For all experiments related to pre-training, 
I strictly follow the setups described in ~\citep{wang2020unsupervised}.
I also use SI284 as the source of unlabeled data for improving phone-based and character-based CTC systems trained on SI84.
The input feature is $40D$ log mel filter bank energy (LFBE) with a window size of $25$ms and hop size of $10$ms.
Per-speaker mean (but not variance) normalization is applied for WSJ.
Besides SI284, I also pre-train an acoustic model on LibriSpeech ~\citep{panayotov2015librispeech},
a dataset consisting of $960$ hours of training data.
I do not use any information other than the audio of LibriSpeech.
Every $3$ consecutive frames are stacked after data augmentation (if applied) to accelerate the training.
I investigate the effect of different pre-training approaches on CTC systems that further fine-tune on SI284 and/or SI84.
I use the Kaldi s5 recipe ~\citep{povey2011kaldi} to generate a set of $351$ position-dependent phones, which are the labels for the phone-based CTC system.
I use $60$ characters for the character-based system, including the alphabet, digits, and punctuation symbols. 
I use the beam search algorithm (no language model) implemented in TensorFlow ~\citep{Abadi2016TensorFlowAS}, with a beam size of $20$, for evaluating phone/character error rates.

%----------------------------------------------------------------------------------------
%       2. Semi-supervised learning
%----------------------------------------------------------------------------------------

\section{Recurrent semi-supervised acoustic feature learning}
\label{sec:recurrent-semi-supervised}

In this section, I explore using basic RecRep-Pyramid-MT for semi-supervised representation learning on WSJ.
The labeled training set is SI84 ($15$ hours of speech) while the unlabeled training set is SI284 partition ($80$hour speech).
The baseline supervised recognizer is a character-based CTC recognizer.
The recognizer consists of one bidirectional LSTM on top of a neural network identical to the encoder of RecRep-Pyramid. The RecRep-Pyramid encoder consists of another two bidirectional LSTM layers followed by linear transformation layers.
I train RecRep-Pyramid-MT  by minimizing the loss:

\begin{eqnarray}
(1-\alpha) \mathbb{E}_{ x_{1:T} \in SI84 } \Big\{ \log{ p( l_{1:M} \vert R\left( x_{1:T} \right) ) } \Big\} - \alpha \mathbb{E}_{ x'_{1:T'} \in SI284 } \Big\{ ELBO(x'_{1:T'}) \Big\} 
\label{eqn:recrep-semi-joint-loss}
\end{eqnarray}

Note that the only difference compared with the loss used for multitask learning (Equation ~\ref{eqn:recrep-elbo-pyramid}) is the ELBO term $ \mathbb{E}_{ x'_{1:T'} \in SI284 } \Big\{ ELBO(x'_{1:T'}) \Big\}$.
Unlike in multitask learning, the training samples for this term are from SI284 partition rather than SI84. 
I implement it by calculating both the cross-entropy loss and negative ELBO for all samples from SI84 partition in one batch,
then randomly select a few samples from SI284,
%and only calculate the negative ELBO for them.
for which I only calculate the negative ELBO for them.
The per-epoch loss is then the sum of the three terms' expectation: discriminative loss, ELBO on SI84, and ELBO on SI284.
Depending on the batch size used for SI84 and SI284,
we may not finish iterating both SI84 and SI284 partitions simultaneously.
In practice, I say one epoch is finished when the training has iterated over all samples of SI84 once. \par

\begin{figure*}[htbp]
 \centering
 \includegraphics[width=1.0\textwidth]{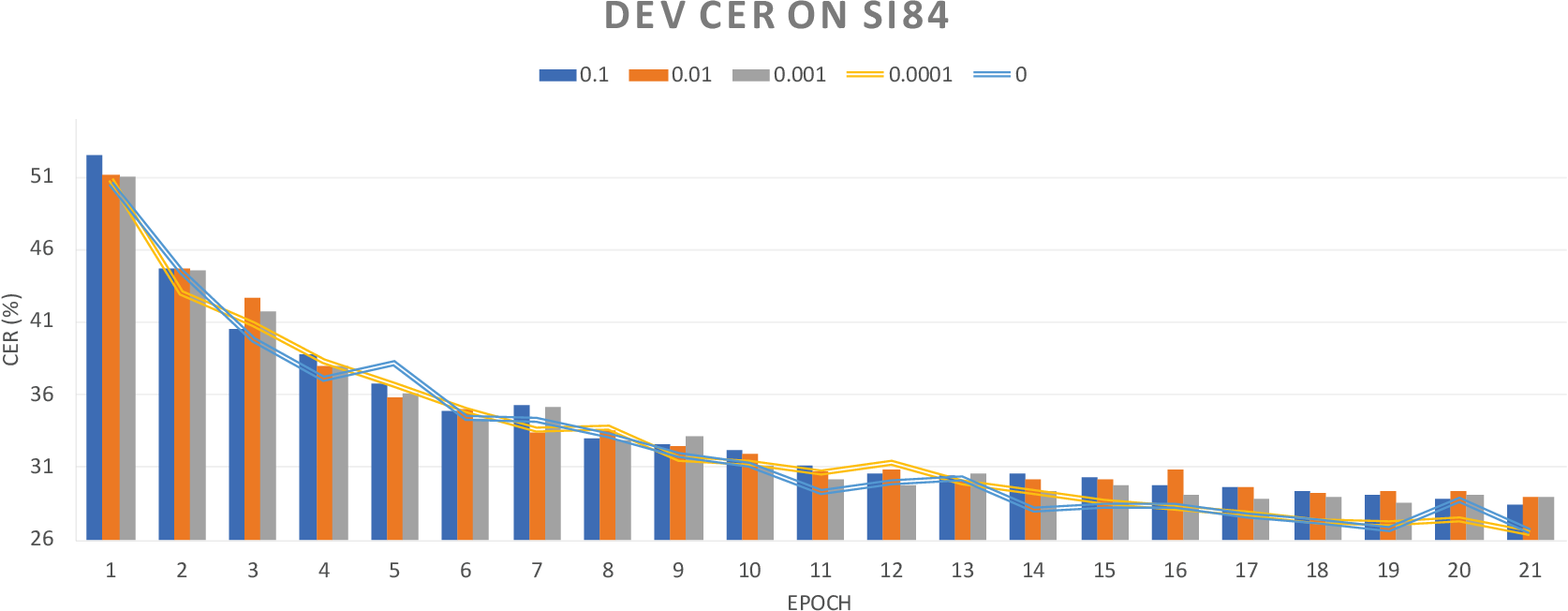}
 \caption{Dev CER ($\%$) of baseline CTC and CTC jointly trained with RecRep-Pyramid (RecRep-Pyramid-MT).
 The RecRep-Pyramid-MT is trained via semi-supervised learning.
SI84 is the labeled training set while $SI284$ is the source for training ELBO. 
$\alpha$ is the weight of negative ELBO term in the joint loss.
$\alpha=0$ and $\alpha=10^{-4}$ are shown by lines in the plot while the other $\alpha$s are shown as bars.}
 \label{fig:wsj-cmp2}
\end{figure*}

I vary the trade-off  hyper-parameter $\alpha$ from among $\{0.0, 10^{-4}, 10^{-3}, 10^{-2}, 10^{-1}\}$ to see how the loss on the unlabeled data affects the performance of the CTC recognizer. 
Unfortunately, our experiments suggest that it is very difficult to outperform the CTC baseline recognizer. 
As shown in Figure ~\ref{fig:wsj-cmp2}, only when $\alpha=10^{-4}$ does RecRep-Pyramid-MT achieve similar dev set CER to the baseline ($\alpha=0.0$). \par

\begin{figure*}[htbp]
 \centering
 \includegraphics[width=1.0\textwidth]{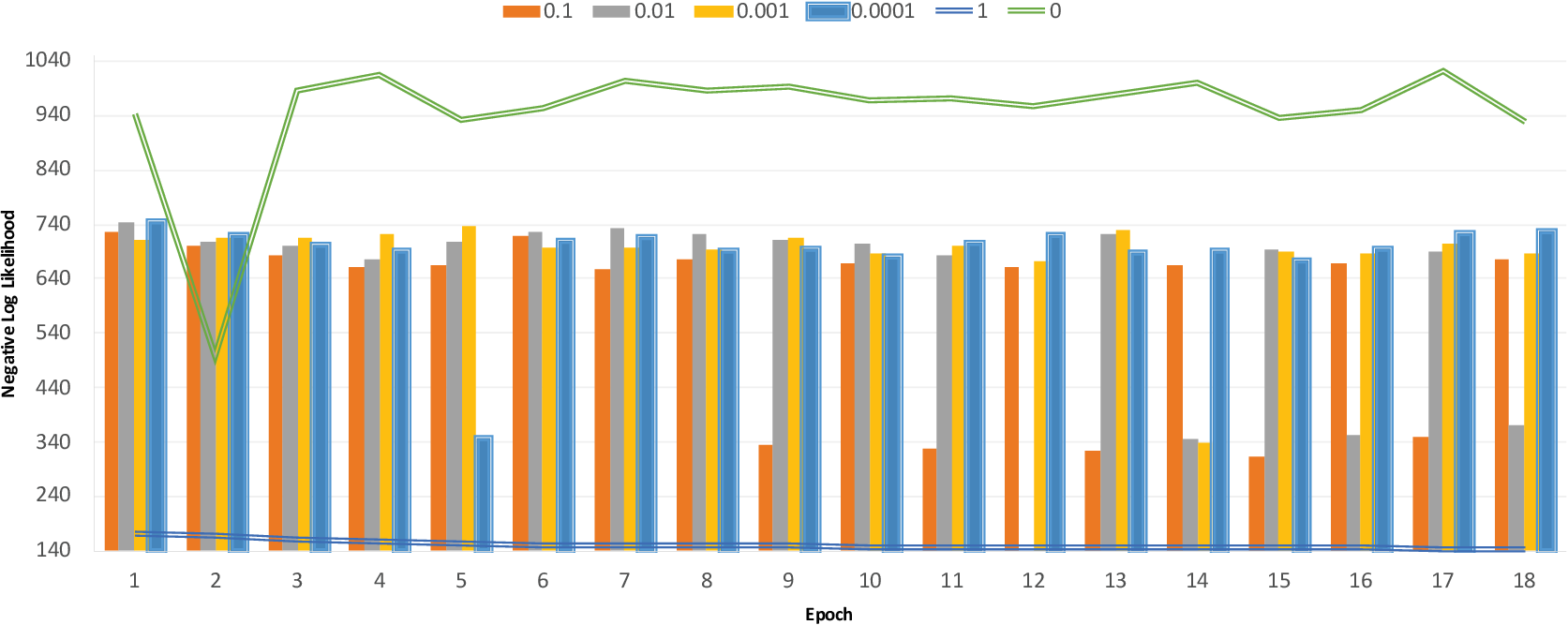}
 \caption{The plot shows the averaged negative log likelihood ($-\frac{1}{\floor{\frac{T}{2}}} \sum_{k=1}^{\floor{\frac{T}{2}}}\log{p_{\theta}([x_{2k}||x_{2k+1}]|z_{2k})}$ on SI284 for models with different $\alpha$s.
 $\alpha=0$ and $\alpha=1$ are shown as lines while all the other cases are shown as bars. }
 \label{fig:semi-loss}
\end{figure*}

I also examine the averaged negative log likelihood on the dev set (dev93) as shown in Figure ~\ref{fig:semi-loss}.
In this figure, the cases where $\alpha=0$ (CTC loss only) and $\alpha=1$ (RecRep-Pyramid loss only) are plotted as lines. 
These two cases indicate the upper and lower bounds on the averaged negative log likelihood ($-\frac{1}{\floor{\frac{T}{2}}} \allowbreak \sum_{k=1}^{\floor{\frac{T}{2}}}\log{p_{\theta}([x_{2k}||x_{2k+1}]|z_{2k})}$) \footnote{$||$ means concatenation here} .
All of the other cases are shown as bars. \par

As expected, I observe that larger $\alpha$ leads to lower loss (better reconstruction).
However, I also observe one interesting phenomenon: when the weight on ELBO is no longer negligible (e.g., when $\alpha=0.1$ or $\alpha=0.01$), 
the quality of the reconstruction fluctuates substantially.
To stabilize the training, I need to use relatively small $\alpha$ (e.g., $10^{-3}$ and $10^{-4}$). 
Unfortunately, smaller $\alpha$s do not reduce CER on the dev set significantly. \par

This conflict does not only exist in semi-supervised learning; 
it is also observed in multitask learning; that is, training RecRep-Pyramid-MT without using additional unlabeled data.
For example, as we can see from the Figure ~\ref{fig:wsj-layer-cmp} of Section ~\ref{sec:reducing-conflict}, 
compared with Baseline (a CTC system), 
RecRep-Pyramid-MT shows more significant fluctuation in terms of per-epoch dev set phone error rate.
The fluctuation indicates the conflict between the per-frame reconstruction loss and CTC loss, presumably due to the different preferences of the two losses.
As blank label is dominant in the outputs (before removing redundant labels) of a well-trained CTC recognizer,
such a label distribution implicitly encourages the encoder's LSTM layers, especially those closer to the output, to be more predictable to blank labels.
However, when doing per-time-step reconstruction, 
the latent representations need to capture the phoneme information and temporal information rather than simply favoring blank labels. \par

A potential solution to this conflict is to introduce an auxiliary latent variable as described in Section ~\ref{subsec:auxiliary}.
As the auxiliary latent variable can focus on the reconstruction task, the primary latent variable can be more effectively adapted to the values favored by CTC loss.
Another potential solution could be to use more private layers for the CTC system, which is carefully studied in Section ~\ref{sec:reducing-conflict}.
In this way, the representation layers are less affected by the CTC output as they are now further away.
In Chapter ~\ref{cha:recurrent}, I showed that either of these options helps RecRep(-Pyramid) to learn representations more suitable for CTC-based ASR in the multitask learning setting.
As we are more interested in how to learn good speech representations rather than alleviating the conflict between CTC loss and frame reconstruction loss in this chapter,
I do not conduct other large-scale semi-supervised experiments using the techniques introduced in Character ~\ref{cha:recurrent}. 
In the next section, I explore unsupervised speech representation learning, where the representation learner is not jointly trained with any ASR system. 
Thus the unsupervised training objective is not affected by the different preferences of the discriminative recognizer loss during training. \par

%----------------------------------------------------------------------------------------
%       3. Unsupervised learning; Reconstruction -> Prediction
%----------------------------------------------------------------------------------------

\section{Unsupervised acoustic feature learning}
\label{sec:recurrent-unsupervised}

I explore unsupervised acoustic feature learning in this section using XRMB with setups described in Section ~\ref{sec:data},
and also WSJ with setups described in Section ~\ref{sec:pre-training-data}.
I first train VAE, RecRep-Pyramid, and Contrastive Predictive Coding (CPC, ~\citep{oord2018representation}) on SI284.
I then use the pre-trained encoders to transform the surface features into representations and train character-based CTC recognizers using SI84, which is much smaller than SI284.

On the multitask learning setting from Section ~\ref{sec:reducing-conflict},
I observe that ``predicting unseen content" does not outperform ``reconstructing what we have seen";
however, I was using one-layer LSTM in our experiments, and I simply encouraged our model to predict next-time-step unseen frame.
%works slightly better than ``predicting unseen content" in multitask learning setting in Section ~\ref{sec:reducing-conflict},
%this observation may not generalize to an unsupervised learning setting when label information is not available during pre-training.
I revisit the topic of comparing ``reconstructing what we have seen" (e.g., RecRep-Pyramid and VAE) and ``predicting unseen content" (e.g., CPC) in this section for unsupervised learning settings.
Our experimental results suggest that the representations learned by CPC yield the best CTC recognizer. \par

I explained the use of VAE (with feedforward encoder) and RecRep-Pyramid for representation learning in Chapter ~\ref{cha:feedforward} and Chapter ~\ref{cha:recurrent} respectively.
See Figure ~\ref{fig:feedforward} to review how we use feedforward VAE for representation learning, and see Equation ~\eqref{eqn:recrep-elbo-pyramid} in Chapter ~\ref{cha:recurrent} for the loss to minimize to train RecRep-Pyramid.
Before I describe my experiments, I first briefly explain Contrastive Predictive Coding (CPC, ~\citep{oord2018representation}).

\subsection{Contrastive Predictive Coding}
\label{subsec:cpc}

The key insight of Contrastive Predictive Coding (CPC) is to learn representations that can predict the future in the latent space.
To induce the latent space to encode information which is maximally useful to predict future samples, 
the authors try to maximize the mutual information between current step representations and future-time-step observations.
More specifically, given a sequence input $x_{1:T}$, a representation $c_t$ for time step $t$ can be learned by maximizing a lower bound on the mutual information $ I \left( x, c_t \right)$, 
where $x \in \left\{ x_{t+1}, x_{t+2}, \cdots, x_{t+K} \right\}$ and $K$ is the number of steps in the future to be predicted. \par

Several approaches ~\citep{belghazi2018mine, hjelm2018learning, oord2018representation} have been proposed for estimating the mutual information.
The particular estimate used by CPC ~\citep{oord2018representation}, called ``InfoNCE", is:
\begin{equation}
I \left( x_{t+k}, c_t \right) \geq - \mathbb{E}_{X} \left\{ \frac{ f(x_{t+k},c_t) }{ \sum_{m=1}^{N} f(x_{j_m},c_t) } \right\}
\label{eqn:cpc}
\end{equation}
where $X = \left\{ x_{j_1}=x_{t+k}, x_{j_2},\cdots,x_{j_{N}} \right\}$ 
consists of one positive sample $x_{j_1}=x_{t+k}$ and $N-1$ random samples from $x_{1:T}$ other than $x_{t+k}$ for $1 \leq k \leq K$, or from utterances other than $x_{1:T}$.
$f(x_{j_m},c_t)$ is a log bilinear function $e^{z_{j_m} W_k c_t}$ where $z_{j_m}$ is the latent representation of $x_{j_m}$. 
$c_{t}$ is the output of an autoregressive model summarizing $z_{\leq t}$.

\subsection{Experiments}
\label{subsec:cpc-exp}

I compare CPC, RecRep-Pyramid, and non-sequential VAE (feedforward neural network encoder) using XRMB and WSJ.
When working on WSJ,
all the three unsupervised feature learning approaches are trained on SI284, 
then the encoders are used to transform the samples of SI84 before recognizer training. 
For the remaining setups on the two datasets, please see Section ~\ref{sec:data}.
Below I describe some implementation and experimental details:

\begin{itemize}
\item[VAE:] I use $3-$layer feedforward neural networks ($1024$ units ReLU) followed by one linear transformation to generate a $k-$dimensional vector for each $7-$frame window centered at each frame of every utterance.
The $k-$dimensional vector is the mean of a multivariate Gaussian.
Here $k=70$ for XRMB and $k$ is selected from among $\{120, 150\}$ for WSJ. 
I use batch size $1024$, dropout $0.2$, and $\beta=1.0$.
All models are trained up to $20$ epochs.
\item[RecRep-Pyramid:] The encoder is a $2-$layer bidirectional LSTM ($256$ hidden units per direction) followed by linear transformation.
The dimensionality of the per-time-step latent variable is either $120$ or $150$ when working on WSJ, and $70$ when working on XRMB.
The weight of $\beta$ is chosen from $\{ 1.0, 0.1, 0.01, 0.001\}$;
dropout is selected from $\{0.0, 0.2, 0.4\}$. 
I use batch size $16$ and train each model up to $40$ epochs. 
I infer the mean value of per-time-step Gaussian posteriors as the sequential representations, which are the input to the downstream tasks.
More specifically, given a RecRep-Pyramid model we have trained, I use it to infer $\{\mu_{2k}\}_{k \in 1,\cdots,\floor{\frac{T}{2}}}$ for each input $x_{1:T}$. 
\item[CPC:] When implementing CPC,
I use $2-$layer feedforward neural networks ($1024$ units ReLU) followed by one linear transformation to learn $z_{j_m}$.
I use unidirectional $3-$layer LSTMs followed by a linear transformation layer to learn $W_k c_t$, with the dimension of all hidden units being $256$.
CPC features are then extracted using the linear transformation on top of the $3-$layer unidirectional LSTMs. 
The input of the CPC model is the concatenation of $3$ consecutive frames to shorten the length of the input sequence and accelerating training.
\item[CTC Recognizer:] For downstream tasks, I train a $3-$layer bidirectional LSTM CTC recognizer, taking surface features or learned representations as input.
For each pre-trained model, the epoch with the best validation loss is saved for generating representations.
All the CTC recognizers are trained using Adam optimizer with the initial learning rate being $0.005$.
All models are trained to convergence (e.g., No improvement observed on the dev set in $5$ epochs).
For each pre-training method, the model with the best dev set CER is then reported in Table ~\ref{tab:compare-cpc}. 
For remaining setups (e.g., train/dev/test partitions information, other hyperparameters for training downstream CTC recognizers), please refer to Section ~\ref{sec:pre-training-data}. 
\end{itemize}

\begin{table} [htbp]
\centering
\begin{tabular}{| l | r | r | r | r | r | r |}
 \hline
 Models & $\frac{1}{16}$ SI84 & $\frac{1}{8}$ SI84 & $\frac{1}{4}$ SI84 & $\frac{1}{2}$ SI84 & XRMB \\
 \hline \hline
 1. Baseline & 53.0 & 48.6 & 38.7 & \textbf{27.1} & 11.3 \\
 2. VAE & 51.5 & 46.3 & 38.1 & 34.9 & 9.6 \\
 3. RecRep-Pyramid & 55.6 & - & - & - & - \\
 4. CPC & \textbf{51.0} & \textbf{44.1} & \textbf{35.6} & 31.3 & \textbf{8.9}\\
 \hline
\end{tabular}
\caption{Comparison of several representation learning approaches on the dev set of WSJ and on XRMB. 
I report character error rate (CER) on WSJ and $6-$fold averaged test set phone error rate (PER) on XRMB. 
All recognizers are trained to convergence.
For each trained CTC recognizer, the epoch with lowest dev set error rate is reported.}
\label{tab:compare-cpc}
\end{table}

From Table ~\ref{tab:compare-cpc}, 
I observe that CTC recognizers trained using VAE and CPC features both significantly outperform the baseline recognizer (trained directly using the surface feature) on XRMB, with the recognizer trained using CPC features significantly outperforming that trained using VAE features.
Note that, on XRMB, 
the recognizers corresponding to CPC take the concatenation of MFCCs and features learned by CPC as input.
On WSJ, when using $\frac{1}{16}$, $\frac{1}{8}$ and $\frac{1}{4}$ of SI84 to train recognizers, 
the CPC-based recognizer significantly outperforms both VAE and baseline.
However, recognizers trained using RecRep-Pyramid features can outperform neither VAE-feature-based recognizers nor baseline recognizers, even when the baseline recognizer is trained using only $\frac{1}{16}$ of the SI84.
When we use more labeled training data, e.g., when using the complete set of SI84 for recognizer training,
CPC-based and VAE-based recognizers all fail to outperform the baseline recognizer.
This observation again is consistent with our finding in Section ~\ref{sec:15w-exp} 
that the relative amount of unlabeled data matters for unsupervised representation learning. \par

Note that, although VAE, RecRep-Pyramid, and CPC use different encoder architectures,
VAE is comparable to RecRep-Pyramid in terms of the number of model parameters, 
%VAE is comparable to RecRep-Pyramid, 
and they all have more parameters than CPC. 
From this perspective, I believe that the comparison shown in Table ~\ref{tab:compare-cpc} is still fair.

\subsection{Potential Problems Preventing RecRep(-Pyramid) from learning good representations}
\label{sec:recurrent-limitation}

Our experimental study reveals the difficulty of learning good representations using RecRep-Pyramid under both unsupervised and semi-supervised acoustic feature learning settings.
In Section ~\ref{sec:recurrent-semi-supervised}, I discussed that different preferences of CTC loss and reconstruction loss prevent RecRep-Pyramid from learning ASR-favored representations in the semi-supervised learning setting.
Here, I discuss two potential reasons why RecRep-Pyramid cannot learn useful representations for both unsupervised and semi-supervised learning. \par

\paragraph{The Reconstruction unit matters}
In Table ~\ref{tab:compare-cpc}, 
VAE clearly outperforms RecRep-Pyramid on learning representations for speech recognition. \par
VAE takes a $7-$frame concatenated window $\{x_{t-3},\cdots,x_t,\cdots,x_{t+3}\}$ as input, 
and uses inferred latent posterior $q_{\phi}\left(z_t \vert \big[ x_{t-3}, \cdots, x_t, \cdots, x_{t+3} \big]\right)$ to approximate ground truth posterior $p_{\theta}\left(z_t \vert \big[ x_{t-3}, \cdots, x_t, \cdots, x_{t+3} \big]\right)$. \par

However, the RecRep-Pyramid model described in Figure ~\ref{fig:pyramid} uses $q_{\phi} \left( z_t|h_t \right) $ to approximate $p_{\theta}\left(z_t \vert [x_{t-1}, x_{t}]\right)$. 
That is, maximizing the ELBO encourages the inferred posterior $q_{\phi}\left( z_t|h_t \right)$ to capture and store local information (e.g. the information needed for reconstructing $[x_{t-1}, x_{t}]$) in $z_t$,
rather than the contextual information. 
As $z_t$ is only used for reconstructing the time step $t$
in the RecRep-Pyramid model,
there is no other direct driving force to encourage $z_t$ to be more contextually aware.
\par

This issue is more severe in unsupervised learning settings than in multitask learning settings.
In a multitask learning setting, the RecRep-Pyramid per-time-step ELBOs serve as regularizers of the discriminative model,
as I have shown in Figure ~\ref{fig:timit-cmp1} of Section ~\ref{subsubsec:recurrent-asr}.
In unsupervised learning, the only loss I have is the ELBO.
If ELBO does not encourage latent variables to capture contextual information,
the latent variables presumably will only capture very local information that is insufficient to benefit many downstream discriminative tasks. \par

\paragraph{Reconstruction vs Predicting Unseen Content}
Recently, many successful unsupervised representation learning methods were proposed across the computer vision, natural language processing, and speech processing communities.
I listed a few closely related to speech processing in the introduction of this chapter.
Unlike RecRep(-Pyramid), which tries to reconstruct each time step given the knowledge of the whole input sequence,
these methods all fall under the umbrella of ``predicting the unseen content".
Like the CPC investigated in this section, it maximizes a lower bound on the mutual information between the current-time-step representation and future unseen frames.
I hypothesize that predicting unseen content forces the encoder to infer the missing/proceeding frames in a context-aware manner.
If we imagine the input sequence of frames as a legitimate language, 
predicting the masked or future content forces the model to understand what an appropriate spectrogram looks like and encodes the needed information.
The way RecRep(-Pyramid) learns representations lacks this ``inferring the missing information" part when encoding information into the representations.
However, I am not trying to draw the conclusion that ``predicting unseen content" is always a better way than ``reconstructing what we have seen" for learning acoustic representations; the conclusion may vary between problems.
Based on the literature as well as our experimental observation
%and what we observe in experiments 
that predicting unseen content is the more promising approach, I will use it exclusively in the remainder of this section.
%more promising,
%we would shift to ``predicting unseen content" in the remaining sections of this chapter.

\paragraph{Potential solutions for improving RecRep(-Pyramid)}
Before moving to the next section, I would like to discuss potential solutions for improving RecRep(-Pyramid) for unsupervised and semi-supervised representation learning.
As discussed in Chapter ~\ref{cha:recurrent}, in multitask learning scenarios,
the success of RecRep(-Pyramid) relies heavily on extra supervision or heuristic guidance measures such as better priors, context-aware per-time-step unit reconstruction, and auxiliary latent variables used for information decomposition.
%, and auxiliary latent variables. 
For example, in Chapter ~\ref{cha:recurrent} which focuses on multitask learning, 
the basic version of RecRep(-Pyramid-)MT (e.g., without self prior updating or auxiliary latent variables) fails to outperform the baseline.
Tables ~\ref{tab:ner+chunking} for the NER and chunking task and ~\ref{tab:timit-multitask} for speech recognition on TIMIT both show that all
basic RecRep(-Pyramid) models without self prior updating and auxiliary latent variables fail to outperform the baseline.
Similarly, in Table ~\ref{tab:recrep-win}, 
RecRep(-Pyramid) itself does not improve upon the baseline unless we change the reconstruction unit $u_t$ to a window with proper window size.\par

As discussed in Section ~\ref{sec:recurrent-semi-supervised},
using a shallower representation encoder and auxiliary latent variables will presumably help RecRep-Pyramid-MT in the semi-supervised learning scenario.
I suspect that the two methods, specifically designed for when labeled information is available, are not applicable in the unsupervised learning scenario -- we need extra inductive bias (instead of labels) to encourage the primary and auxiliary latent variables to capture different information without labels. 
However, self prior updating (see our discussion on the advantage of prior updating in Section ~\ref{subsec:self-prior-updating}) and reconstructing complex per-time-step units may be helpful.
In particular, reconstructing complex per-time-step units tackles the first problem we discussed preventing RecRep(-Pyramid) from learning useful information.
As we now understand the limitations of vanilla RecRep-Pyramid for learning contextual-aware representations compared to predicting-unseen-content approaches, we move to the predicting-unseen-content learning paradigm from the next section. \par

%----------------------------------------------------------------------------------------
%       4. Masked Reconstruction
%----------------------------------------------------------------------------------------

\section{Masked Reconstruction}
\label{sec:mask}

%% Need some glue to move to this section
We compared RecRep-Pyramid and CPC for unsupervised representation learning in Section ~\ref{sec:recurrent-unsupervised}.
We have seen the superior performance of CTC recognizers based on CPC-learned representations versus recognizers trained with surface features and features learned by other approaches.
We discussed the potential difficulties for basic RecRep-Pyramid in learning representations in unsupervised learning scenarios and decided to switch to the learning style of ``predicting unseen content". 
%In this section, we still study representation learning without using label information.
%However, in Section ~\ref{sec:recurrent-unsupervised} 
Unlike in previous section where we used a pre-trained encoder as a feature extractor to generate representations of surface features subsequently used as input to the downstream tasks, 
here we fine-tune the pre-trained encoder using a smaller labeled dataset rather than only using the pre-trained encoder as feature extractor.
As CPC (and some closely related work such as wav2vec) relies on predicting future content, the pre-trained encoder is typically unidirectional and is challenging to be used to initialize a recognizer with a bidirectional RNN encoder.
In this section, we explore ``predicting unseen content using bidirectional RNN" rather than ``predicting future observations using unidirectional RNN." \par

Inspired by BERT-style pre-training (i.e., inferring the missing content from the remaining content), 
in the paper ~\citep{wang2020unsupervised},
the authors propose masked reconstruction for speech signals (e.g., spectrograms).
Due to the difference between speech and text input (e.g., the speech signal is continuous and typically with much finer granularity),
it does not make sense to directly apply the masking strategy of BERT to a spectragram.
Motivated by the success of SpecAugment ~\citep{Park2019SpecAugmentAS}, 
the authors propose to mask a few channels across all time steps and a few segments (e.g., contiguous frames) of the input sequence. 
As we focus on speech signals throughout this chapter, for simplicity, we refer to this masked reconstruction method as ``\textbf{BERT}" in this section. \par

%% Need some introduction to the organization of this section
We explore a few extensions of ``BERT" in this section, including:
\begin{itemize}
\item[1] \textbf{Reconstructing the Central Region (BERT-Half)}: Unlike BERT, which is trained to predict the entire missing regions, we tried \textbf{BERT-Half}, which only tries to predict the central part of the masked area of the input. We experiment with this idea in Section ~\ref{subsec:bert-half}.

\item[2] \textbf{Masked Contrastive Predictive Coding (BiCPC)}: We combine CPC and BERT and propose to make the learned representations maximally useful for predicting the regions covered by the masks. We discuss more details in Section ~\ref{subsec:contrastive}.

\item[3] \textbf{Multiview Masked Reconstruction}: We also extend BERT to multi-view BERT, that is, we treat an utterance with different masks applied as two different views of the same utterance, and encourage the learned representations of the two views to be similar to each other. 
We discuss this approach in more depth in Section ~\ref{subsec:consistency}.
\end{itemize}

\subsection{Reconstructing the Central Region}
\label{subsec:bert-half}

\begin{figure*}[htbp]
 \centering
 \includegraphics[scale=0.5]{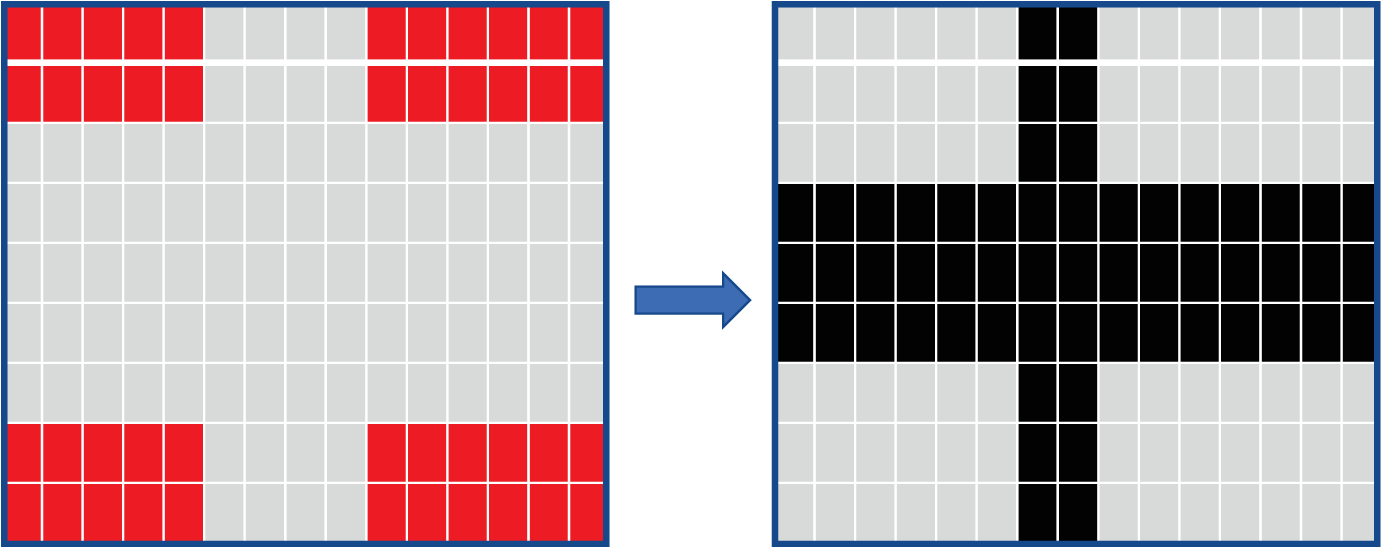}
 \caption{Illustration of BERT-Half. 
The left part indicates a spectrogram after a binary mask $M$ is applied (gray cells for unseen/missing part). The right part indicates the central region of the unseen region that BERT-Half tries to predict.}
 \label{fig:Bert-Half}
\end{figure*}

\begin{figure*}[htbp]
 \centering
 \includegraphics[scale=0.4]{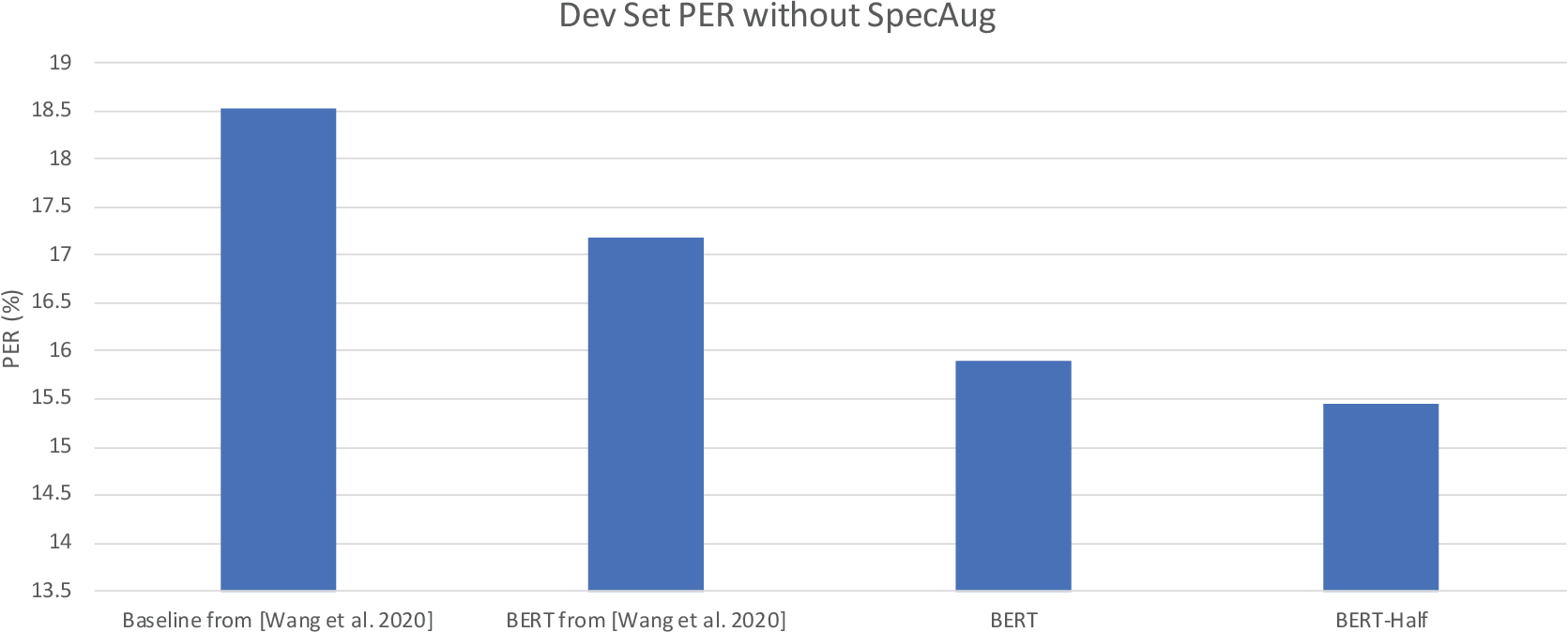}
 \caption{Phone error rate on SI84.
 Performance of BERT (masked reconstruction) vs BERT-Half (Reconstructing center of the masked region) vs Baseline. 
 Note that, in ~\citep{wang2020unsupervised}, the authors show results with and without using SpecAugment ~\citep{Park2019SpecAugmentAS} during CTC training; in this figure, we only show the results without using SpecAugment.}
 \label{fig:performance1}
\end{figure*}

Given an utterance $X$ and a mask $M$ (a binary matrix with the same shape as $X$), 
the input to the encoder $f$ is $X \odot M$, 
and decoder $g$ attempts to recover the missing region $(1-M) \odot X$. 
Unlike many recent powerful pre-training works ~\citep{Ling2020DeCoAR2D, baevski2020wav2vec} whose encoders are based on a transformer ~\citep{vaswani2017attention}, we use a stacked bidirectional LSTM as the encoder in this thesis as most works predate the transformer era. \footnote{The work of this thesis was mainly done before the transformer era, and this thesis represents a line of work done before most of the recent wave of transformer models and self-supervised speech representation work. In fact, my work represents some of the early work on unsupervised neural sequence representation learning. If this work were done today, it would likely use different types of models and different experimental settings. Many of the ideas in the thesis, such as the losses used and aspects of variational models we developed, may well apply to newer models as well, although we haven't tested them in that context.}
The decoder $g$ is a feedforward neural network.
The loss is 

\begin{equation}
L(X,M;f,g) = \norm{(1-M) \odot [ X - g(f(M \odot X)) ] }_{\text{Fro}}^{2}
\label{eqn:masked_recon}
\end{equation}

Given a mask $M$,
due to the temporal coherence, the observed context provides the most information regarding the frames near the boundaries of missing regions.
Thus, it is intuitive that
the frames located near the boundaries are more easily reconstructed
%much easier to be reconstructed 
than those in the central part of the missing region.
To make the masked reconstruction (Equation ~\ref{eqn:masked_recon}) more difficult,
we can only reconstruct the time steps and frequency banks centered over the region being masked.
We illustrate this idea in Figure ~\ref{fig:Bert-Half}. We denote the corresponding binary matrix for BERT-Half as $M_{\text{central}}$. Then the loss for BERT-Half is shown as below:
\begin{equation}
L(X,M,M_{\text{central}};f,g) = \norm{(1-M_{\text{central}}) \odot [ X - g(f(M \odot X)) ] }_{\text{Fro}}^{2}
\label{eqn:masked_recon_half}
\end{equation}

In Figure ~\ref{fig:performance1},
we compare the two reconstruction strategies: 1) completely reconstructing the masked region (BERT) and 2) only reconstructing the contiguous central frames/banks of the masked region (BERT-Half).
Both encoders are $4-$layer bidirectional LSTM,
with $512$ units per direction.
The pre-training is done on si284. Then the pre-trained model is used to warm-start a phone-based CTC recognizer trained on si84.
The baseline recognizer is also a $4-$layer bidirectional CTC recognizer with random weight initialization.
As shown in Figure ~\ref{fig:performance1},
masked reconstruction improves the performance (which aligns with the findings in ~\citep{wang2020unsupervised}), and only reconstructing the central region improves the performance futher,
% further helps,
which meets our expectations. \par
%All results are without SpecAug during recognizer training.

%%%%%%%%%%%%%%%%%%%%%%%%%%%%%%%%%%%%%%%%%%%%%%%%%%%%%%%%%%%%%%%%%
\subsection{Masked Contrastive Predictive Coding}
\label{subsec:contrastive}

Based on the success observed on WSJ using CPC for acoustic feature learning and BERT for pre-training,
we propose to combine the concepts of masked reconstruction and contrastive predictive coding for representation learning.
One naive way of combining the two approaches is to replace the reconstruction loss used in masked reconstruction with InfoNCE loss.
We refer to such a model as ``\textbf{Masked Contrastive Predictive Coding}".
Rather than directly reconstructing the missing region,
this modified approach tries to maximize a lower bound on mutual information (MI) between the representations of the observed context and the missing frames.
From the perspective of contrastive predictive coding,
masked contrastive predictive coding extends the ``unidirectional" CPC to a bidirectional version,
thus we also name ``masked contrastive predictive coding" as ``\textbf{BiCPC}" for short.
 \par

\begin{figure*}[htbp]
 \centering
 \includegraphics[scale=0.47]{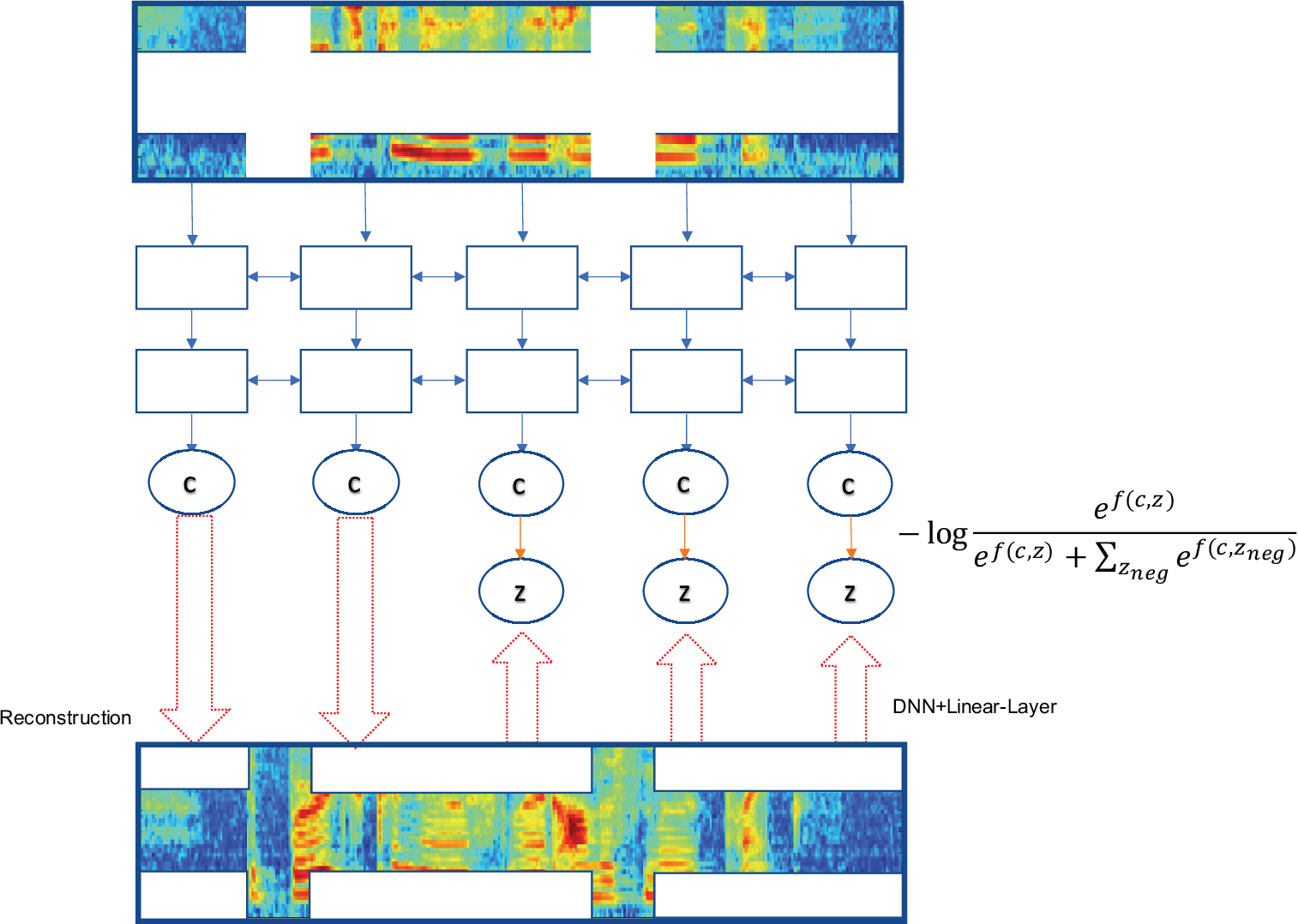}
 \caption{Comparison of BERT (masked reconstruction) and BiCPC (masked contrastive predictive coding). 
BERT directly constructs the missing regions of the input spectrogram,
\textbf{while BiCPC tries to predict the missing regions in the latent space} via the InfoNCE loss proposed in the paper ~\citep{oord2018representation}.}
 \label{fig:recon+contrastive}
\end{figure*}

\begin{figure*}[htbp]
 \centering
 \includegraphics[scale=0.4]{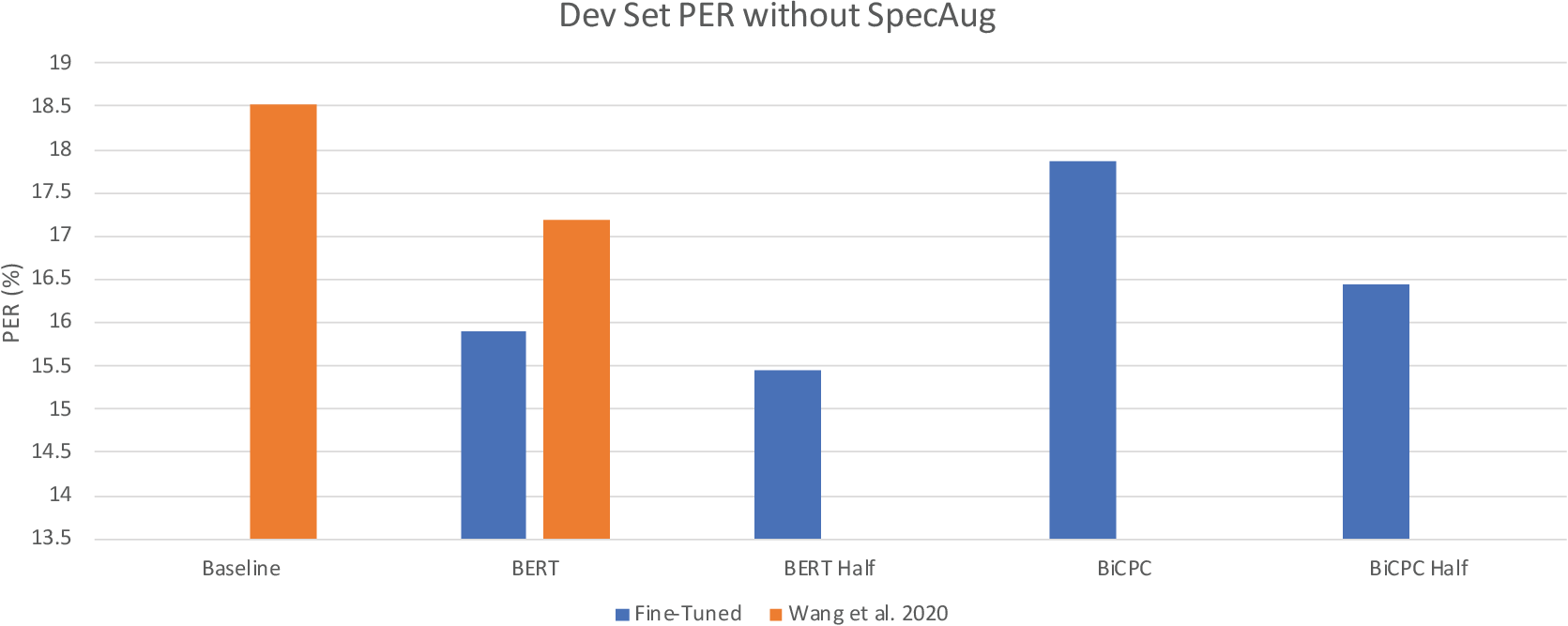}
 \caption{Phone error rate (PER) on SI84. 
We compare the PER of the baseline CTC recognizer and CTC recognizers pre-trained by 1) BERT, 2) BERT-Half, 3) BiCPC, and 4) \textbf{BiCPC-Half}.
Please note, ``BiCPC-Half" takes the central region of $X \odot (1-M)$ as input to learn both positive and negative samples in the latent space.
The orange bars are numbers taken from the paper ~\citep{wang2020unsupervised}.}
 \label{fig:performance2}
\end{figure*}

We use Figure ~\ref{fig:recon+contrastive} to illustrate ``Masked Contrastive Predictive Coding (BiCPC)". 
As shown in the right part of the Figure ~\ref{fig:recon+contrastive}, 
we first use a DNN (a few ReLU layers each with 1024 units followed by linear transformation) to transform masked region into latent space,  

\begin{equation}
Z^{\text{pos}} = DNN_{\phi}(X \odot (1-M))
\label{eqn:latent}
\end{equation}

Please note, if $X$ is of length $T$, $Z^{\text{pos}}$ is also of length $T$. 
We denote $i$-th element of $Z^{\text{pos}}$ as $z_i^{\text{pos}}$. To obtain negative samples of $Z^{\text{pos}}$,
we first randomly shuffle $X \odot (1-M)$ to get $N$ shuffled copies, denoted as $S^{(1:N)}$;
Then, for each shuffled version of $X \odot (1-M)$, we also obtain latent representations in the latent space as:

\begin{equation}
Z^{(i)} = DNN_{\phi}(S^{(i)}) \quad \text{for $1\leq i \leq N$} 
\end{equation}

The context encoder we use is a $4-$layer bidirectional LSTM ($512$ units per direction), which takes $X$ and $M$ as input.
The outputs of this $4-$layer bidirectional LSTM (Equation ~\ref{eqn:context}) are considered as contextual representations that are used for predicting missing frames in the latent space.

\begin{equation}
C = BiLSTM_{\theta}(X \odot M)
\label{eqn:context}
\end{equation}

We then have InfoNCE loss, which tries to predict the masked regions in the latent space as
\begin{equation}
L(X,M,S^{(1:N)},\phi,\theta) = \sum_{i=1}^{T} l(c_i,z^{\text{pos}}_i,z^{(1:N)}_i)
\end{equation}
where
\begin{equation}
l(c_i,z^{\text{pos}}_i,z^{(1:N)}_i) = -\log \frac{\text{exp}(c_i z^{\text{pos}}_i)}{\text{exp}(c_i z^{\text{pos}}_i) + \sum_{j=1}^N \text{exp}(c_i z^{(j)}_i) }
\end{equation}

\subsubsection{Model details and hyper-parameter tuning}
\label{subsubsec:cha6-hyper-tune}

As in the previous section, we also pre-train our BiCPC (masked contrastive predictive coding) models on SI284.
We do not use the DNN encoder (with parameter $\phi$, used for learning latent representations ($z$)) when training the downstream CTC recognizer.
We directly use the pre-trained context encoder ($4-$layer bidirectional LSTM, with parameter $\theta$) to initialize the CTC system.
We compare BiCPC with BERT and other baselines in Figure ~\ref{fig:performance2}. \par

All recognizers are $4-$layer BiLSTM phone-based CTC recognizers trained on SI84. 
These recognizers are trained up to $50$ epochs, and the epoch with the best error rate on the validation data set (dev93) are used for test set (eval92) evaluation.
We use batch size $4$ while the dropout is chosen from among $\{0.0, 0.1, 0.2, 0.3 , 0.4 \}$.
The optimizer we used is Adam, with initial learning rate selected from $\{0.001, 0.0005, 0.0001 \}$. \par

When doing pre-training, 
the most important hyper-parameter is the mask.
The number of masks across channels is selected from among $\{1,2,3\}$,
with the maximum size of each mask being either $8$, $16$, or $24$.
The number of masks in the time domain is either $1$ or $2$,
with the maximum size of each mask selected from among $\{ 8, 16, 24\}$.
We use batch size $16$ for pre-training in this section.
The optimizer we used for pre-training is also ADAM, with the initial learning rate being $0.0005$. 

\subsubsection{Key observations from experiments}
\label{subsubsec:cha6-key-observation}
\begin{itemize}
\item[1] From Figure ~\ref{fig:performance2}, we can see though BiCPC (masked contrastive predictive coding) improves over the baseline system, it does not outperform BERT (masked reconstruction).

\item[2] As shown in Figure ~\ref{fig:performance2}, ``BiCPC-Half" also clearly improves over ``BiCPC". 
As BERT-Half and BiCPC-Half all improve over their counterparts, this consistency suggests that predicting the central region could be a technique that generally fits other losses and encoders. 

%\item[3] We also do an ablation study to investigate the number of layers used in $\text{DNN}_{\phi}$. 
%Our observation is that a shallower $\text{DNN}_{\phi}$ yields the best downstream task performance.
%More specifically, when $\phi$ (the encoder for learning $z$) consists of only one linear layer, the resulting contextual encoder (with parameter $\theta$) works best for the downstream speech recognition task.
%This observation suggests the possibility that using a less flexible feature encoder (e.g., the DNN with parameter $\phi$ in Equation ~\eqref{eqn:latent}) might benefit masked contrastive predictive coding. 
%This observation motivates us to explore a variant of BiCPC in which $\phi$ is part of the contextual encoder, which is discussed in Appendix Section ~\ref{sec:cha6-better-architecture}.
\item[3] We also try to augment the training dataset by repeating each utterance $3$ times, such that each utterance is trained with a larger set of masks. 
This can somewhat improves masked contrastive predictive coding, but the benefit does not carry over to the case of masked reconstruction.
\end{itemize}

%%%%%%%%%%%%%%%%%%%%%%%%%%%%%%%%%%%%%%%%%%%%%%%%%%%%%%%%%%%%%%%%%
\subsection{Multi-view Masked Reconstruction}
\label{subsec:consistency}

Taking inspiration from multi-view representation learning, 
specifically its use of the second view to learn representations better suited for downstream tasks,
we investigate the combination of multi-view representation learning, masked reconstruction, and contrastive predictive coding.
We first use a stochastic process to select channels across all time steps and a few time-domain segments given one sequence.
We then mask the selected channels and time-domain segments.
A sequence with such different kinds of data augmentation operators applied naturally forms correlated views.
In this way, we can force the representations of different masked inputs of the same object to be similar to each other. 
The resulting consensus representation would be more robust to the different types of noise introduced by different operators. \par

Without loss of generality,
we only consider using two correlated views of the same sequence in this section.
For each of the noisy versions of the same sequence,
we minimize masked reconstruction loss (Equation ~\eqref{eqn:masked_recon}) to reconstruct the missing part respectively.
In addition to the two masked reconstruction losses, we also use a consistency loss to minimize the difference between the representations of the two noisy versions.
We explore two different choices of consistency losses:

\begin{itemize}
\item[1] \textbf{Consistency constraint on contextual representations}:
The contextual representations from two views are forced to be similar to each other by minimizing \textbf{Mean absolute error (MAE)}, \textbf{Mean squared error (MSE)} or \textbf{Contrastive loss}.

\item[2] \textbf{Cross-view noisy frame prediction}:
The contextual representation (e.g., $c_t$) of one view is used to predict the latent representation $z_t$ of the other view. We use InfoNCE loss to achieve this.
\end{itemize}

\subsubsection{Consistency constraint on contextual representations}
\label{subsubsec:contextual-representation}

\begin{figure*}[htbp]
  \centering
  \includegraphics[scale=0.46]{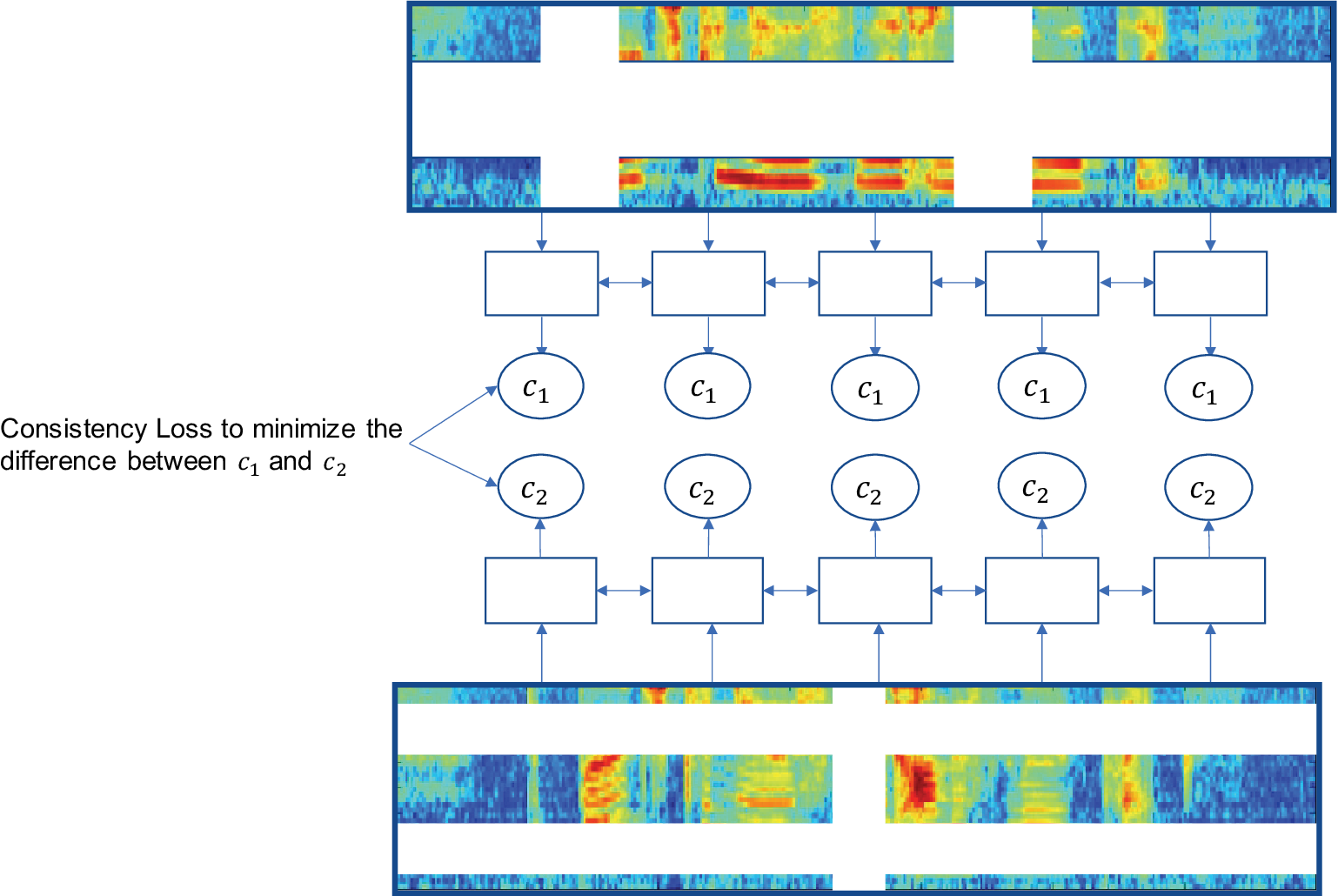}
  \caption{Illustration of encouraging consistency among contextual representations of different masked inputs.
The encoder is stacked BiLSTM, but we only show one layer in this figure for simplicity.}
  \label{fig:consistency}
\end{figure*}

As shown in Figure ~\ref{fig:consistency}, 
given a sequence $X$ and two masks $M_1$ and $M_2$, 
we first obtain the contextual representations of $X \odot M_1$ and $X \odot M_2$ via
\begin{eqnarray}
C_1 &=& \text{BiLSTM}_{\theta}(X \odot M_1) \nonumber \\
C_2 &=& \text{BiLSTM}_{\theta}(X \odot M_2)
\label{eqn:pre-training-multiview-contexual}
\end{eqnarray}

Decoder $g$ consists of stacked ReLU layers ($1024$ units) followed by a linear layer. 
We use a reconstruction loss for each of the two masked inputs, as shown in the below equations: % ~\eqref{eqn:pre-training-multiview-reconstruction}:
\begin{eqnarray}
L_{\text{recon1}} &=& \norm{(1-M_1) \odot [ X - g(C_1) ] }_{\text{Fro}}^{2} \nonumber \\
L_{\text{recon2}} &=& \norm{(1-M_2) \odot [ X - g(C_2) ] }_{\text{Fro}}^{2}
\label{eqn:pre-training-multiview-reconstruction}
\end{eqnarray}

The final loss is the linear combination of the reconstruction losses (i.e. Equation ~\eqref{eqn:pre-training-multiview-reconstruction}) and the consistency loss:
\begin{eqnarray}
\alpha (L_{\text{recon1}} + L_{\text{recon2}}) + (1-\alpha) L_{\text{consistency}}(C_1,C_2)
\label{eqn:consistency}
\end{eqnarray}

In this section, we investigate MAE and contrastive loss to minimize the difference between the contextual representations $C_1$ and $C_2$. 
The contrastive loss we use is:

\begin{eqnarray}
L_{\text{consistency}}(C_1,C_2)  &=& \sum_{i=1}^T  -\log{\frac{\text{exp}(C_{1,i}C_{2,i}) }{\text{exp}(C_{1,i}C_{2,i}) + \sum_{j=1}^N \text{exp}(C_{1,i}C_{2,j}) } }  \nonumber \\
&+& \sum_{i=1}^T  - \log{\frac{\text{exp}(C_{2,i}C_{1,i}) }{\text{exp}(C_{2,i}C_{1,i}) + \sum_{j=1}^N \text{exp}(C_{2,i}C_{1,j}) }} 
\label{eqn:cha6-contrastive-loss}
\end{eqnarray}
where $C_{1,i}$ and $C_{2,i}$ are the $i-$th time step of the BiLSTM output of the two views, respectively, and are positive samples of each other. 
To perform contrastive learning, we randomly sample $N$ negative samples from time steps other than $i$ from the other view. \par

For simplicity, we use \textbf{Multi-View-MAE} to denote Multi-view masked reconstruction with MAE as the consistency loss and \textbf{Mulit-View-Contrast} to denote the model using the contrastive loss (Equation ~\ref{eqn:cha6-contrastive-loss}) as consistency loss.

%Please note, given $\alpha=1.0$, the Equation ~\eqref{eqn:consistency} still has some minor difference comparing to equation ~\eqref{eqn:masked_recon}. 
%That is, when using the same size mini-batch, the number of reconstructions in Equation ~\eqref{eqn:consistency} is two times of the number of Equation ~\eqref{eqn:masked_recon},
%and thus the inter-batch variance would also be smaller for Equation ~\eqref{eqn:consistency}.

\subsubsection{Cross-view noisy frame prediction}
\label{subsubsec:cross-view-noisy-prediction}

\begin{figure*}
 \centering
 \includegraphics[scale=0.36]{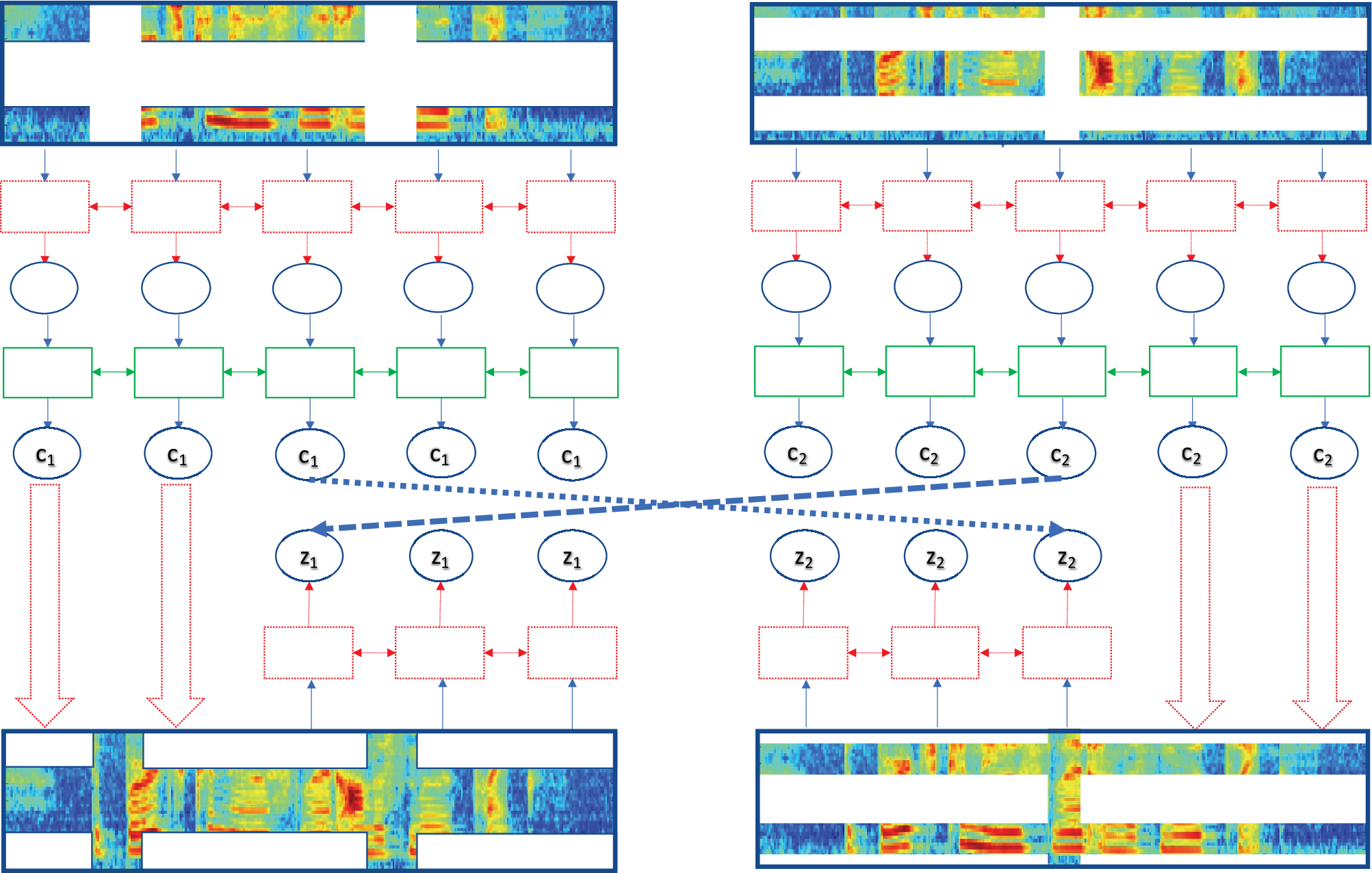}
 \caption{Illustration of the use of infoNCE loss for cross-view prediction.
To distinguish the latent representations ($z$) and contextual representations ($c$) of the two views, we use indices $1$ and $2$ respectively.
We ignore the time step index in the figure.
As shown, both views share the latent (red) and contextual (green) encoders.
Contextual representation $C_1$ is used to predict $Z_2$ (e.g., the $t^{th}$ context representation of the first view is used to predict the $t^{th}$ latent representation of the second view), while $C_2$ is used to predict $Z_1$.}
 \label{fig:consistency2}
\end{figure*}

Besides directly applying a consistency loss to encourage $C_1$ and $C_2$ to be similar to each other,
we also investigate cross-view predictive coding.
Figure ~\ref{fig:consistency2} illustrates the idea of cross-view contrastive prediction.
The encoder (shown as red in Figure ~\ref{fig:consistency2}) for learning the features $Z$ is a one-layer BiLSTM:
\begin{eqnarray}
Z^{1} &=& \text{BiLSTM}_{\phi}(X \odot (1-M_1)) \nonumber \\
Z^{2} &=& \text{BiLSTM}_{\phi}(X \odot (1-M_2))
\end{eqnarray}

\weiran{why shuffling for N times? cannot you just randomly sample?}
Assume $Z^{1}$ is of length $T$. Randomly shuffling $Z^{1}_{1:T}$ N times, 
we obtain sequences $S^{1,(1:N)}$.
Each $S^{1,(i)}_t$ for $1\leq i \leq N$ is considered to be one of the negative samples of $Z^{1}_t$.
Similarly, we also shuffle $Z^{2}$ to obtain $N$ shuffled sequences $S^{2,(1:N)}$ as the source of negative samples for $Z^{2}$.

As shown in Figure ~\ref{fig:consistency2}, 
a BiLSTM contextual encoder further takes $X \odot M_1$ and $X \odot M_2$ as input to generate the contextual representations of the two views, respectively:
%$Z^{1}$ and $Z^{2}$ are further transformed to contextual representations via contextual encoder (stacked BiLSTM, indicated as green color in the figure):
\begin{eqnarray}
C_1 &=& \text{BiLSTM}_{\theta}(X \odot M_1) \nonumber \\
C_2 &=& \text{BiLSTM}_{\theta}(X \odot M_2)
\end{eqnarray}

Given $C_1$ and $C_2$, we then have reconstruction losses for both masked inputs. 
The decoder $g$ is again a feedforward neural network with the last layer being linear, same as that described in subsection ~\ref{subsubsec:contextual-representation}:
\begin{eqnarray}
L_{\text{recon1}} &=& \norm{(1-M_1) \odot [ X - g(C_1) ] }_{\text{Fro}}^{2} \nonumber \\
L_{\text{recon2}} &=& \norm{(1-M_2) \odot [ X - g(C_2) ]}_{\text{Fro}}^{2}
\end{eqnarray}

To enforce $C$ to be predictive of the latent representations $z$ in the other view, we use the cross-view contrastive loss: 
\begin{equation}
L_{\text{Contrastive}} = \sum_{i=1}^{T} \Big\{ l(c_{1,i},z^{2}_i,s^{2,(1:N)}_i) + l(c_{2,i},z^{1}_i,s^{1,(1:N)}_i) \Big\}
\label{eqn:cross-view-predictive-coding}
\end{equation}
where
\begin{equation}
l(c,z,s^{(1:N)}) = -\log \frac{\text{exp}(c z)}{\text{exp}(c z) + \sum_{j=1}^N \text{exp}(c s^{(j)})}
\end{equation}

The complete loss is then
\begin{equation}
\alpha (L_{\text{recon1}} + L_{\text{recon2}}) + (1-\alpha) L_{\text{Contrastive}}
\end{equation}

For simplicity, we denote this multi-view masked reconstruction model as ``\textbf{Cross-View-BERT}".

\subsubsection{Experimental Setups for Multi-view Masked Reconstruction}
\label{subsubsec:exp_multi_masked}

We compare the proposed multi-view masked reconstruction methods with other pre-training methods in three scenarios: 
1) Pre-training is done on SI284, and recognizers are fine-tuned on SI84 ($SI284 \rightarrow SI84 $ for short),
2) Pre-training is done on LibriSpeech, and recognizers are fine-tuned on SI284 ($Libri \rightarrow SI284$ for short) and 3) Pre-training is done on LibriSpeech and recognizers are trained using SI84 ($Libri \rightarrow SI84$ for short).
As in Section ~\ref{sec:pre-training-data}, we train phone-based and character-based CTC recognizers.
We train all the recognizers with a $4-$layer BiLSTM encoder either randomly initialized (for baseline) or initialized with the weights of a pre-trained encoder.
For more details regarding the setup, please refer to Section ~\ref{sec:pre-training-data}. \par

\begin{figure*}[htbp]
 \centering
 \includegraphics[scale=0.42]{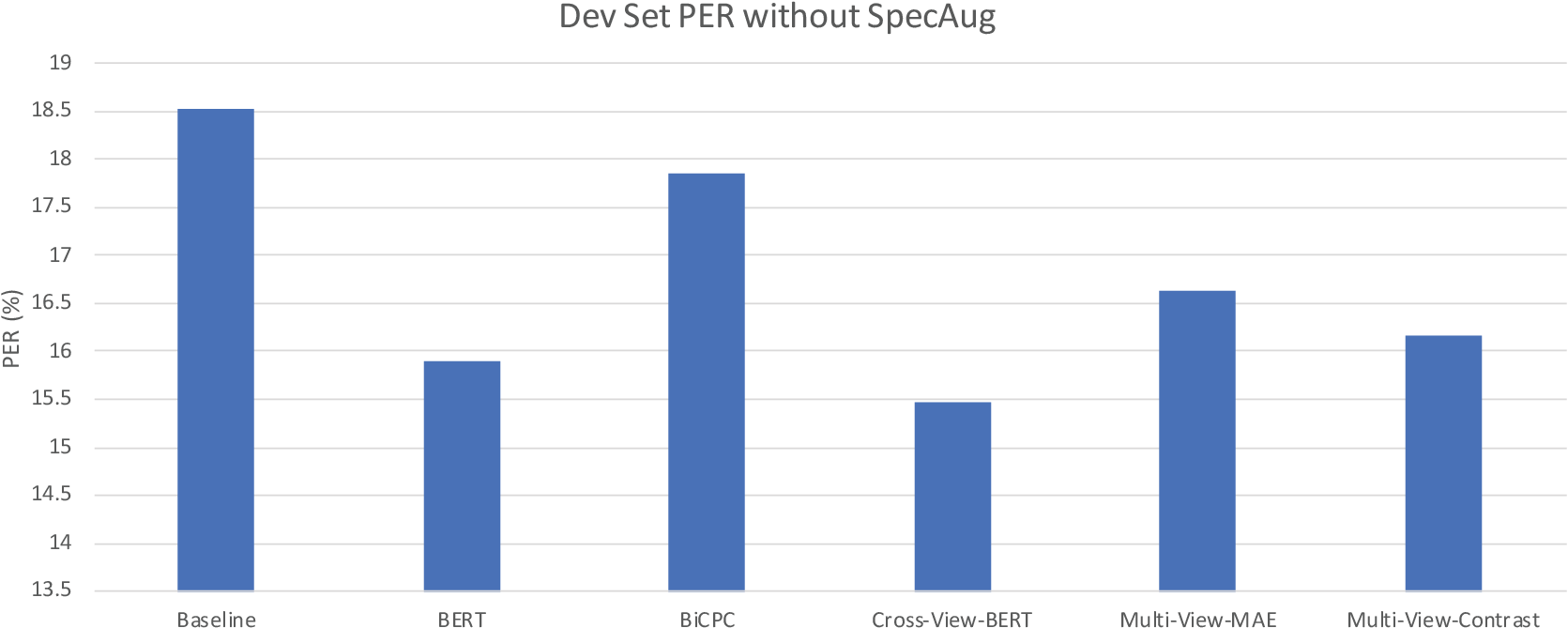}
 \caption{Comparison of a few phone-based CTC recognizers on dev93.
 All recognizers are fine-tuned on SI84.
 All but the baseline recognizer are initialized with the weights of encoders pre-trained on SI284 using BERT, BiCPC or Multi-view masked reconstruction.}
 %Please note, ``Two-View Masked Reconstruction+MAE" and ``Two-View Masked Reconstruction + Contrast" refer to the multi-view pre-training using MAE and Contrastive loss as consistency loss, respectively.
 %``Two-View Masked Reconstruction + Predicting 2nd-view content" refers to the multi-view pre-training approach described in Section ~\ref{subsubsec:cross-view-noisy-prediction}.}
 \label{fig:performance3}
\end{figure*}

We first compare the multi-view masked reconstruction methods with single-view methods like BERT (masked reconstruction) and BiCPC (masked contrastive predictive coding) in scenario one ($SI284 \rightarrow SI84 $).
We use batch size $16$ for all models during pre-training, and batch size $4$ for all CTC recognizer training.
For all of the multi-view reconstruction methods, we select $\alpha$ from $\{ 0.5, 0.3, 0.1, 0.01\}$.
The setup used for recognizer training is described in detail in Section ~\ref{subsubsec:cha6-hyper-tune}. \par

Figure ~\ref{fig:performance3} shows that CTC recognizers with encoders pre-trained using multi-view masked reconstruction methods outperform the baseline and masked contrastive predictive coding approaches.
One of the multi-view learning approaches (referred to as ``Cross-View-BERT") outperforms all of the other approaches,
which suggests that multi-view learning produces better suited representations for downstream tasks. \par

\begin{figure*}[htbp]
 \centering
 \includegraphics[scale=0.43]{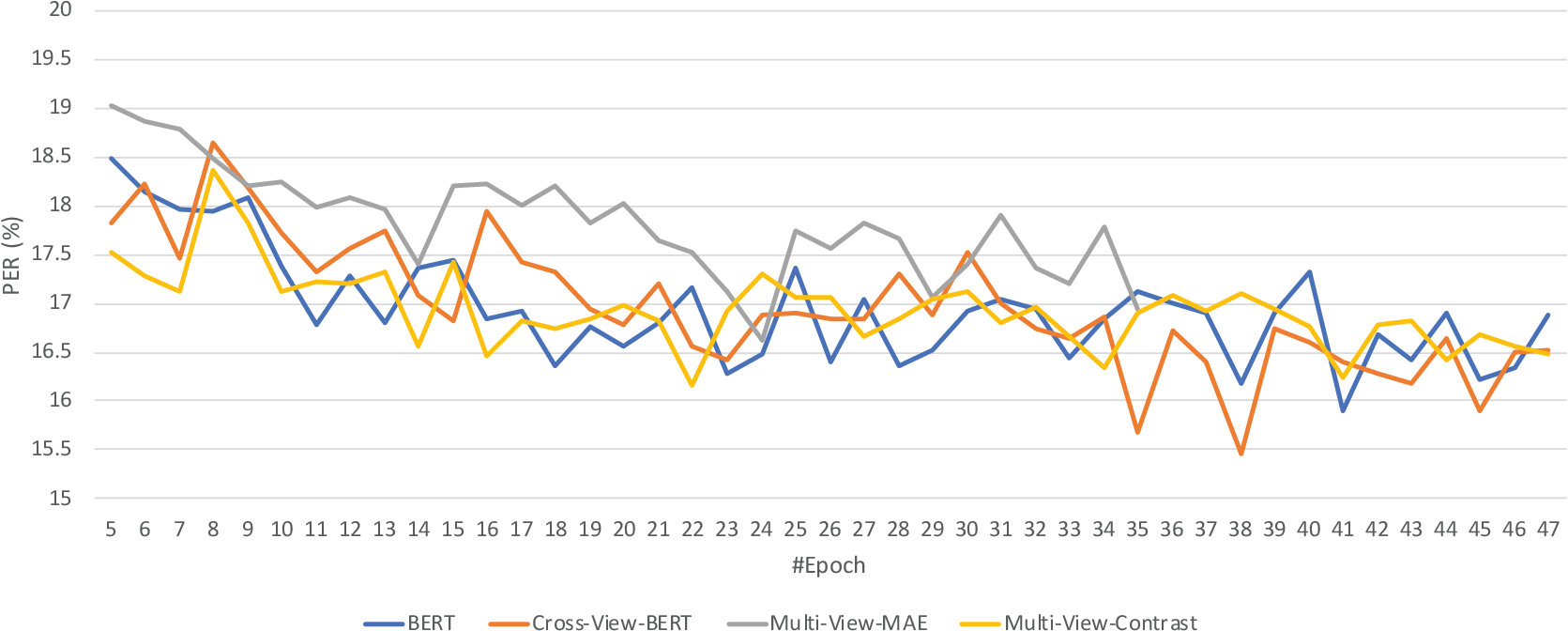}
 \caption{Per-epoch PER on dev 93. 
We compare ``BERT", ``Cross-View-BERT", ``Multi-View-MAE" and ``Multi-View-Contrast" in the scenario $SI284 \rightarrow SI84 $. }
 \label{fig:multiview-per}
\end{figure*}

\begin{figure*}[htbp]
 \centering
 \includegraphics[scale=0.43]{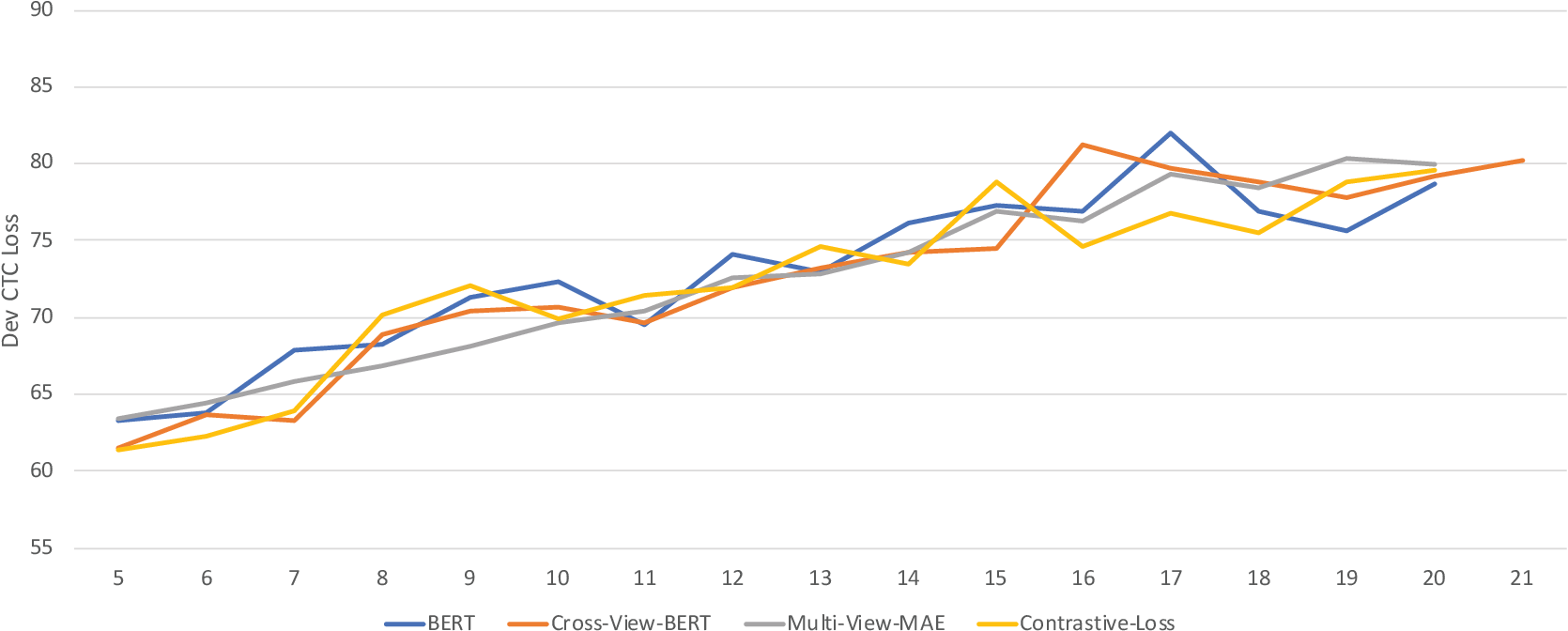}
 \caption{Per-epoch CTC loss on dev 93. 
  We compare ``BERT", ``Cross-View-BERT", ``Multi-View-MAE" and ``Multi-View-Contrast" in the scenario $SI284 \rightarrow SI84 $.
 Note that CTC phone recognizers trained on SI84 usually achieve minimum dev loss around epoch $5$; the dev loss then increases and fluctuates.
 This aligns with the observation shown in ~\citep{wang2020unsupervised}.}
 \label{fig:multiview-loss}
\end{figure*}

Surprisingly, two multi-view masked reconstruction approaches (referred to as Multi-View-MAE and Multi-View-Contrast) perform worse than the masked reconstruction approach.
The two methods minimize the loss shown in Equation ~\eqref{eqn:consistency}, with ``MAE" and ``Contrastive loss" as the consistency loss, respectively.
Note that although we use the same size of mini-batch to train both single-view methods (BERT and BiCPC) and multi-view masked reconstruction methods,
the multi-view methods address doubled masked reconstruction tasks within one mini-batch.
We suspect that we can improve the performance of Multi-View-MAE and Multi-View-Contrast by carefully tuning hyper-parameters.
For example, if we had used $\alpha=1.0$ and batch size half of that used for single-view methods, the two multi-view approaches would almost become a single-view approach,
thus should have performed similarly to single-view masked reconstruction. \par
%When using $\alpha=0$, 
%Equation ~\eqref{eqn:consistency} becomes the loss of masked reconstruction.
%However, compared with vanilla BERT,
%Equation ~\eqref{eqn:consistency} would sum over more reconstruction loss terms and lead to smaller per-batch variance. 

We also check the per-epoch phone error rate and per-epoch CTC loss of the CTC recognizers presented in Figure ~\ref{fig:performance3},
as shown in Figure ~\ref{fig:multiview-per} and Figure ~\ref{fig:multiview-loss} respectively.
As we can see, 
both ``Multi-View-Contrast" and ``Cross-View-BERT" achieve lower dev set PER and lower dev set loss than BERT (masked reconstruction) in the beginning.
This observation suggests that pre-training provides meaningful initialization for training downstream CTC recognizers.
However, as training continues,
BERT finally catches up to and outperforms Multi-View-Contrast. 
Cross-View-BERT achieves a lower validation set PER as training goes on. \par

\begin{figure*}[htbp]
 \centering
 \includegraphics[scale=0.43]{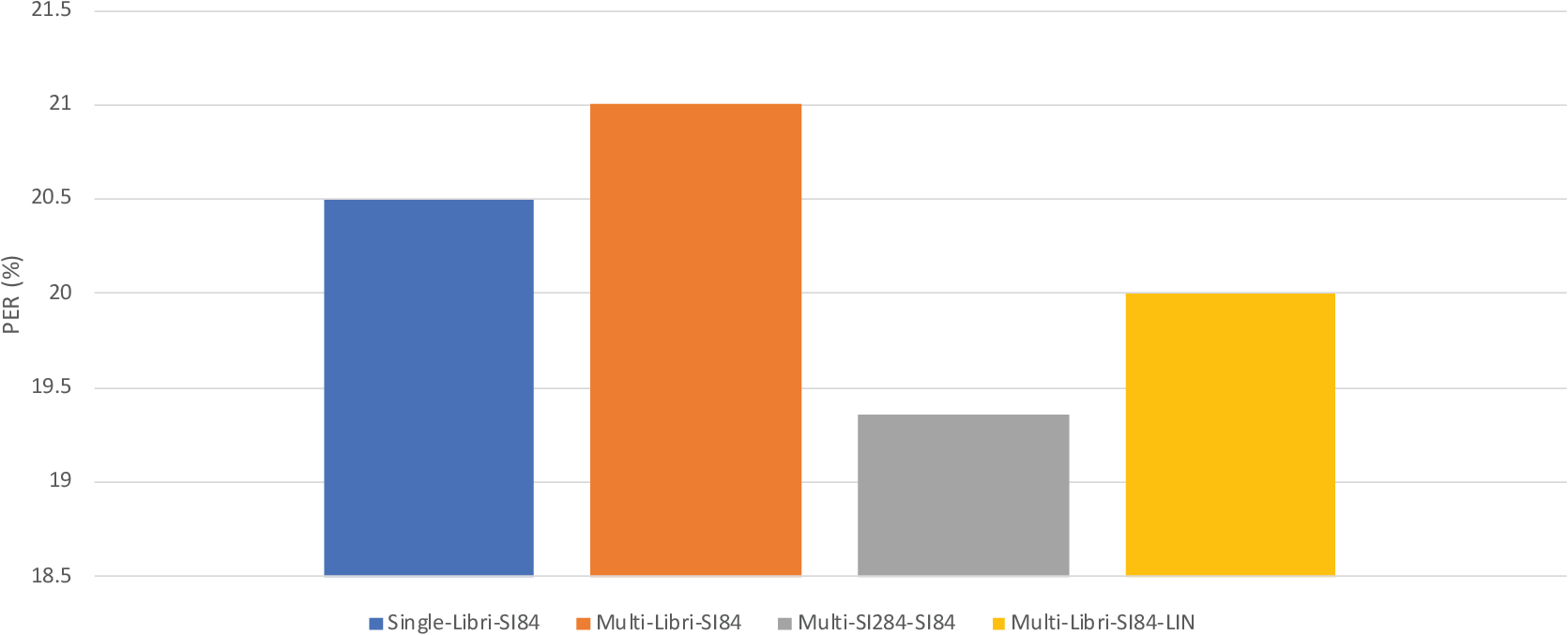}
 \caption{Domain effect.
 All recognizers are phone-based CTC recognizers and are all fine-tuned using SI84.
 ``Single-Libri-SI84", ``Multi-Libri-SI84" and ``Multi-Libri-SI84-LIN" are pre-trained using LibriSpeech.
 ``Multi-SI284-SI84" is pre-trained using SI284.
 For ``Multi-Libri-SI84", ``Multi-SI284-SI84", and ``Multi-Libri-SI84-LIN",
 the encoders are pre-trained using Cross-View-BERT. 
 ``Multi-Libri-SI84-LIN" has one linear layer for domain adaptation.
 All evaluation is done on dev93.}
 \label{fig:performance-4}
\end{figure*}

So far, we have shown that multi-view masked reconstruction methods can outperform single-view masked reconstruction methods when pre-training is done on SI284 and fine-tuning is done on SI84.
There is little domain difference between SI284 and SI84.
We now investigate the scenario that pre-training and fine-tuning are done in two data sets with bigger domain differences.
We first pre-train encoders on LibriSpeech; afterwards, we fine-tune the pre-trained encoders on SI284 or SI84 respectively.
%can help speech recognizers.fine-tuned on SI284 and SI84, respectively.
For pre-training on LibriSpeech, we use batch size $48$ for BERT, and use batch size $24$ for multi-view masked reconstruction methods.
\textbf{For the remaining part of this section, we focus on a single multi-view method ``Cross-View-BERT". 
All discussion of the ``multi-view masked reconstruction" approach will refer specifically to the Cross-View-BERT model.}
%Whenever we say a ``multi-view masked reconstruction" approach, we mean a Cross-View-BERT model}.
As the length of utterances in LibriSpeech are typically longer than the sentences of SI284,
we use more and wider masks in the time domain.
Specifically,
we either use three masks of width less than or equal to 24,
or use two masks of width less than or equal to 16. \par

%We first examine how encoders pre-trained on LibriSpeech can reduce Character error rate (CER) of speech recognizers trained using SI84.
We compare the encoders pre-trained on LibriSpeech and SI284.
We examine the phone error rate (PER) of speech recognizers trained on SI84, when starting from models pre-trained on different datasets.
%how model pre-trained on different datasets can help reduce Character error rate (CER) of speech recognizers trained on SI84.
As shown in Figure ~\ref{fig:performance-4}, 
pre-training on SI284 is more helpful than pre-training on LibriSpeech for reducing the PER of recognizers fine-tuned on SI84.
We hypothesize that this discrepancy in performance arises due to the domain difference.
To address the domain difference between WSJ and LibriSpeech,
we use a linear input network (LIN, ~\citep{Neto1995SpeakeradaptationFH, Yao2012AdaptationOC}) for domain adaptation, which inserts an additional identity mapping-initialized linear transformation layer between the input and the BiLSTM encoder initialized by a pre-trained model.
This simple approach reduces the domain gap between WSJ and LibriSpeech,
as we can see that ``Multi-Libri-SI84-LIN" performs better than ``Multi-Libri-SI84", though still slightly worse than ``Multi-SI284-SI84". \par

We now present the experimental results for scenario $LibriSpeech \rightarrow SI284$. When training a phone-based or character-based CTC recognizer using SI284, we fix the batch size to be $16$. \par

\begin{table} [htbp]
\centering
\begin{tabular}{| l | r | r | r | r | r | r | r | }
 \hline
 & Baseline & +LIN & BERT & +LIN & Multi-view & +LIN \\
 \hline \hline
 1. Best & 6.72 & 5.79 & 6.58 & 5.76 & 6.58 & 5.88 \\
 2. 2nd & 6.75 & 5.86 & 6.65 & 5.77 & 6.62 & 5.91 \\
 3. 3rd & 6.85 & 6.20 & 6.67 & 5.97 & 6.64 & 5.95 \\
 4. Best CTC Loss & 29.90 & 26.41 & 29.43 & 26.73 & 29.10 & 25.98 \\
 \hline
\end{tabular}
\caption{Character error rate (CER) and best CTC loss on dev93. 
We compare the baseline CTC recognizer, a CTC recognizer pre-trained using masked reconstruction (``BERT", ``BERT+LIN"), and CTC recognizers pre-trained using multi-view masked reconstruction (``Multi-view", ``Multi-view+LIN"). 
For multi-view masked reconstruction, we use the variant described in paragraph ~\ref{subsubsec:cross-view-noisy-prediction}. 
We present the three models with the lowest dev set CER in each column of the table as well as the model with the lowest CTC loss.}
%Please note, for each trained CTC recognizer, the epoch of the best CTC loss typically does not correspond to the epoch with the best character error rate.}
\label{tab:char-recognizer}
\end{table}

Table ~\ref{tab:char-recognizer} shows that character-based CTC recognizers do achieve lower CTC loss on the validation set when initialized with models pre-trained using multi-view masked reconstruction methods.
The multi-view masked reconstruction method also outperforms both the single-view masked reconstruction methods and the baseline recognizer trained with random initialization, in terms of CTC loss.
To overcome the domain difference between WSJ and LibriSpeech depicted in Figure ~\ref{fig:performance-4},
we add a LIN layer to train stronger recognizers (e.g. ``BERT+LIN" and ``Multi-view+LIN").
``BERT+LIN" and ``Multi-view+LIN" clearly and consistently outperform their counterparts, respectively.
For a fair comparison,
we also add LIN layer to train another baseline,
referred to as ``Baseline+LIN" in the table.
Surprisingly, ``Baseline+LIN" is comparable to both ``Multi-view+LIN" and ``BERT+LIN".

\begin{figure*}[htbp]
 \centering
 \includegraphics[scale=0.43]{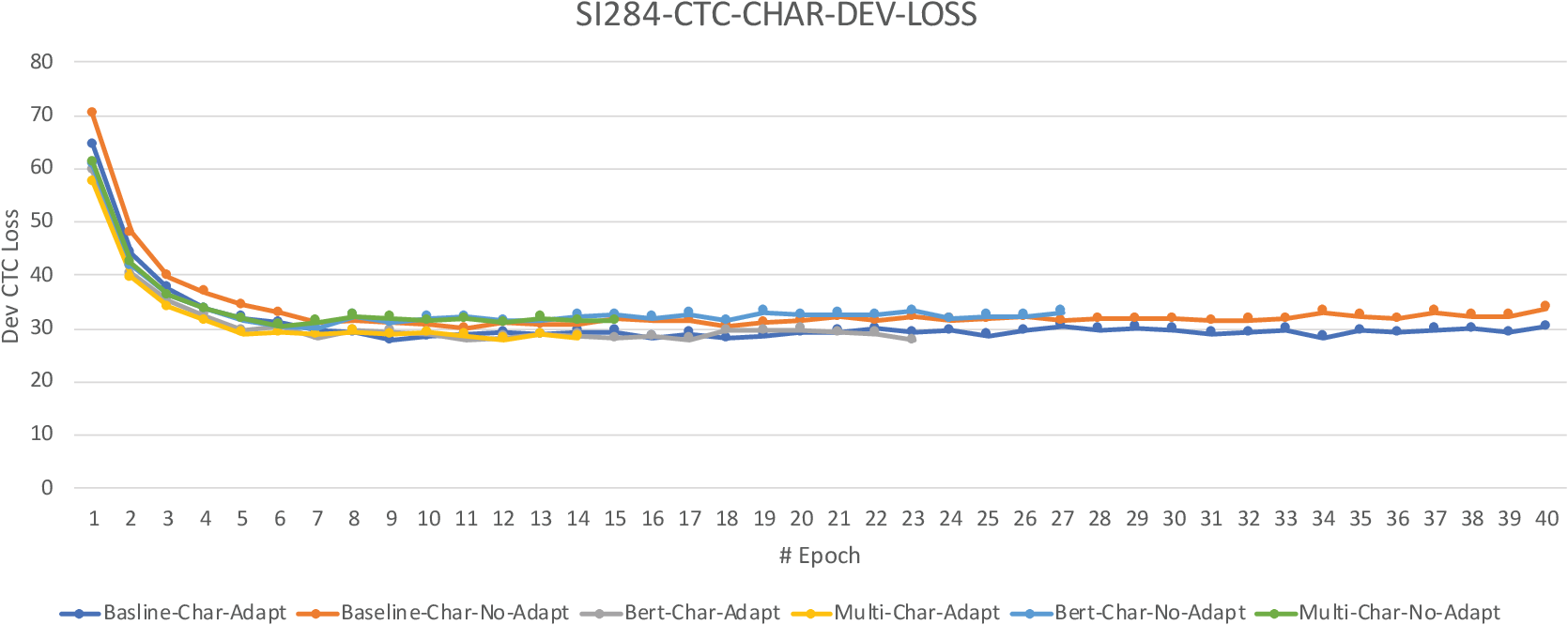}
 \caption{Per-epoch CTC loss on dev 93 for recognizers trained on SI284.
We compare Baseline recognizers with/without LIN layer (``Baseline-Char-Adapt", ``Baseline-Char-No-Adapt"), recognizers with encoder pre-trained on LibriSpeech with masked reconstruction with/without LIN layer (``BERT-Char-Adapt", ``BERT-Char-No-Adapt"), and recognizers with encoder pre-trained on LibriSpeech with multi-view masked reconstruction with/without LIN layer (``Multi-Char-Adapt", ``Multi-Char-No-Adapt").
 All recognizers are char-based CTC recognizers.}
 \label{fig:multiview-loss-2}
\end{figure*}

We further examine the per-epoch CTC loss of CTC recognizers trained with different initializations, as shown in Figure ~\ref{fig:multiview-loss-2}.
We find that pre-trained CTC recognizers initially tend to exhibit lower CTC loss,
which suggests that pre-trained model weights contain useful information.
However, as the recognizer training continues,
the baseline recognizer with one more LIN layer gradually catches up.
Regardless, Table ~\ref{tab:char-recognizer} indicates that the CTC
recognizer initialized by a model pre-trained using multi-view masked reconstruction still achieves lower CTC loss than recognizer initialized by a single-view pre-trained model on the dev set. \par

\begin{table} [htbp]
\centering
\begin{tabular}{| l | r | r | r | r | r | r | r | }
 \hline
 & Baseline & +LIN & BERT & +LIN & Multi-view & +LIN \\
 \hline \hline
 1. Best & 8.73 & 7.32 & 8.19 & 7.26 & 7.91 & 7.16 \\
 2. 2nd & 8.95 & 7.51 & 8.34 & 7.35 & 7.92 & 7.26 \\
 3. 3rd & 9.11 & 7.82 & 8.40 & 7.38 & 7.96 & 7.34 \\
 4. Best CTC Loss & 36.08 & 28.62 & 30.30 & 27.92 & 29.44 & 27.33 \\
 \hline
\end{tabular}
\caption{Phone error rate (PER) and best CTC loss on dev93. 
We compare the baseline CTC recognizer, CTC recognizers pre-trained using masked reconstruction (``BERT", ``BERT+LIN"), and CTC recognizers pre-trained using multi-view masked reconstruction (``Multi-view", ``Multi-view+LIN"). 
Besides having used a phone-based CTC recognizer, all other details are the same as presented in Table ~\ref{tab:char-recognizer}.}
\label{tab:phone-recognizer}
\end{table}

We also compare the phone-based CTC recognizers as shown in Table ~\ref{tab:phone-recognizer}.
We find a clear advantage to using
recognizers initialized with pre-trained weights.
For example, the best validation losses
are $36.08$, $30.30$ and $29.44$ for baseline CTC, CTC recognizer pre-trained with masked reconstruction, and CTC recognizer pre-trained with multi-view masked reconstruction method, respectively.
Even when a LIN layer is used, 
recognizers initialized with pre-trained weights still have lower CTC loss and a slightly lower phone error rate.
Multi-view pre-trained models also show a clear advantage over the single-view pre-trained models when used for fine-tuning phone-based CTC recognizers.

%%%%%%%%%%%%%%%%%%%%%%%%%%%%%%%%%%%%%%%%%%%%%%%%%%%%%%%%%%%%%%%%%
\section{Summary}
\label{sec:pre-training-summary}

We have studied semi-supervised representation learning, unsupervised acoustic feature learning and self-supervised learning in this chapter. We offer the following conclusions and observations:

\begin{itemize}
\item[1] \textbf{Different layers capture diverse information}: We find that it is difficult to learn representations suitable for speech recognition tasks using RecRep-$\allowbreak$Pyramid-MT, a variational sequential semi-supervised learning framework.
We attribute this difficulty to the difference in preferred representations between the CTC loss and per-frame reconstruction loss.
This observation aligns our findings in Chapter ~\ref{cha:recurrent} where RecRep-Pyramid-MT does not work well without extra treatments. 
%RecRep-Pyramid-MT becomes more powerful when CTC recognizer and the RecRep encoder only share a single BiLSTM layer (low-level information).

\item[2] \textbf{Difficult tasks help representation learning}: 
We also find (non-sequential) VAE and CPC dramatically outperform RecRep-Pyramid in learning acoustic representations in unsupervised learning scenarios.
We hypothesize that reconstructing the input sequence on top of stacked BiLSTM is a too simple task, thus without learning context-aware representations the model can easily reconstruct the sequence. 
%two reasons for this observation:
%1) the reconstruction targets may mislead the posterior learning,
%and 2) the learning strategy (reconstructing frames that have observed) may be less context-aware than predicting unseen content/frames. 
Adopting techniques used in Chapter ~\ref{cha:recurrent} (e.g., self-prior-updating and constructing complex per-time-step unit)
may improve the learning performance of RecRep(-Pyramid) in both semi-supervised and unsupervised learning scenarios. 
However, as we have realized that using a more difficult task on top of bidirectional encoder is crucial for learning acoustic representations for downstream ASR tasks,
we switch to exploring masked reconstruction, contrastive predictive coding, and autoregressive prediction in this chapter, and leave further exploration of variational sequential semi-supervised and unsupervised learning as future work.
When we are exploring masked reconstruction, we also find that another example showing that a more difficult task could benefit representation learning -- We find that if we only reconstruct the central part of the masked region (i.e., focusing on the more difficult region), the learned representation could achieve lower error rate compared with vanilla masked reconstruction in terms of speech recognition.

\item[3] \textbf{Robust representations benefit downstream tasks}: 
Our experimental study shows that making representations invariant to different domains and robust to distortions is crucial.
We find that, adding a simple linear layer as a domain adaptor can help masked reconstruction learn domain-invariant representations and thus significantly boost the performance on downstream tasks.
%Contrastive Predictive Coding (CPC) works well in unsupervised acoustic feature learning (i.e., the CPC-extracted features yield stronger ASR models than surface features). 
%The learned representations are helpful when no more than $50\%$ of the training samples of SI84 are used in downstream speech recognizer fine-tuning. \weiran{If you have not used augmentation/masking for CPC, it is better to merge the discussion on CPC with previous point. And make masking/predicting unseen frames a separate point to emphasize.}
%We also observe masked reconstruction and its variants (e.g. only reconstructing the central region of the missing region) learn representations that are suitable for speech recognition. However, the combination of CPC and masked reconstruction (which we refer to as masked contrastive predictive coding, BiCPC) does not outperform masked reconstruction.
We also explore multi-view masked reconstruction techniques in this chapter, and observe consistent improvement across three scenarios, including pre-training on SI284 and training recognizers on SI84 (SI284 $\rightarrow$ SI84), LibriSpeech $\rightarrow$ SI84 and LibriSpeech $\rightarrow$ SI284.

\end{itemize}

%These observations demonstrate the impact of learning representations in a self-supervised manner, with little assistance from using label information.

%----------------------------------------------------------------------------------------
%      Future works
%----------------------------------------------------------------------------------------
\chapter{Contributions and Future Work}
\label{cha:future}

In this chapter, we summarize the contributions and main implications of this work, and discuss possible future work. 
The work described in this thesis is unique \footnote{Please note, as most of the work described in this thesis is done before March of 2019, the new ideas highlighted below may not be novel compared to literature after 2019.}in the following ways:

\begin{itemize}
\item[1] \textbf{A broad study of representation learning.} Unlike most other works that focus on one or a few learning problems, this thesis studies supervised learning with an auxiliary loss (Chapter ~\ref{cha:recurrent}), unsupervised learning (Chapter ~\ref{cha:feedforward} and ~\ref{cha:semi-and-pre-training}), semi-supervised learning (Chapter ~\ref{cha:semi-and-pre-training}) and multi-view learning (Chapter ~\ref{cha:multiview}). Besides different learning problems, this thesis also explores multiple approaches for representation learning. Below are some of the main outcomes:
\begin{itemize}
\item[a.] We verified that autoencoding approaches can learn representations that are beneficial for downstream phoneme recognition in unsupervised and multi-view learning scenarios. \
\item[b.] We also found that, under the umbrella of autoencoding approaches, variational models outperformed their non-variational counterparts. We hypothesized that this is because both the reparameterization trick and the KL divergence term have regularization effects.
%We also compare autoencoding and a simplified bidirectional autoregressive prediction approach (i.e., predicting both one forward and backward frames) in Chapter ~\ref{cha:semi-and-pre-training} in supervised learning with an auxiliary loss, and found that the two approaches achieve similar performance. 
\item[c.] We further study masked reconstruction and contrastive prediction approaches in Chapter ~\ref{cha:semi-and-pre-training}. In our experiments, we found that masked reconstruction outperforms both autoencoding and uni-directional contrastive prediction approaches. 
\item[d.] We also combined the masked reconstruction and contrastive prediction approach, wherein we encourage the consistency between the representations of two corrupted inputs (with different masks applied). More specifically, besides performing masked reconstruction, the contextual representation of each corrupted input is also enforced to be more predictive to the latent representations of the other view via using InfoNCE loss. We found that encouraging the consistency between representations of different corrupted versions of the same input can make the representations more useful for downstream phoneme and character recognition tasks.
\end{itemize}

\item[2] \textbf{Good prior distributions can help autoencoding representation learning.} Another unique contribution of this work is using ``prior updating (or posterior-as-prior)" to benefit representation learning. 
%Long words in short, prior updating uses approximate posterior of one sample to serve as the sample-specific prior instead of $\mathcal{N}(0,I)$. 
%We found that prior updating clearly help with optimization, i.e., able to learn representations that achieve better downstream PER on phoneme recognition tasks. For example, without using prior updating, VCCA learns worse (in terms of downstream task PER) acoustic representation using consecutive $31$ frames as input than using $15$ frames; In the contrary, when extending VCCA with prior updating, VCCA is capable to achieve superior performance using consecutive $31-$frame input.
%Similarily, when training a speech character recognizer jointly with vanilla RecRep (no prior updating), the CER is even higher than the baseline (supervised recognizer without auxiliary loss). However, RecRep with prior updating changes the story -- The character recognizer jointly trained with RecRep with prior updating steadily outperforms the baseline. 
Though learning a good prior distribution is crucial for sequence generation, it is less emphasized and explored in representation learning research. Our pioneering study (in 2017 and 2018) shows the potential of using good prior distributions for learning higher-quality audio representations, and we believe it could also benefit other learning approaches for representation learning.

\item[3] \textbf{Exploring bidirectional contextual encoder for learning speech representations.} This work is also pioneering in that it explores how to learn sequential speech representations using a bidirectional encoder using an autoencoding approach. 
At the time this work was done, 
most works on self-supervised audio representation learning in the literature were using uni-directional encoders. We found:
%, wherein CPC ~\citep{oord2018representation}, APC ~\citep{Chung2019AnUA}, Deep InfoMax ~\citep{hjelm2018learning} and Wav2vec ~\citep{schneider2019wav2vec} are among the most successful ones.
%Unlike the four works which either rely on contrastive learning or autoregressive prediction, we aim at summarizing the whole input surface feature as a sequence of representations, where the representation of time step $t$ (e..g, $z_t$) is used to reconstruct the surface feature $x_t$ or a few consecutive time steps centered at $t$. 
%We found this approach works well only when the RecRep and downstream recognizer are trained jointly and only share one bidirectional layer. Doing a reflection on our ablation studies, we have two hypothesis. 
\begin{itemize}
\item[a.] \textbf{Posterior Collapse.} Basically, the deep bidirectional encoder is too powerful while reconstructing a single frame $x_t$ is not challenging enough. Thus, this sequential variational model spends too much effort on minimizing the KL term and suffers from the common challenge, posterior collapse, when optimizing a variational model. Especially, vanilla RecRep uses $\mathcal{N}(0,I)$ as a prior for latent variable $z_t$ of time step $t$; This makes the learned representations not useful for downstream recognition tasks.
We hypothesize the posterior collapse issue can be alleviated when we encourage RecRep to reconstruct a more complex target (e.g., a few time steps centered at $t$) instead of $x_t$, or when we use our posterior-as-prior method to replace $\mathcal{N}(0,I)$ by a more informative sample-specific prior. 
We jointly train RecRep (with and without the aforementioned treatments) with a speech recognizer. 
We see that the joint training is helpful when and only when one or more of the treatments are applied to RecRep.
However, we haven't done further experimental study in self-supervised learning scenario. 
\item[b.] \textbf{High-level layers may not be good choices for CTC recognizer.} Another barrier that prevents RecRep to help downstream tasks could be ``we are using the wrong layer". In our self-supervised learning experiments, we pre-train a two-layer bidirectional LSTMs using RecRep, and then an extra bidirectional LSTM followed by CTC loss is added on top of the pre-trained two-layer LSTMs, and fine-tuning is conducted. 
%We have similar setup in our semi-supervised learning setup, where the only difference is that RecRep and the recognizer are trained jointly in semi-supervised learning. 
We realized that CTC loss and per-frame reconstruction loss favor different representations; Thus we hypothesize that RecRep's low-level representation layers may better help with downstream CTC recognizer. 
% when RecRep and CTC recognizer only share low-level representation layers.
Actually, as we mentioned earlier, we did verify this hypothesis in Chapter ~\ref{cha:recurrent} in supervised learning with auxiliary task setup. 

\end{itemize}
\end{itemize}

\section{Future work}

We have summarized the unique contributions of this work. We now list a few possible future directions.

\begin{itemize}
\item[1] \textbf{Model architecture.} At the time this work was done, LSTMs were still the most popular model architecture in the field of audio representation learning. 
In this work, we use stacked bidirectional LSTM as the encoder of RecRep and its variants. 
We also found that having one layer of the bidirectional LSTM be shared by both reconstruction and CTC loss,
leaving the other layers to be private to CTC loss, produced the best results in supervised learning with an auxiliary loss, as shown in Section ~\ref{sec:reducing-conflict}.
%However, this observation does not necessarily imply that shallow RecRep can also learn representations more suitable for speech recognition tasks than deeper models in self-supervised learning.
One could for instance explore deeper stacked LSTM encoders and Transformer-based encoders for RecRep and its variants,
and perform ablation studies to understand how different layers (or blocks) can benefit downstream speech recognition tasks.
\item[2] \textbf{More on variational models.} We have shown that a pre-trained variational model can outperform its non-variational counterparts. 
There are at least two possible directions that deserve further investigation. 
\begin{itemize}
\item[a.] Our experiments on comparing variational and non-variational models were done using relatively small data set(s) (e.g., XRMB and TIMIT). 
It would be interesting to thoroughly compare variational and non-variational models using much larger scale datasets and see if the conclusions still hold.
\item[b.] One could also investigate variational methods for autoregressive models, which is an unexplored direction.
All of the techniques that we have discussed in Chapter ~\ref{cha:recurrent} could be applied towards this new direction.
\end{itemize}
\item[3] \textbf{More on data priors and reconstruction target.} One could also revisit the ``self prior updating" approach for RecRep (see Section ~\ref{subsec:self-prior-updating}), especially in an unsupervised learning scenario. 
%As we discussed in Section ~\ref{subsec:self-prior-updating},
%it is possible to enhance approximate prior either through better heuristics or by better approximating the marginal of all posteriors (e.g., by normalizing flow as described in Section ~\ref{subsec:appendix_flow}).
%In Section ~\ref{subsubsec:prior},
%we also proposed to use a Gaussian mixture of posteriors of contiguous frames as the prior for learning new posteriors.
%That is, we can use a mixture distribution with components $\{ q_{\phi}(z_{t-K}|h_{t-K}), \allowbreak ..., q_{\phi}(z_{t}|h_t), \allowbreak ..., q_{\phi}(z_{t+K}|h_{t+K})\}$ as the new prior for learning posterior $q_{\phi}(z_{t}|h_t)$ in the later optimization.
%As each acoustic unit typically spans a couple of frames (or even longer), this kind of mixture prior could potentially help RecRep become more aware of the duration of the underlying acoustic unit. \par
Because the latent variable $z_t$ encodes not only information needed for reconstructing frame $x_t$ but also the contextual information,
one could also explore using a more complex reconstruction target in RecRep for unsupervised acoustic feature learning.
Intuitively,
$z_t$ should encode more information for frames near time step $t$ than for frames far away from $t$.
It may make sense to design a per-time-step reconstruction target to be a weighted window centered at the current time step,
as we proposed in Table ~\ref{tab:reconstruction-target}.
%\item[4] \textbf{More on masked contrastive predictive coding. } One possible direction is to understand what prevents the proposed ``masked contrastive predictive coding" from learning good representations for speech recognition tasks.
%As we described in Section ~\ref{sec:mask},
%although ``masked contrastive predictive coding" does learn a good initialization for fine-tuning phone-based and character-based CTC systems on WSJ,
%the fine-tuned models are not comparable to those pre-trained using ``masked reconstruction" (~\citep{wang2020unsupervised}).
%One could conduct ablation studies to understand why this happens.
%In fact, in the recently proposed pre-training method wav2vec 2.0 ~\citep{baevski2020wav2vec}, 
%one loss (Equation (3)) also takes masked latent representations as input and solves a contrastive task similar to our ``masked contrastive predictive coding".
%However, their contrastive task is defined over a quantized codebook rather than continuous representations.
\end{itemize}

%----------------------------------------------------------------------------------------
%	Bib
%----------------------------------------------------------------------------------------
\newpage
\bibliography{mybib}
\bibliographystyle{plainnat}
%\bibliographystyle{abbrv}

%----------------------------------------------------------------------------------------
%      Appendix
%----------------------------------------------------------------------------------------
\chapter{Appendix}
\label{cha:appendix}

%\section{Feedforward Unsupervised Acoustic Feature Learning on TIMIT}
%\label{sec:TIMIT}

%In Table ~\ref{tab:compare-framework-timit}, 
%we present the experimental results of unsupervised acoustic feature learning experiments on TIMIT mentioned
%in Section ~\ref{sec:data} of the main text. 
%We follow the standard train/dev/test split for TIMIT, 
%and use all $3696$ training utterance for unsupervised feature learning
%and $400$ dev utterances for early stopping in feature learning. 
%The learned inference networks are then used to transform all the 3696/400/192 train/dev/test utterances into new features before training a speech recognizer. 
%The speech recognizer training still uses the 3696/400/192 train/dev/test utterances.\par

%%%%%%%%%%%%%%%%%%%%%%%%%%%%%%%%%%%%%%%%%%%%%%%%%%%%%%%%%%%%%%%%%%%%%%%%%%%%%%%%%%%%%%%%%%%%%%%%%%%%%%%%%%%%%%%%%%%%%%%%%%%%%%%%%%%%%%%%%%%%%%%%%%

\section{Effect of $\alpha$s in Label Embedding}
\label{sec:alpha-effect}

% To understand the effect of hyper-parameter $\alpha_1$ on the performance of 
The label embedding method (see Equation ~\eqref{eqn:lb-loss} and Figure ~\ref{fig:lb} in the main text) has two important hyper-parameters $\alpha_1$ and $\alpha_2$.
In order to understand how the two hyper-parameters would affect the performance of label embedding,
we choose different combinations of $\alpha_1$ and $\alpha_2$, 
and experiment on phone classification task using XRMB and TIMIT respectively.
We choose $\alpha_1$ from among \allowbreak$(0.0, 0.1, 0.3, 0.5, 0.7, 0.9, 0.99, 1.0)$, and set $\alpha_2 = \frac{1-\alpha_1}{2}$.
For this ablation study, the similarity loss we used is CCA loss. \par
%we vary the trade-off hyper-parameter $\alpha_1$ from $0$ to $1.0$, and set $\alpha_2 = \frac{1-\alpha_1}{2}$) to check how $\alpha_1$ affects the phone classification performance on XRMB and TIMIT respectively. \
%Note that the similarity loss we used is CCA loss. 
The trends on XRMB and TIMIT are consistent. 
When $\alpha_1=0$, we are not training the discriminative network, and thus the performance is terrible. 
When $\alpha_1=1$, we are training the discriminative model only, which is our baseline. 
The trend seems to be: When increasing the $\alpha_1$ from $0$, 
the joint model shown in the Figure ~\ref{fig:lb} gets better and better performance.
When $\alpha_1$ gets very close to $1$ (e.g., $0.99$), 
the label embedding achieves the best performance, 
and the performance would worsen as further increasing $\alpha_1$.

\begin{figure*}[t]

\begin{minipage}{1.0\textwidth}
\begin{subfigure}{1.0\textwidth}
  \centering
  \includegraphics[width=0.70\textwidth]{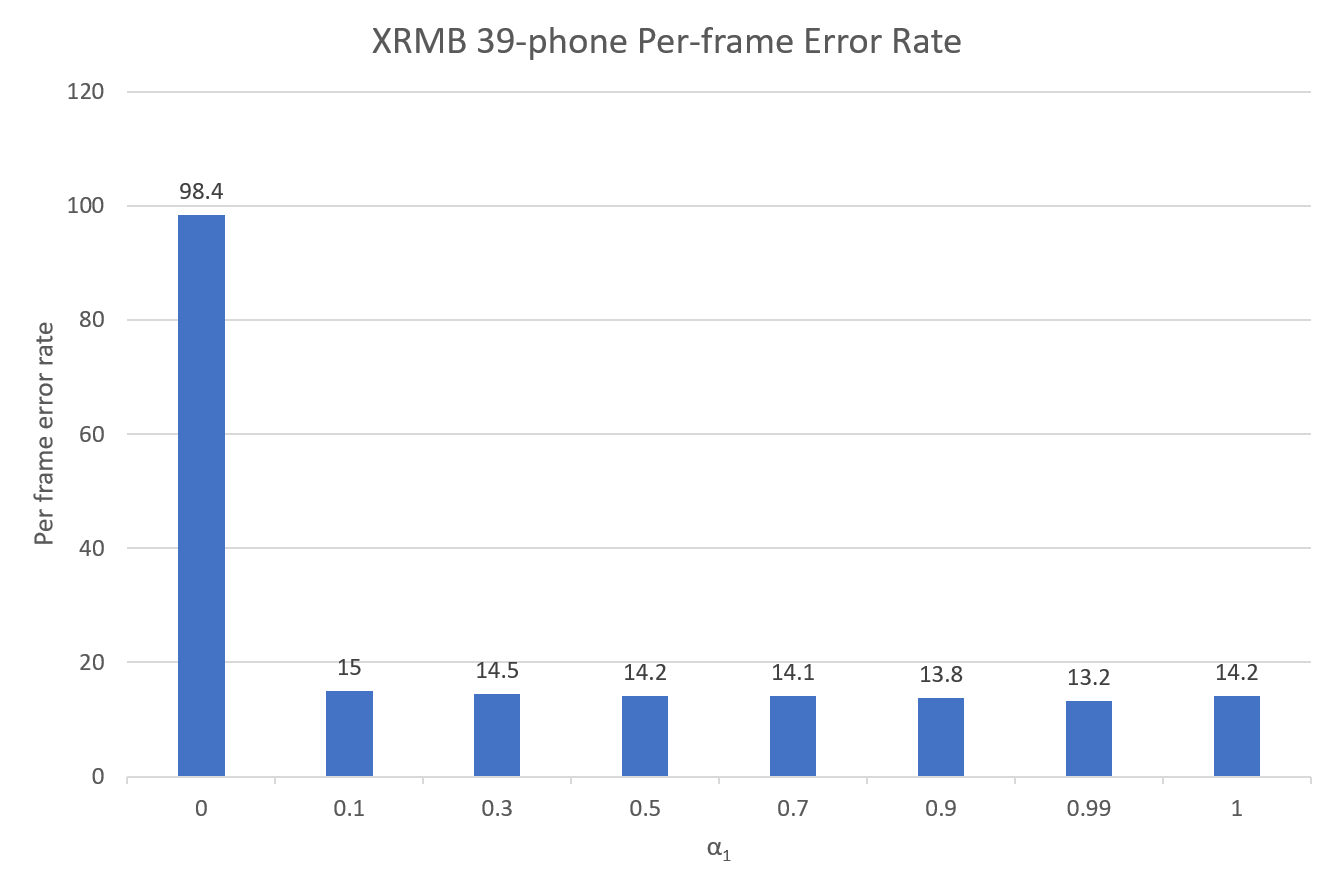}
  \caption{The effect of $\alpha = \alpha_1 + \alpha_2 $ on label embedding method on XRMB. }
  \label{fig:xrmb_alpha}
\end{subfigure}
\end{minipage}

\begin{minipage}{1.0\textwidth}
\begin{subfigure}{1.0\textwidth}
  \centering
  \includegraphics[width=0.70\textwidth]{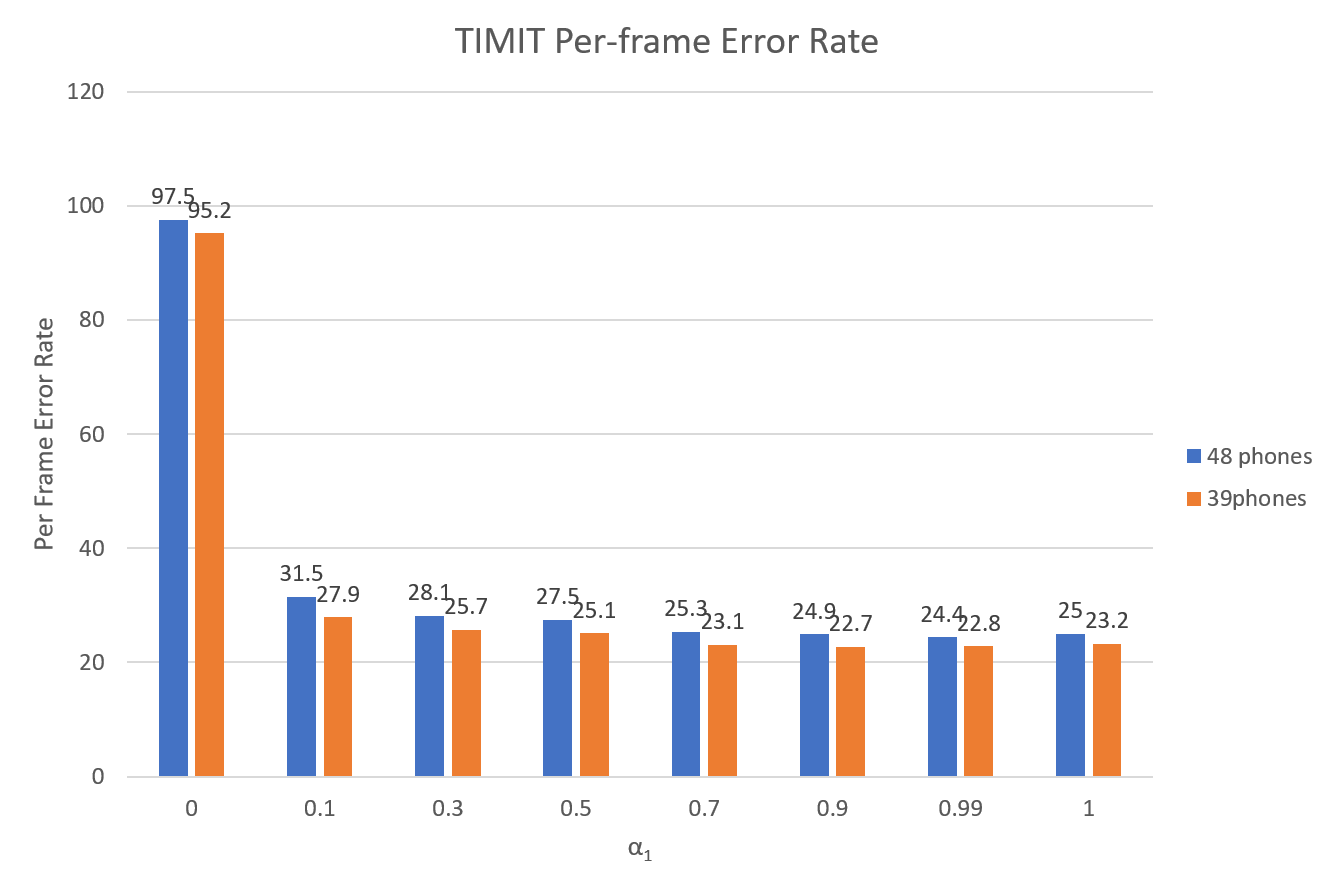}
  \caption{The effect of $\alpha = \alpha_1 + \alpha_2$ on label embedding method on TIMIT. }
  \label{fig:timit_alpha}
\end{subfigure}
\end{minipage}

\end{figure*}

%%%%%%%%%%%%%%%%%%%%%%%%%%%%%%%%%%%%%%%%%%%%%%%%%%%%%%%%%%%%%%%%%%%%%%%%%%%%%%%%%%%%%%%%%%%%%%%%%%%%%%%%%%%%%%%%%%%%%%%%%%%%%%%%%%%%%%%%%%%%%%%%%%

\section{Derivation of ELBOs}

\subsection{ELBO of $\log{p_{\theta}(x)}$}
\label{sec:derive-elbo}

\begin{eqnarray}
\log p_{\theta}(x) &=& \log \frac{p_{\theta}(x|z)p_{\theta}(z)}{p_{\theta}(z|x)} \nonumber \\
&=& \log \frac{p_{\theta}(x|z)p_{\theta}(z)q_{\phi}(z|x)}{q_{\phi}(z|x)p_{\theta}(z|x)} \nonumber \\
&=& \mathbb{E}_{q_{\phi}(z|x)} \log \frac{p_{\theta}(x|z)p_{\theta}(z)q_{\phi}(z|x)}{q_{\phi}(z|x)p_{\theta}(z|x)} \nonumber \\
&=& \mathbb{E}_{q_{\phi}(z|x)} \bigg\{ \log \frac{q_{\phi}(z|x)}{p_{\theta}(z|x)} + \log \frac{p_{\theta}(z)}{q_{\phi}(z|x)} + \log p_{\theta}(x|z) \bigg\} \nonumber \\
&=& \mathbb{E}_{q_{\phi}(z|x)} \big\{ \log p_{\theta}(x|z) \big\} - \KL{q_{\phi}(z|x)}{p_{\theta}(z)} + \KL{q_{\phi}(z|x)}{p_{\theta}(z|x)} \nonumber \\
&\geq & \mathbb{E}_{q_{\phi}(z|x)} \big\{ \log p_{\theta}(x|z) \big\} - \KL{q_{\phi}(z|x)}{p_{\theta}(z)}
\label{eqn:elbo-derive}
\end{eqnarray}

%--------------------------------------------------------------------------------------------------------------------------------------------------------------------------------------------------------------------------------------------------------------------------------------------------------------------------------------------------------%

\subsection{ELBO of $\log{p_{\theta}(x,y)}$}
\label{sec:derive-elbo-vcca}

\begin{eqnarray}
\log p_{\theta}(x,y) &=& \log \frac{p_{\theta}(x,y|z)p_{\theta}(z)}{p_{\theta}(z|x,y)} \nonumber \\
&=& \log \frac{p_{\theta}(x,y|z)p_{\theta}(z)q_{\phi}(z|x)}{q_{\phi}(z|x)p_{\theta}(z|x,y)}  \label{eqn:depend-on-single-view} \\
&=& \mathbb{E}_{q_{\phi}(z|x)} \log \frac{p_{\theta}(x,y|z)p_{\theta}(z)q_{\phi}(z|x)}{q_{\phi}(z|x)p_{\theta}(z|x,y)} \nonumber \\
&=& \mathbb{E}_{q_{\phi}(z|x)} \log \frac{p_{\theta}(x|z)p_{\theta}(y|z)p_{\theta}(z)q_{\phi}(z|x)}{q_{\phi}(z|x)p_{\theta}(z|x,y)} \label{eqn:factorization-on-gm} \\
&=& \mathbb{E}_{q_{\phi}(z|x)} \bigg\{ \log \frac{q_{\phi}(z|x)}{p_{\theta}(z|x,y)} + \log \frac{p_{\theta}(z)}{q_{\phi}(z|x)} + \log p_{\theta}(x|z) + \log p_{\theta}(y|z) \bigg\} \nonumber \\
&=& \mathbb{E}_{q_{\phi}(z|x)} \big\{ \log p_{\theta}(x|z) \big\} + \mathbb{E}_{q_{\phi}(z|x)} \big\{ \log p_{\theta}(y|z) \big\} - \KL{q_{\phi}(z|x)}{p_{\theta}(z)} + \KL{q_{\phi}(z|x)}{p_{\theta}(z|x,y)} \nonumber \\
&\geq & \mathbb{E}_{q_{\phi}(z|x)} \big\{ \log p_{\theta}(x|z) \big\} + \mathbb{E}_{q_{\phi}(z|x)} \big\{ \log p_{\theta}(y|z) \big\} - \KL{q_{\phi}(z|x)}{p_{\theta}(z)}
\label{eqn:elbo-derive-vcca}
\end{eqnarray}

At test time, we sometimes only have access to one view. Thus, we learn the approximate posterior only using $x$, as $q_{\phi}(z|x)$, which leads to Equation ~\eqref{eqn:depend-on-single-view}. According to the graphical model shown in Figure ~\ref{fig:vccap}, we have $p_{\theta}(x,y|z)=p_{\theta}(x|z)p_{\theta}(y|z)$, which leads to Equation ~\eqref{eqn:factorization-on-gm}.

%--------------------------------------------------------------------------------------------------------------------------------------------------------------------------------------------------------------------------------------------------------------------------------------------------------------------------------------------------------%

\subsection{ELBO of $\log{p_{\theta}(x,y)}$ when having private latent variables}
\label{sec:derive-elbo-vccap}

We have shared latent variable $z$, and private variable $h_1$ of $x$, and private variable $h_2$ of $y$. The marginal $p_{\theta}(x,y)$ thus can be written as

\begin{equation}
p_{\theta}(x,y) = \int_{z,h_1,h_2} p_{\theta}(x,y,z,h_1,h_2)
\label{eqn:vccap_marginalization}
\end{equation}

According to the graphical model shown in Figure ~\ref{fig:vccap}, we have 

\begin{eqnarray}
p_{\theta}(x,y,z,h_1,h_2) &=& p_{\theta}(x,y|z,h_1,h_2) p_{\theta}(z, h_1,h_2) \nonumber \\
&=& p_{\theta}(x|z,h_1)p_{\theta}(y|z,h_2)p_{\theta}(z)p_{\theta}(h_1)p_{\theta}(h_2)
\label{eqn:vccap_factorization}
\end{eqnarray}

\begin{equation}
q_{\phi}(z,h_1,h_2|x,y) = q_{\phi}(z|x)q_{\phi}(h_1|x)q_{\phi}(h_2|y)
\label{eqn:inference_factorization}
\end{equation}

\begin{eqnarray}
\log p_{\theta}(x,y) &=& \log \frac{p_{\theta}(x,y|z,h_1,h_2)p_{\theta}(z)p_{\theta}(h_1)p_{\theta}(h_2)}{p_{\theta}(z,h_1,h_2|x,y)} \nonumber \\
&=& \log \frac{p_{\theta}(x,y|z,h_1,h_2)p_{\theta}(z)p_{\theta}(h_1)p_{\theta}(h_2)q_{\phi}(z,h_1,h_2|x,y)}{q_{\phi}(z,h_1,h_2|x,y)p_{\theta}(z,h_1,h_2|x,y)}  \nonumber \\
&=& \mathbb{E}_{q_{\phi}(z,h_1,h_2|x,y)} \log \frac{p_{\theta}(x,y|z,h_1,h_2)p_{\theta}(z)p_{\theta}(h_1)p_{\theta}(h_2)q_{\phi}(z,h_1,h_2|x,y)}{q_{\phi}(z,h_1,h_2|x,y)p_{\theta}(z,h_1,h_2|x,y)}  \nonumber \\
&=& \mathbb{E}_{q_{\phi}(z,h_1,h_2|x,y)} \bigg\{ \log \frac{q_{\phi}(z,h_1,h_2|x,y)}{p_{\theta}(z,h_1,h_2|x,y)} + \log \frac{p_{\theta}(z)p_{\theta}(h_1)p_{\theta}(h_2)}{q_{\phi}(z,h_1,h_2|x,y)} + \log p_{\theta}(x,y|z,h_1,h_2) \bigg\} \nonumber \\
&\geq& \mathbb{E}_{q_{\phi}(z,h_1,h_2|x,y)} \bigg\{ \log p_{\theta}(x,y|z,h_1,h_2) \bigg\} -\KL{q_{\phi}(z,h_1,h_2|x,y)}{p_{\theta}(z)p_{\theta}(h_1)p_{\theta}(h_2)} \nonumber \\
&=& \mathbb{E}_{q_{\phi}(z|x)q_{\phi}(h_1|x)} \bigg\{ \log p_{\theta}(x|z,h_1) \bigg\} + \mathbb{E}_{q_{\phi}(z|x)q_{\phi}(h_2|y)} \bigg\{ \log p_{\theta}(y|z,h_2) \bigg\} \nonumber \\
&& -\KL{q_{\phi}(z|x)}{p_{\theta}(z)} -\KL{q_{\phi}(h_1|x)}{p_{\theta}(h_1)} -\KL{q_{\phi}(h_2|x)}{p_{\theta}(h_2)}
\label{eqn:elbo-derive-vccap}
\end{eqnarray}

%--------------------------------------------------------------------------------------------------------------------------------------------------------------------------------------------------------------------------------------------------------------------------------------------------------------------------------------------------------%

\subsection{ELBO of $\log{p_{\theta}(x,y)}$ when using sample specific priors}
\label{subsec:vccap-prior-updating}

In Section ~\ref{sec:pu}, 
we described a technique called ``prior updating".
Given a context window $\left\{x_{t-K},\cdots,x_t,\cdots,x_{t+K} \right\}$,
denote the learned posterior of this window as $q(z|x_{t-K:t+K})$,
when learning from a larger context window $\left\{ x_{t-W},\cdots,x_t,\cdots,x_{t+W} \right\}$ ($W>K$),
we could use $q(z|x_{t-K:t+K})$ as the prior specific to this $2W+1$ window, instead of using $\mathcal{N}(0,I)$.
In the main text, 
we showed the modified ELBO for VCCA, 
below is the ELBO for VCCAP (with private latent variables) when using ``prior updating"

\begin{eqnarray}
&& \mathbb{E}_{q_{\phi} \left( z|x_{t-W:t+W} \right)  q_{\phi} \left( h_1|x_{t-W:t+W} \right)} \Big\{ \log{p_{\theta} \left( x_{t-W:t+W}|z,h_1 \right) }  \Big \} \nonumber \\
&+& \mathbb{E}_{q_{\phi} \left( z|x_{t-W:t+W} \right)  q_{\phi} \left( h_2|y_{t-W:t+W} \right)} \Big\{ \log{p_{\theta} \left( y_{t-W:t+W}|z,h_2 \right) }  \Big \} \nonumber \\
&-& \KL{q_{\phi}(z|x)}{q \left( z|x_{t-K:t+K} \right) } \nonumber \\
&-& \KL{q_{\phi}(h_1|x)}{q \left( h_1|x_{t-K:t+K} \right) } \nonumber \\
&-& \KL{q_{\phi}(h_2|x)}{q \left( h_2|y_{t-K:t+K} \right) }
\label{eqn:elbo-vccap-prior-updating}
\end{eqnarray}

%%%%%%%%%%%%%%%%%%%%%%%%%%%%%%%%%%%%%%%%%%%%%%%%%%%%%%%%%%%%%%%%%%%%%%%%%%%%%%%%%%%%%%%%%%%%%%%%%%%%%%%%%%%%%%%%%%%%%%%%%%%%%%%%%%%%%%%%%%%%%%%%%%

\section{Multiview variational information bottleneck}
\label{sec:multi-vib}

The graphical model is shown in Figure ~\ref{fig:vcca}. We prove that the Equation ~\eqref{eqn:vib-lowerbound} defined below.

\begin{equation}
\mathbb{E}_{p(x,y)} \Bigg \{  \mathbb{E}_{q_{\phi}(z|x)} \Big\{  \log{\big[ p_{\theta}(x,y|z) \big]}  \Big\}  - \beta \KL{q_{\phi}(z|x)}{p_{\theta}(z)} \Bigg\}
\label{eqn:vib-lowerbound}
\end{equation}

is the lower-bound of

\begin{equation}
I_{\phi}(Z;X,Y) - \beta I_{\phi}(Z;X)
\end{equation}

First, the derivation below shows that Equation ~\eqref{eqn:mutual-lower} is one lower bound of $I_{\phi}(Z;X,Y)$, the mutual information between $Z$ and $(X,Y)$.

\begin{eqnarray}
I_{\phi}(Z;X,Y) &=& \int q_{\phi}(x,y,z)\log \big[\frac{q_{\phi}(x,y,z)}{q_{\phi}(x,y)q_{\phi}(z)} \big] {dzdxdy} \nonumber \\
&=& \int q_{\phi}(x,y,z)\log \big[\frac{q_{\phi}(x,y,z)}{q_{\phi}(z)} \big]{dzdxdy} - \int q_{\phi}(x,y,z)\log \big[ q_{\phi}(x,y) \big] {dzdxdy} \nonumber \\
&=& \int q_{\phi}(x,y,z)\log \big[\frac{q_{\phi}(x,y,z)}{q_{\phi}(z)} \big] {dzdxdy} - \int q_{\phi}(x,y)\log \big[ q_{\phi}(x,y) \big] {dxdy} \nonumber \\
&:=& \int q_{\phi}(x,y,z)\log \big[\frac{q_{\phi}(x,y,z)}{q_{\phi}(z)} \big] {dzdxdy} - \int p(x,y)\log \big[ p(x,y) \big] {dxdy} \nonumber \\
&=& \int q_{\phi}(x,y,z)\log \big[\frac{q_{\phi}(x,y,z)}{q_{\phi}(z)} \big] {dzdxdy} + H(x,y) \nonumber \\
&:=& H(x,y) + \int \Bigg\{ p(x,y) \int \Big\{ q_{\phi}(z|x) \log \big[\frac{q_{\phi}(x,y,z)}{q_{\phi}(z)} \big] dz \Big\} dxdy \Bigg\} \nonumber \\
&\geq& H(x,y) + \int \Bigg\{ p(x,y) \int \Big\{ dz q_{\phi}(z|x) \log \big[ p_{\theta}(x,y|z) \big] \Big\} dxdy \Bigg\} \nonumber \\
&\geq& \int \Bigg\{ p(x,y) \int \Big\{ dz q_{\phi}(z|x) \log \big[ p_{\theta}(x,y|z) \big] \Big\} dxdy \Bigg\} \nonumber \\
&=& \mathbb{E}_{p(x,y)} \Bigg \{  \mathbb{E}_{q_{\phi}(z|x)} \Big\{  \log{\big[ p_{\theta}(x,y|z) \big]}  \Big\}  \Bigg\} \label{eqn:mutual-lower}
\end{eqnarray}

Due to the fact that $p_{\phi}(z)$ = $\arg min_{p_{\theta}}\big( \frac{q_{\phi}(x,z)}{q_{\phi}(x)p_{\theta}(z)}\big)$, Equation ~\eqref{eqn:bottle-lower} defined below is one lower bound of $-\beta I_{\phi}(Z;X)$.

\begin{eqnarray}
-\beta I_{\phi}(Z;X) &:=& -\beta \int \bigg\{ q_{\phi}(x,z) \log \big[ \frac{q_{\phi}(x,z)}{q_{\phi}(x)q_{\phi}(z)} \big] dxdz \bigg\} \nonumber \\
&\geq& -\beta \int \bigg\{ q_{\phi}(x) q_{\phi}(z|x) \log \big[ \frac{q_{\phi}(x,z)}{q_{\phi}(x)p_{\theta}(z)} \big] dzdx \bigg\} \nonumber \\
&=& -\beta \int \bigg\{ p(x) q_{\phi}(z|x) \log \big[ \frac{q_{\phi}(z|x)}{p_{\theta}(z)} \big] dz dx \bigg\} \nonumber \\
&-& -\beta \mathbb{E}_{p(x)} \Bigg \{  \KL{q_{\phi}(z|x)}{p_{\theta}(z)}  \Bigg \}\label{eqn:bottle-lower}
\end{eqnarray}

Thus, combining Equation ~\eqref{eqn:mutual-lower} and ~\eqref{eqn:bottle-lower}, we complete the proof. \par

%%%%%%%%%%%%%%%%%%%%%%%%%%%%%%%%%%%%%%%%%%%%%%%%%%%%%%%%%%%%%%%%%%%%%%%%%%%%%%%%%%%%%%%%%%%%%%%%%%%%%%%%%%%%%%%%%%%%%%%%%%%%%%%%%%%%%%%%%%%%%%%%%%

\section{Losses towards enhancing similarity of two vectors}
\label{sec:similarity}

In this section, we present the different choices of similarity losses of the model illustrated in Figure ~\ref{fig:lb}. \par

If we use cosine distance loss, then the $L_{similarity}$ is written as
\begin{equation}
L_{similarity}(\mu, \hat{\mu}) = -\frac{\mu \cdot \hat{\mu} }{\norm{\mu}\norm{\hat{\mu}}}
\label{eqn:cosine}
\end{equation}

If we use contrastive loss, then $L_{similarity}$ is written as
\begin{equation}
L_{similarity}(\mu, \hat{\mu}) = \max \left\{ \big\{ \frac{ \mu \cdot \hat{\mu}' }{\norm{\mu}\norm{ \hat{\mu}' } } - \frac{\mu \cdot \hat{\mu} }{\norm{\mu}\norm{\hat{\mu}}} + m \big\} , 0 \right\}
\label{eqn:contrastive}
\end{equation}

Where $\hat{\mu}'$ is a negative sample (e.g., representation learned using a permutation of $l_{1:15}$), and the margin $m$ is a non-negative number.

Now we discuss CCA loss. Assume the batch size is $N$, and representation we obtained are $E_1^{N\times d}$ and $E_2^{N \times d}$ respectively, where $d$ is the dimension of the representation. Then the CCA loss is

\begin{equation}
L_{similarity} = -\frac{\trace{E_1^T E_2}}{N}
\end{equation}

with respect to
\begin{eqnarray}
\frac{E_1^T E_1}{N} + r_x I_{d\times d} &=& I_{d \times d} \nonumber \\
\frac{E_2^T E_2}{N} + r_y I_{d\times d} &=& I_{d \times d} \nonumber
\end{eqnarray}
Where $r_x$ and $r_y$ are typically small positive numbers.
Strictly speaking, the CCA loss implemented here is not the CCA loss presented in the paper ~\citep{andrew2013deep}. We need to introduce one Lagrangian multiplier $\lambda$ to incorporate the two constraints when applying this CCA loss. \par

%%%%%%%%%%%%%%%%%%%%%%%%%%%%%%%%%%%%%%%%%%%%%%%%%%%%%%%%%%%%%%%%%%%%%%%%%%%%%%%%%%%%%%%%%%%%%%%%%%%%%%%%%%%%%%%%%%%%%%%%%%%%%%%%%%%%%%%%%%%%%%%%%%

\section{Hyperparameter Tuning for multitask speech recognition on TIMIT}
\label{sec:tune-on-timit}

We describe our hyper-parameter tuning corresponding to experiments of Section ~\ref{subsubsec:recurrent-asr} in the main text. \par

The latent variable $z$ is either with dimension $120$ or $150$, 
while latent variable $h$ (if used) is of size $30$ or $50$. 
The dropout is chosen from $\{0.3, 0.4, 0.5, 0.6 \}$ and the optimizer is ADAM, 
with initial learning rate $0.0005$, 
and with learning rate decay $0.85$ starting from the $21$ epoch. 
The $\beta$ is selected among $\{0.001, 0.01, 0.1\}$, 
the $\kappa$ is chosen from $\{0, 0.001, 0.01, 0.1, 1.0\}$, 
and the trade-off hyper-parameter $\alpha$ is chosen from $\{0.1, 0.3, 0.5, 0.6\}$. \par

%%%%%%%%%%%%%%%%%%%%%%%%%%%%%%%%%%%%%%%%%%%%%%%%%%%%%%%%%%%%%%%%%%%%%%%%%%%%%%%%%%%%%%%%%%%%%%%%%%%%%%%%%%%%%%%%%%%%%%%%%%%%%%%%%%%%%%%%%%%%%%%%%%

\section{One-layer generative model for multitask learning on TIMIT}
\label{sec:one-for-multitask}

We run RecRep-Pyramid-MT-Onelayer on the $11$ best models of RecRep-Pyramid-MT.
All experimental settings are identical to that for running RecRep-Pyramid-MT. The only difference is that only one recurrent layer is used for representation learning, and the CTC part has two private recurrent layers.
Clear improvements are observed for each of the $11$ settings from Table ~\ref{tab:onelayer-timit}.

\begin{table} [htbp]
\centering
\begin{tabular}{| c | c | c | c | c | r | r |}
 \hline
 $Dim(z)$ & $\alpha$ & $\beta$ & $\kappa$ & dropout & Dev PER ($\%$) $2-$layers & Dev PER ($\%$) $1-$layer \\
 \hline \hline
 150 & 0.5 & 0.01 & 0.1 & 0.4 & 18.2 & 16.9 \\
 \hline
 150 & 0.5 & 0.01 & 1.0 & 0.4 & 18.2 & 17.2 \\
 \hline
 150 & 0.5 & 0.01 & 0.0 & 0.4 & 17.9 & 16.9 \\
 \hline
 250 & 0.5 & 0.01 & 0.1 & 0.4 & 18.1 & 16.9 \\
 \hline
 150 & 0.6 & 0.01 & 0.1 & 0.4 & 18.0 & 16.8 \\
 \hline
 150 & 0.4 & 0.01 & 0.1 & 0.4 & 18.0 & 17.0 \\
 \hline
 150 & 0.3 & 0.01 & 0.1 & 0.4 & 18.0 & 16.9 \\
 \hline
 150 & 0.2 & 0.01 & 0.1 & 0.4 & 18.2 & 17.1 \\
 \hline
 150 & 0.5 & 0.001 & 1.0 & 0.4 & 17.9 & 16.9 \\
 \hline
 150 & 0.5 & 0.01 & 0.0 & 0.4 & 17.8 & 16.7 \\
 \hline
 150 & 0.5 & 0.01 & 1.0 & 0.5 & 17.2 & 17.0 \\
 \hline
\end{tabular}
\caption{We pick up the best $11$ models in our hand for vanilla RecRep-Pyramid-MT (with CTC and generative model share 2 layers).
We run RecRep-Pyramid-MT-Onelayer on all the $11$ models.
The column ``Dev PER ($\%$) $2-$layers'' is the vanilla RecRep-Pyramid-MT results.
The column ``Dev PER ($\%$) $1-$layers'' is the RecRep-Pyramid-MT-Onelayer results.
The last model shown in this table uses $\mathcal{L}_2$ weight $10^{-7}$, 
and all other models do not use $\mathcal{L}_2$ regularization.
$\alpha$ is the weight on the ELBO as shown in Equation ~\eqref{eqn:recrep-joint-loss},
$\kappa$ (introduced in Section ~\ref{sec:basic-model}) is the hpyerparameter we introduced to control how much uncertainty we would use for discriminative task.}
\label{tab:onelayer-timit}
\end{table}

\section{Approximate Prior}
\label{sec:approx}
In Section ~\ref{subsec:self-prior-updating}, we propose ``prior updating" for sequence representation learning. 
As we further described in Section ~\ref{subsubsec:prior},
the simple ``prior updating" heuristically estimates the marginal prior $p'(z)$ for each sample using one Gaussian component of $p'(z)$.
This section described two possible solutions for better approximating $p'(z)$.

\subsection{Multiple Gaussian Components}
\label{subsec:appendix_multiple}

We can assume that there is a set of Gaussian components, 
denoted as $\mathcal{C}=\{ q^{(i)}_{\phi^{(c-1)}}\}$ for $1\leq i \leq |\mathcal{C}|$, 
which are similar enough to $q_{{\phi}}(z_t|h_t)$.
We can then approximately calculate the KL divergence between $q_{{\phi}}(z_t|h_t)$ and $p'(z)$ as

\begin{equation}
\KL{q_{{\phi}}(z_t|h_t)}{\frac{1}{|\mathcal{C}|}\sum_{i=1}^{|\mathcal{C}|} q^{(i)}_{\phi^{(c-1)}}}
\end{equation}

Now the question is how can we easily define this set $\mathcal{C}$.
Current ``prior updating" actually defines $\mathcal{C}=\{q_{{\phi}^{(c-1)}}(z_t|h_t)\}$.
A better solution might be defining $\mathcal{C}$ to be the collection of posteriors of a few consecutive frames,
and calculate the KL divergence between $q_{{\phi}^{(c)}}(z_t|h_t)$ and $p^{\text{window}}(z)$ (Equation ~\eqref{eqn:context-prior}) during optimization.

\begin{eqnarray}
p^{\text{window}}(z) =  \frac{1}{2K+1}\sum_{k=t-K}^{t+K} q_{\phi^{(c-1)}}(z_k|h_k)
\label{eqn:context-prior}
\end{eqnarray}

\subsection{Flow Prior}
\label{subsec:appendix_flow}

It worths to mention that it is possible to more accurately calculate $\log p'(z)$ for each $z \sim q_{\phi^{(c)}}(z_t|h_t)$ via using normalizing flow ~\citep{rezende2015variational, kingma2016improved}.

Normalizing-flow-based deep generative models 
transform a simple probability distribution (e.g., $\mathcal{N}(0,I)$) into a complex distribution by using a sequence of differentiable and invertible mappings.
Simply speaking, consider a smooth and invertible mapping $f: \mathcal{R}^d \rightarrow \mathcal{R}^d$ such that $f^{-1}\circ f (z) = z$. 
The one-step invertible mapping used in normalizing flow can be described as
\begin{equation}
p(z') \equiv p(f(z)) = p(z) \begin{vmatrix*}[r] \det \frac{\partial f^{-1}}{\partial z'} \end{vmatrix*} = p(z) { \begin{vmatrix*}[r] \det \frac{\partial f}{\partial z} \end{vmatrix*} }^{-1}
\label{eqn:change-of-variable}
\end{equation}

Based on Equation ~\eqref{eqn:change-of-variable}, we can obtain a complex density $p_K(z)$ by iteratively (invertiblely) transforming a random variable $z_0 \sim \mathcal{N}(0,I)$,
\begin{equation}
z_K = f_K \circ f_{K-1} \circ ... \circ f_2 \circ f_1(z_0)
\label{eqn:chain}
\end{equation}

And we can calculate the log density for $z_K$ via
\begin{equation}
\log p_K(z_K) = \log p(z_0) - \sum_{k=1}^K \log \det \begin{vmatrix*}[r] \frac{\partial f_k}{\partial z_k} \end{vmatrix*}
\label{eqn:cal-density}
\end{equation}

Coming to our problem, we can learn $K$ invertible transformations, such that for $z_K \sim p'(z)= \mathbb{E}_{x} \Bigg\{ \mathbb{E}_{1 \leq t \leq |x|} \Big\{ q_{{\phi}^{(c-1)}}(z_t|h_t)  \Big\} \Bigg\} $,
we have $z_0 = f_1^{-1} \circ f_{2}^{-1} \circ ... \circ f_{K-1}^{-1} \circ f_K^{-1}(z_K) \sim \mathcal{N}(0,I)$. 
In order to draw samples from $p'(z)$, we first draw one utterance $x$ from $\mathcal{D}$ and draw one time step $t$. 
Then we can draw sample $z_K$ from the posterior $q_{\phi^{(c-1)}}(z_t|h_t)$.  
To learn the transformation $f_1^{-1} \circ f_{2}^{-1} \circ ... \circ f_{K-1}^{-1} \circ f_K^{-1}$, 
we need to enforce the transformed samples (e.g. $z_0$ follows $\mathcal{N}(0,I)$). \par

Given the $K$ invertible transformations, we can analytically calculate $\KL{q_{{\phi}}(z_t|h_t)}{p'(z)}$ as below:
\begin{itemize}
\item[1] We first sample $z_K \sim q_{{\phi}^{(c)}}(z_t|h_t)$. For each $z_K$, we can use the $K$ invertible transformations to invert $z_K$ to $z_0$ via $z_0 = f_1^{-1} \circ f_{2}^{-1} \circ ... \circ f_{K-1}^{-1} \circ f_K^{-1}(z_K)$
\item[2] For the $z_0$, calculate its density $p(z_0)$. Here we typically use $p(z)=\mathcal{N}(0,I)$.
\item[3] We can then calculate $\log p'(z_K) \equiv \log p_K(z_K) = \log p(z_0) - \sum_{k=1}^K \log \det \begin{vmatrix*}[r] \frac{\partial f_k}{\partial z_k} \end{vmatrix*} $
\item[4] The $\KL{q_{{\phi}}(z_t|h_t)}{p'(z)}$ then can be calculated analytically.
\end{itemize}

Besides providing a theoretical framework to explain and generalize ``prior updating'', ``flow prior" also has the potential advantage of delivering ``global regularization" (as mentioned in ~\citep{hoffman2016elbo}).
As we can see, $p'(z)$ is a mixture of many Gaussian components. 
When we do ``prior updating", we use one or a few Gaussian components to guide the optimization heuristically and thus can overly ignore the global structure of the prior. \par

\end{document}